\documentclass[a4paper,11pt]{article}
\usepackage{jheppub} % for details on the use of the package, please
\usepackage[T1]{fontenc}
\usepackage[utf8]{inputenc}
\usepackage[english]{babel}
\usepackage{amsmath,bbm}
\usepackage{bm}
\usepackage{amsfonts}
\usepackage{amssymb}
\usepackage{graphicx}
\usepackage[normalem]{ulem}
\usepackage{tabularx}
\usepackage{slashed}
\usepackage[compat=1.1.0]{tikz-feynman}
\usepackage{tikz}
\usetikzlibrary{decorations.pathreplacing,calligraphy} % Decorations for tikz

\usepackage{braket}

\usepackage{subcaption}
\hypersetup{unicode,psdextra}
\usepackage{scalerel}
\usepackage{mathabx} % \drsh down right arrow for chemical reactions
\title{Charm associated production of the pseudoscalar Higgs boson $h''$
  in the MCPM' at the LHC}

\author[a]{R.~ Maciu{\l}a,}
\author[b]{M.~Maniatis,}
\author[c]{O.~Nachtmann,}
\author[a,d]{A.~Szczurek}
\affiliation[a]{Institute of Nuclear Physics Polish Academy of Sciences, ul. Radzikowskiego 152, PL-31-342 Krak\'ow, Poland}
\affiliation[b]{Centro de Ciencias Exactas,  Universidad del B\'io-B\'io,  Avda. Andres Bello 720,  Chill\'{a}n, Chile}
\affiliation[c]{Institut für Theoretische Physik, Universit\"at Heidelberg,
Philosophenweg 16, D-69120 Heidelberg, Germany}
\affiliation[d]{College of Mathematics and Natural Sciences, University of Rzeszów,
ul. Pigonia 1, PL-35-310 Rzeszów, Poland}

 \abstract{We discuss charm associated production of the pseudoscalar Higgs boson $h''$ in the MCPM'.
In our analysis we assume $m_{h''}$ = 95.4 GeV, which corresponds
to an enhancement observed by the CMS collaboration in the
$\gamma \gamma$ channel. As discussed recently, the MCPM' is consistent with the CMS enhancement.
In this model the $h''$ Higgs boson of this mass decays dominantly
into $c \bar c$ jets. The cross section for the associated $h''$
production in $p p \to h'' c \bar c$ is found to be almost as big 
as that for the Drell-Yan (DY) $c \bar c \to h''$ production, much larger
than that for two-gluon fusion $g g \to h''$.
Since in the MCPM' the dominant $h''$ decay is into the $c \bar c$ channel
we have to deal with the $p p \to c \bar c c \bar c$ process
with $c$ and $\bar c$ jets. In addition to the signal processes,
we have sizeable single parton scattering (SPS) and double parton scattering (DPS) nonresonant background.
In our analysis we include diagrams of gluon-gluon fusion $g g \to h''$
which lead to the $c \bar c c \bar c$ final state.
We discuss also the possible role of the reaction $p p \to h'' h''$.
Assuming 100\% c and $\bar c$ jet tagging we discuss in detail how
to improve the signal-to-background ratio.
Our analysis shows promising resonance-like enhancements
and relatively large cross sections in the $c \bar c c \bar c$
channel within the MCPM'. In an appendix we discuss the total cross section for $h''$ production at the LHC. We also compare distributions for the reactions $p p \to h'' X$ and $p p \to Z X$ with $h''$ and $Z$ decaying to $c\bar c$ and $\mu^{+}\mu^{-}$. We hope that the LHC collaborations will in the future perform
the experimental studies corresponding to the theoretical studies
discussed here.
 }
\begin{document} 

\maketitle
\flushbottom
% =============================================================================
\section{Introduction}
\label{sec:intro}
%=============================================================================
In this paper we are studying consequences of the model MCPM', which was introduced in~\cite{Maniatis:2023aww}. The MCPM' is a two-Higgs-doublet model (2HDM) that, in our view, exhibits several noteworthy features when compared with experimental observations.
 The basis of the MCPM' is the maximally CP symmetric two-Higgs-doublet model, MCPM, which was introduced in~\cite{Maniatis:2007de} and a short description of it is given in~\cite{Maniatis:2010sb}. The MCPM has four generalized CP (GCP) symmetries: see Table~1 of~\cite{Maniatis:2023aww}. These symmetries lead to highly constrained Yukawa interactions. In the strict symmetry limit only the third generation fermions, $\tau$, $t$, $b$, are massive and they couple to the first Higgs doublet $\varphi_1$. The fermions $\mu$, $c$, $s$ of the second family are massless and couple to the second Higgs doublet~$\varphi_2$. The fermions $e$, $u$, $d$ are uncoupled to the Higgs bosons. Phenomenological predictions/estimates in the MCPM have been worked out in~\cite{Maniatis:2007de, Maniatis:2009vp, Maniatis:2009by,Maniatis:2010sb,Brehmer:2012hh}. However, clearly the highly symmetric MCPM can only be an approximate representation of what is observed in nature. 
Therefore, in~\cite{Maniatis:2023aww} a realistic model, the MCPM', was constructed. The key idea there was to generate masses for fermions of the second and first family by effective interactions originating from integrating out high-mass particles having suitable couplings. In this way, masses for all fermions except neutrinos were generated. That is, in the MCPM' we work with massless neutrinos which should be alright for the reactions which we want to study. In addition, electroweak symmetry breaking (EWSB) and a realistic, nontrivial, Cabibbo-Kobayashi-Maskawa (CKM) matrix are generated in the MCPM'. 

In the MCPM', as in any 2HDM, we have five Higgs bosons which we denote as follows:
\begin{align}
& \text{neutral bosons:} \qquad && \rho', h', h''\;,\\
& \text{charged boson pair:} && H^\pm \; .
\end{align}
The $\rho'$ behaves very much like the Standard-Model~(SM) Higgs boson. Therefore, we set~\cite{ParticleDataGroup:2024cfk}
\begin{equation} \label{eq:1.2}
m_{\rho'} = 125.20 \pm 0.11~\text{GeV.}
\end{equation}
The Higgs bosons $h'$ and $h''$ have a scalar and pseudoscalar coupling to fermions, respectively. The main coupling of $h'$ and $h''$ is to charm quarks; see (A.81) of~\cite{Maniatis:2023aww}. The couplings of the Higgs bosons of the MCPM' among themselves and to gauge bosons are as in the MCPM and are listed in appendix~A of~\cite{Maniatis:2009vp}. The Yukawa couplings of the Higgs bosons in the MCPM' are given in appendix~A of~\cite{Maniatis:2023aww}; see (A50)--(A58) and (A81) there. 

In~\cite{Maniatis:2023aww} predictions of the MCPM' were worked out concerning the following reaction at the LHC:
\begin{equation} \label{eq:1.3}
p + p \to (h'' \to \gamma\gamma) + X\;.
\end{equation}
The comparison with the CMS results~\cite{CMS:2023yay}, where a possible $\gamma\gamma$ signal at 95.4~GeV was reported, is quite encouraging. Therefore, in~\cite{Maniatis:2023aww} tentatively the mass of the $h''$ boson was set to
\begin{equation} \label{eq:1.4}
m_{h''} = 95.4 \text{ GeV.}
\end{equation}
In~\cite{Maniatis:2023aww} also the leptonic decays of the mesons $D^\pm$ and $D_s^\pm$ were studied. It was shown that the contributions of the charged Higgs bosons~$H^\pm$ of the MCPM' lead to deviations from the predictions of the Standard Model (SM). From comparison with experiment it was concluded that the $H^\pm$ mass $m_{H^\pm}$ should be around 300~GeV. 

\begin{figure}[!ht]
\centering
\begin{subfigure}{0.45\textwidth}
\centering
\begin{tikzpicture}
  \begin{feynman}
  %place first vertices
    \vertex (i1) at (0,1.5);
    \vertex (i2) at (0,-1.5);
     \vertex (a) at (2.0, 1.5) ;
     \vertex (b) at (2.0, -1.5) ;
     \vertex (c1) at (3, 0.75);
     \vertex (c) at (3, 0);     
     \vertex (c2) at (3, -0.75);
     \vertex (h1) at (5, 0.75)  {\( c \)};
     \vertex (h) at (4,0);
     \vertex (h2) at (5, -0.75) {\( \bar{c} \)};
     \vertex (f1) at (4.5, 1.5)  {\( c \)};
     \vertex (f2) at (4.5, -1.5)  {\( \bar{c} \)};
     \vertex (X1) at (4, 1.8) ;
     \vertex (X1a) at (4, 2.0) ;
     \vertex (X1b) at (4, 1.6) ;
     \vertex (X2) at (4, -1.8) ;
     \vertex (X2a) at (4, -1.6) ;
     \vertex (X2b) at (4, -2.0) ;
    \diagram* {
      (i1) -- [plain, very thick, edge label=\(p\)] (a),
      (i2) -- [plain, very thick,edge label'=\(p\)] (b),
      (a) -- [gluon, thick, edge label'=\(G\)] (c1),
      (b) -- [gluon, thick, edge label=\(G\)] (c2),
      (f2) -- [fermion, thick, label=\(\bar{c}\)] (c2),
      (c2) -- [fermion, thick] (c),
      (c) -- [fermion, thick] (c1),
      (c1) -- [fermion, thick] (f1),
      (c) -- [scalar, thick, edge label=\(h''\)] (h),
      (h) -- [fermion, thick] (h1),
      (h2) -- [fermion, thick] (h),
      (a) -- [plain, thick] (X1),
      (a) -- [plain, thick] (X1a),
      (a) -- [plain, thick] (X1b),
      (b) -- [plain, thick] (X2),      
      (b) -- [plain, thick] (X2a),      
      (b) -- [plain, thick] (X2b),      
    };
   	\draw [fill, gray] (a) circle (.2);
 	\draw [fill, gray] (b) circle (.2);
 	\draw [fill] (c) circle (.05);
  \end{feynman}
\end{tikzpicture}
   \caption{}
    \label{fig:feyn1a}
 \end{subfigure}
  \begin{subfigure}{0.45\textwidth}
\centering
\begin{tikzpicture}
  \begin{feynman}
  %place first vertices
    \vertex (i1) at (0,1.5);
    \vertex (i2) at (0,-1.5);
     \vertex (a) at (2.0, 1.5) ;
     \vertex (b) at (2.0, -1.5) ;
     \vertex (c1) at (3, 0.75);
     \vertex (c) at (3, 0);     
     \vertex (c2) at (3, -0.75);
     \vertex (h1) at (5, 0.75)  {\( c \)};
     \vertex (h) at (4,0);
     \vertex (h2) at (5, -0.75) {\( \bar{c} \)};
     \vertex (f1) at (4.5, 1.5)  {\( c \)};
     \vertex (f2) at (4.5, -1.5)  {\( \bar{c} \)};
     \vertex (X1) at (4, 1.8) ;
     \vertex (X1a) at (4, 2.0) ;
     \vertex (X1b) at (4, 1.6) ;
     \vertex (X2) at (4, -1.8) ;
     \vertex (X2a) at (4, -1.6) ;
     \vertex (X2b) at (4, -2.0) ;
    \diagram* {
      (i1) -- [plain, very thick, edge label=\(p\)] (a),
      (i2) -- [plain, very thick,edge label'=\(p\)] (b),
      (a) -- [gluon, thick, edge label'=\(G\)] (c1),
      (b) -- [gluon, thick, edge label=\(G\)] (c2),
      (f2) -- [fermion, thick, label=\(\bar{c}\)] (c2),
      (c2) -- [fermion, thick] (c),
      (c) -- [fermion, thick] (c1),
      (c1) -- [fermion, thick] (f1),
      (c) -- [gluon, thick, edge label=\(G\)] (h),
      (h) -- [fermion, thick] (h1),
      (h2) -- [fermion, thick] (h),
      (a) -- [plain, thick] (X1),
      (a) -- [plain, thick] (X1a),
      (a) -- [plain, thick] (X1b),
      (b) -- [plain, thick] (X2),      
      (b) -- [plain, thick] (X2a),      
      (b) -- [plain, thick] (X2b),      
    };
   	\draw [fill, gray] (a) circle (.2);
 	\draw [fill, gray] (b) circle (.2);
 	\draw [fill] (c) circle (.05);
  \end{feynman}
\end{tikzpicture}
   \caption{}
    \label{fig:feyn1b}
 \end{subfigure}
\caption{\label{fig:feyn1}
(a): Example of a diagram for charm-associated production of the $h''$ boson which decays into a $c \bar{c}$ pair. (b): Example of a SM diagram contributing to the background.}
\end{figure}
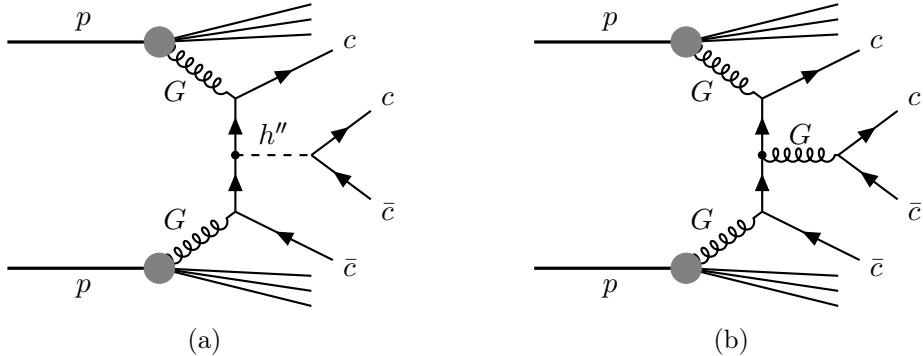

Encouraged by all these findings we make in the present paper a study of the process
\begin{equation} \label{eq:1.5}
p + p \to (h'' \to c + \bar{c}) + c + \bar{c} + X.
\end{equation} 
That is, we consider $h''$ production together with a charm-anticharm quark pair plus a remainder~$X$. 
Note that $h'' \to c + \bar{c}$ is the dominant decay channel of $h''$. The final state in~\eqref{eq:1.5} contains two $c \bar{c}$ pairs. Of course this final state
\begin{equation} \label{eq:1.5b}
p + p \to c + \bar{c} + c + \bar{c} + X
\end{equation} 
can also be produced in the SM. An example of a diagram contributing to~\eqref{eq:1.5} is shown in Fig.~\ref{fig:feyn1a}. In total there are 24 diagrams at tree level for~\eqref{eq:1.5}. An example of a background SM diagram is shown in Fig.~\ref{fig:feyn1b}. Note that in the calculation of the amplitude of~\eqref{eq:1.5b} in the MCPM' we have to take into account all the SM diagrams from QCD and in addition the diagrams which are specific for the MCPM'. 

In the reaction~\eqref{eq:1.5} we have two charm and two anticharm jets. Due to the resonance~$h''$ we expect that in experiment one should find an enhancement in the $c \bar{c}$ mass distribution at $m_{h''}$. Such a study would profit enormously if charm tagging can be realized, a challenging task at the LHC. However, it has been shown to be feasible. ATLAS, for instance, reports a charm-tagging efficiency of about~30\%~\cite{ATLAS:2022qxm}.

\section{Results: generalities and integrated cross sections}
\label{sec:results}
%=============================================================================
%MCPM implementation in MadGraph5
We want to show our results for the calculation of charm-quark
production, $c \bar{c} c \bar{c}$, in proton--proton collisions at the LHC. We compare the predictions of the MCPM' versus those of the SM. We employ the MadGraph5 Monte-Carlo event generator to simulate the events~\cite{Alwall:2011uj}. The MCPM has been implemented in MadGraph5~\cite{Brehmer:2012hh}. We extend this implementation by adding the effective gluon-gluon-$h''$ interaction. The $h''$ boson coupling to the gluons (and photons) is only possible via loops. The calculation of this effective interaction is presented in Appendix~\ref{appA}.

For the process of associated $h''$ charm production we adopt for the center-of-mass energy the current LHC value of 
$\sqrt{s}=$~13.6~TeV. The mass of the pseudoscalar $h''$ boson is chosen to match the possible resonance observed at CMS, that is, $m_{h''} = 95.4$~GeV, \eqref{eq:1.4}. The total width of the $h''$ boson has been calculated in~\cite{Maniatis:2009vp} giving, for the chosen mass value
\begin{equation} \label{eq:1.5c}
\Gamma_{h''} = 5.76 \; \mathrm{GeV}.
\end{equation} 
For the parton densities we use the set MSHT20lo$\!\_\!$as130~\cite{Bailey:2020ooq}, the factorization and the renormalization scales are set to $\mu_F = \mu_R = 95.4$~GeV. 

\begin{figure}[!ht]
\centering
\begin{subfigure}{0.45\textwidth}
\centering
\begin{tikzpicture}
  \begin{feynman}
  %place first vertices
    \vertex (i1) at (0,1.5);
    \vertex (i2) at (0,-1.5);
     \vertex (a) at (2.0, 1.5) ;
     \vertex (b) at (2.0, -1.5) ;
     \vertex (c1) at (3, 0.75);
     \vertex (c) at (3, 0);     
     \vertex (c2) at (3, -0.75);
     \vertex (h) at (4.5, 0.0) {\( \bar{c} \)};
     \vertex (f1) at (4.5, 0.75)  {\( c \)};
     \vertex (f2) at (4.5, -0.75)  {\( h'' \)};
     \vertex (X1) at (4, 1.8) ;
     \vertex (X1a) at (4, 2.0) ;
     \vertex (X1b) at (4, 1.6) ;
     \vertex (X2) at (4, -1.8) ;
     \vertex (X2a) at (4, -1.6) ;
     \vertex (X2b) at (4, -2.0) ;
    \diagram* {
      (i1) -- [plain, very thick, edge label=\(p\)] (a),
      (i2) -- [plain, very thick,edge label'=\(p\)] (b),
      (a) -- [gluon, thick, edge label'=\(G\)] (c1),
      (b) -- [gluon, thick, edge label=\(G\)] (c2),
      (c2) -- [scalar, thick] (f2),
      (c2) -- [gluon, thick, edge label=\(G\)] (c),
      (c) -- [fermion, thick] (c1),
      (c1) -- [fermion, thick] (f1),
      (h) -- [fermion, thick] (c),

      (a) -- [plain, thick] (X1),
      (a) -- [plain, thick] (X1a),
      (a) -- [plain, thick] (X1b),
      (b) -- [plain, thick] (X2),      
      (b) -- [plain, thick] (X2a),      
      (b) -- [plain, thick] (X2b),      
    };
   	\draw [fill, gray] (a) circle (.2);
 	\draw [fill, gray] (b) circle (.2);
 	\draw [fill] (c2) circle (.055);
  \end{feynman}
\end{tikzpicture}
   \caption{}
    \label{fig:feyn2a}
 \end{subfigure}
  \begin{subfigure}{0.45\textwidth}
\centering
\begin{tikzpicture}
  \begin{feynman}
  %place first vertices
    \vertex (i1) at (0,1.5);
    \vertex (i2) at (0,-1.5);
     \vertex (a) at (2.0, 1.5) ;
     \vertex (b) at (2.0, -1.5) ;
     \vertex (c1) at (3, 0.75);
     \vertex (c) at (3, 0);     
     \vertex (c2) at (3, -0.75);
     \vertex (h) at (4.5, 0.0) {\( c \)};
     \vertex (f1) at (4.5, 0.75)  {\( \bar{c} \)};
     \vertex (f2) at (4.5, -0.75)  {\( h'' \)};
     \vertex (X1) at (4, 1.8) ;
     \vertex (X1a) at (4, 2.0) ;
     \vertex (X1b) at (4, 1.6) ;
     \vertex (X2) at (4, -1.8) ;
     \vertex (X2a) at (4, -1.6) ;
     \vertex (X2b) at (4, -2.0) ;
    \diagram* {
      (i1) -- [plain, very thick, edge label=\(p\)] (a),
      (i2) -- [plain, very thick,edge label'=\(p\)] (b),
      (a) -- [gluon, thick, edge label'=\(G\)] (c1),
      (b) -- [gluon, thick, edge label=\(G\)] (c2),
      (c2) -- [scalar, thick] (f2),
      (c2) -- [gluon, thick, edge label=\(G\)] (c),
      (c1) -- [fermion, thick] (c),
      (f1) -- [fermion, thick] (c1),
      (c) -- [fermion, thick] (h),

      (a) -- [plain, thick] (X1),
      (a) -- [plain, thick] (X1a),
      (a) -- [plain, thick] (X1b),
      (b) -- [plain, thick] (X2),      
      (b) -- [plain, thick] (X2a),      
      (b) -- [plain, thick] (X2b),      
    };
   	\draw [fill, gray] (a) circle (.2);
 	\draw [fill, gray] (b) circle (.2);
 	\draw [fill] (c2) circle (.055);
  \end{feynman}
\end{tikzpicture}
   \caption{}
    \label{fig:feyn2b}
 \end{subfigure}
\begin{subfigure}{0.45\textwidth}
\centering
\begin{tikzpicture}
  \begin{feynman}
  %place first vertices
    \vertex (i1) at (0,1.5);
    \vertex (i2) at (0,-1.5);
     \vertex (a) at (2.0, 1.5) ;
     \vertex (b) at (2.0, -1.5) ;
     \vertex (c1) at (3, 0.75);
     \vertex (c) at (3, 0);     
     \vertex (c2) at (3, -0.75);
     \vertex (h) at (4.5, 0.0) {\( c \)};
     \vertex (f1) at (4.5, 0.75)  {\( h'' \)};
     \vertex (f2) at (4.5, -0.75)  {\( \bar{c} \)};
     \vertex (X1) at (4, 1.8) ;
     \vertex (X1a) at (4, 2.0) ;
     \vertex (X1b) at (4, 1.6) ;
     \vertex (X2) at (4, -1.8) ;
     \vertex (X2a) at (4, -1.6) ;
     \vertex (X2b) at (4, -2.0) ;
    \diagram* {
      (i1) -- [plain, very thick, edge label=\(p\)] (a),
      (i2) -- [plain, very thick,edge label'=\(p\)] (b),
      (a) -- [gluon, thick, edge label'=\(G\)] (c1),
      (b) -- [gluon, thick, edge label=\(G\)] (c2),
      (c1) -- [scalar, thick] (f1),
      (c1) -- [gluon, thick, edge label'=\(G\)] (c),
      (c2) -- [fermion, thick] (c),
      (f2) -- [fermion, thick] (c2),
      (c) -- [fermion, thick] (h),

      (a) -- [plain, thick] (X1),
      (a) -- [plain, thick] (X1a),
      (a) -- [plain, thick] (X1b),
      (b) -- [plain, thick] (X2),      
      (b) -- [plain, thick] (X2a),      
      (b) -- [plain, thick] (X2b),      
    };
   	\draw [fill, gray] (a) circle (.2);
 	\draw [fill, gray] (b) circle (.2);
 	\draw [fill] (c1) circle (.055);
  \end{feynman}
\end{tikzpicture}
   \caption{}
    \label{fig:feyn2c}
 \end{subfigure}
  \begin{subfigure}{0.45\textwidth}
\centering
\begin{tikzpicture}
  \begin{feynman}
  %place first vertices
    \vertex (i1) at (0,1.5);
    \vertex (i2) at (0,-1.5);
     \vertex (a) at (2.0, 1.5) ;
     \vertex (b) at (2.0, -1.5) ;
     \vertex (c1) at (3, 0.75);
     \vertex (c) at (3, 0);     
     \vertex (c2) at (3, -0.75);
     \vertex (h) at (4.5, 0.0) {\( \bar{c} \)};
     \vertex (f1) at (4.5, 0.75)  {\( h'' \)};
     \vertex (f2) at (4.5, -0.75)  {\( c \)};
     \vertex (X1) at (4, 1.8) ;
     \vertex (X1a) at (4, 2.0) ;
     \vertex (X1b) at (4, 1.6) ;
     \vertex (X2) at (4, -1.8) ;
     \vertex (X2a) at (4, -1.6) ;
     \vertex (X2b) at (4, -2.0) ;
    \diagram* {
      (i1) -- [plain, very thick, edge label=\(p\)] (a),
      (i2) -- [plain, very thick,edge label'=\(p\)] (b),
      (a) -- [gluon, thick, edge label'=\(G\)] (c1),
      (b) -- [gluon, thick, edge label=\(G\)] (c2),
      (c1) -- [scalar, thick] (f1),
      (c1) -- [gluon, thick, edge label'=\(G\)] (c),
      (c) -- [fermion, thick] (c2),
      (c2) -- [fermion, thick] (f2),
      (h) -- [fermion, thick] (c),

      (a) -- [plain, thick] (X1),
      (a) -- [plain, thick] (X1a),
      (a) -- [plain, thick] (X1b),
      (b) -- [plain, thick] (X2),      
      (b) -- [plain, thick] (X2a),      
      (b) -- [plain, thick] (X2b),      
    };
   	\draw [fill, gray] (a) circle (.2);
 	\draw [fill, gray] (b) circle (.2);
 	\draw [fill] (c1) circle (.055);
  \end{feynman}
\end{tikzpicture}
   \caption{}
    \label{fig:feyn2d}
 \end{subfigure} 
\caption{\label{fig:100}
Examples of diagrams for $p p \to h'' c \bar{c} X$ where the boson $h''$ couples to gluons via a fermionic loop. This is described by the effective $h''GG$ coupling derived in Appendix~\ref{appA}.}
\end{figure}
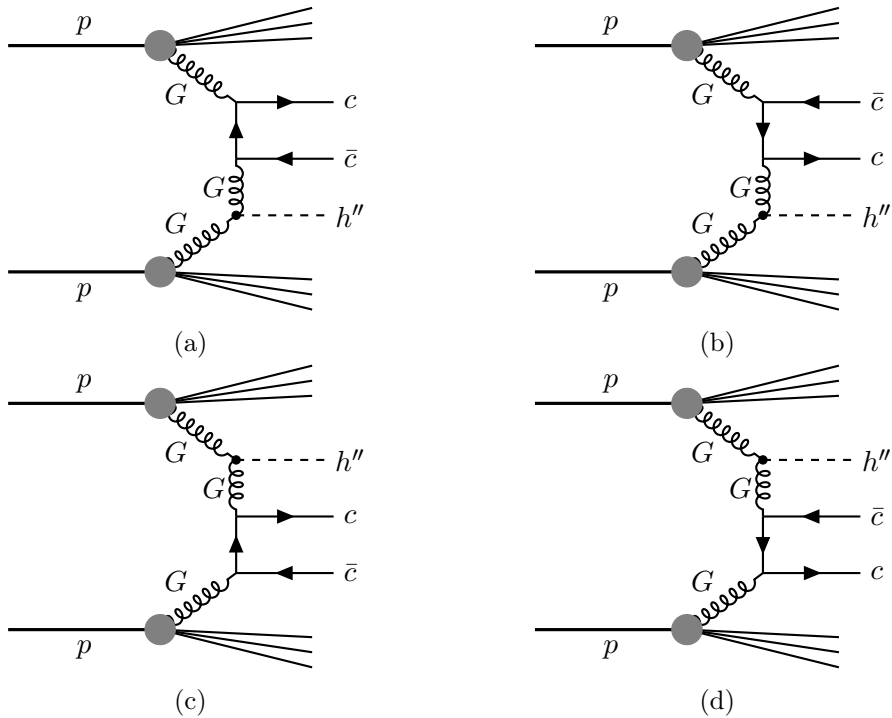

Our first study concerns the total cross section for the reaction
\begin{equation} \label{eq:2.1a}
p + p \to h'' + c + \bar{c} + X.
\end{equation} 
Here we consider all tree-level diagrams of the MCPM' giving a boson $h''$ plus a $c\bar c$ quark-antiquark pair. In addition we consider the diagrams of
Fig.~\ref{fig:100}. In these diagrams the boson $h''$ couples to the gluons via a fermion loop.

From the tree-level diagrams (see Fig.~\ref{fig:feyn1a} for an example) we find
\begin{equation} \label{eq:2.1b}
\left. \sigma_{\mathrm{tot}}(p + p \to h'' + c + \bar{c} + X) \right|_{\text{tree}} = 5433.0 ~\text{pb.}
\end{equation} 
The diagrams with the effective $h''GG$ coupling (Fig.~\ref{fig:100}) give
\begin{equation} \label{eq:2.1c}
\left. \sigma_{\mathrm{tot}}(p + p \to h'' + c + \bar{c} + X) \right|_{\text{h''GG}} = 80.2~\text{pb.}
\end{equation} 
Taking together all diagrams of~\eqref{eq:2.1b} and~\eqref{eq:2.1c} we get  
\begin{equation} \label{eq:2.1d}
\left. \sigma_{\mathrm{tot}}(p + p \to h'' + c + \bar{c} + X) \right|_{\text{SPS}} = 5517.1 ~\text{pb.}
\end{equation} 
We calculated the total cross section ~\eqref{eq:2.1d} in a separate run of our programs. The number given here is not exactly the sum of~\eqref{eq:2.1b} plus~\eqref{eq:2.1c}, due to interference effects between diagrams of the type of Fig.~\ref{fig:feyn1b} and Fig.~\ref{fig:100}.

\begin{figure}[!ht]
\centering
\begin{tikzpicture}
  \begin{feynman}
  %place first vertices
    \vertex (i1) at (0,1.5);
    \vertex (i2) at (0,-1.5);
     \vertex (a) at (2.0, 1.5) ;
     \vertex (b) at (2.0, -1.5) ;

     \vertex (g) at (3.2,-1.2);
     \vertex (g1) at (4.5,-0.8) {\( c \)};
     \vertex (g2) at (4.5,-1.4) {\( \bar{c} \)};

     \vertex (c) at (3, 0);     

     \vertex (h) at (4.5, 0.0) {\( h'' \)};

     \vertex (X1) at (4, 1.8) ;
     \vertex (X1a) at (4, 2.0) ;
     \vertex (X1b) at (4, 1.6) ;
     \vertex (X2) at (4, -1.8) ;
     \vertex (X2a) at (4, -1.6) ;
     \vertex (X2b) at (4, -2.0) ;
    \diagram* {
      (i1) -- [plain, very thick, edge label=\(p\)] (a),
      (i2) -- [plain, very thick,edge label'=\(p\)] (b),
      (a) -- [gluon, thick, edge label'=\(G\)] (c),
      (b) -- [gluon, thick, edge label=\(G\)] (c),

      (b) -- [gluon, thick] (g),

      (g) -- [fermion, thick] (g1),
      (g2) -- [fermion, thick] (g), 

      (h) -- [scalar, thick] (c),

      (a) -- [plain, thick] (X1),
      (a) -- [plain, thick] (X1a),
      (a) -- [plain, thick] (X1b),
      (b) -- [plain, thick] (X2),      
      (b) -- [plain, thick] (X2a),      
      (b) -- [plain, thick] (X2b),      
    };
   	\draw [fill, gray] (a) circle (.2);
 	\draw [fill, gray] (b) circle (.2);
 	\draw [fill] (c) circle (.055);
  \end{feynman}
\end{tikzpicture}
 
\caption{\label{fig:200}
Example of diagram for $p p \to h'' c \bar{c} X$ where the $c\bar c$ pair comes from the fragmentation of one proton remnant.}
\end{figure}
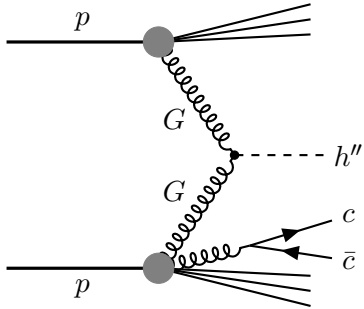

Despite the large number of gluons available in $pp$-collisions we find that the diagrams of Fig.~\ref{fig:100} contribute only at the per-cent level to the total cross section. There are also other diagrams with the effective $h''GG$ vertex. An example is given in Fig.~\ref{fig:200}. This type of diagrams is really part of the Drell-Yan reaction where two gluons fuse to give $h''$. At $\sqrt{s}= 13$ TeV the total cross section $\left. \sigma(p+p\to h'' + X)\right|_{\text{GG}}$ was found to be $210.2 ~\text{pb}$. The fragmentation with a $c\bar c$ pair certainly is only a small part of this and the corresponding total cross section is estimated to be at most at the per-cent level of the cross section of~\eqref{eq:2.1b}.

The numbers for the cross sections in \eqref{eq:2.1b}, \eqref{eq:2.1c}, and \eqref{eq:2.1d} have been obtained in the collinear method using only
single parton scattering (SPS) contributions. Of course, there are also double parton scattering (DPS) contributions possible.
Using the methods for DPS as explained in Section \ref{sec:3.3} below we find the following total cross sections:
\begin{equation} \label{eq:2.1e_dps}
\left. \sigma_{\mathrm{tot}}(p + p \to h'' + c + \bar{c} + X) \right|_{\text{DPS}} = 6159.16 ~\text{pb.}
\end{equation} 
Adding the SPS result \eqref{eq:2.1d} we get
\begin{equation} \label{eq:2.1f_dps}
\left. \sigma_{\mathrm{tot}}(p + p \to h'' + c + \bar{c} + X) \right|_{\text{SPS+DPS}} = 11676.26 ~\text{pb.}
\end{equation}

%=============================================================================
\section{Distributions}
\label{sec_dist}
%=============================================================================
Now we come to the distributions for the reactions \eqref{eq:1.5} and \eqref{eq:1.5b}. The $c$ and $\bar c$ quarks in these reactions will manifest themselves as $c$ and $\bar c$ jets. Therefore, we will now specify the cuts which we apply at parton level in order to obtain distributions which should be also valid at the jet level. In the following we assume that the $c$ and $\bar c$ jets are identified by flavor-tagging. We are aware of the fact that this is a tough requirement for experimental analyses. But, as already mentioned, prospects to achieve this are quite good \cite{ATLAS:2022qxm,CMS-DP-2023-006}. 

%%%%%%%%%%%%%%%%%%%%%%%%%%%%%%%%%%%%%%
\subsection{Cuts}
\label{sub_cuts}

In the generation of the events we apply at
 parton level the following cuts to all jets, that is, to the final state quarks:
\begin{equation} \label{eq:2.1}
\Delta R > 0.4, \qquad |\eta| < 5.0, \qquad p_T > 20~\text{GeV} \;.
\end{equation}
Here, $\Delta R$ is defined as
\begin{equation} \label{eq:2.11}
\Delta R = \sqrt{ (\Delta \eta)^2 + (\Delta \phi)^2}
\end{equation}
with $\Delta \phi$ and $\Delta \eta$ the differences of the azimuthal angles $\phi$ and pseudorapidities~$\eta$ of the members of a quark pair. The transverse momentum of a quark is denoted by $p_T$. The $p_T$ cut in~\eqref{eq:2.1} eliminates events where jets are produced near the proton beams. 
 The $\Delta R$ cut allows the experimentalists to select spatially separated jets in the detector. Furthermore, a Breit-Wigner factor is set defining on-shellness of the $h''$ boson around its mass, which is chosen as the standard value of 15 times its width. We follow here the MadGraph documentation, giving these well established minimal cut values. Later we shall apply additional cuts to enhance the significance of the MCPM' signal over the Standard Model background. 

%%%%%%%%%%%%%%%%%%%%%%%%%%%%%%%%%%%%%%
\subsection{Cross sections for $p + p \to h'' + c + \bar{c} + X$ with $h'' \to c + \bar{c}$}

We study first the signal process 
\begin{equation} \label{eq:h2cc}
p + p \to h'' + c + \bar{c} + X \text{ with } h'' \to c + \bar{c}\;.
\end{equation}
The $h''$ boson couples in particular to the charm quarks with a coupling strength proportional to the top-quark mass. This is a special feature of the MCPM' - eventually originating from the underlying CP symmetry; see~\cite{Maniatis:2023aww}. With the application of the cuts as outlined in Section~\ref{sub_cuts} we find a total cross section of
\begin{equation} \label{eq:2.3a}
\left. \sigma \left( p + p \to (h'' \to c \bar{c}) + c + \bar{c} + X \right) \right|_{\text{with cuts}} =
 99.2~\text{pb.}
 \end{equation}
\begin{figure}
  \begin{subfigure}{0.5\textwidth}
\centering
\includegraphics[width=1. \textwidth]{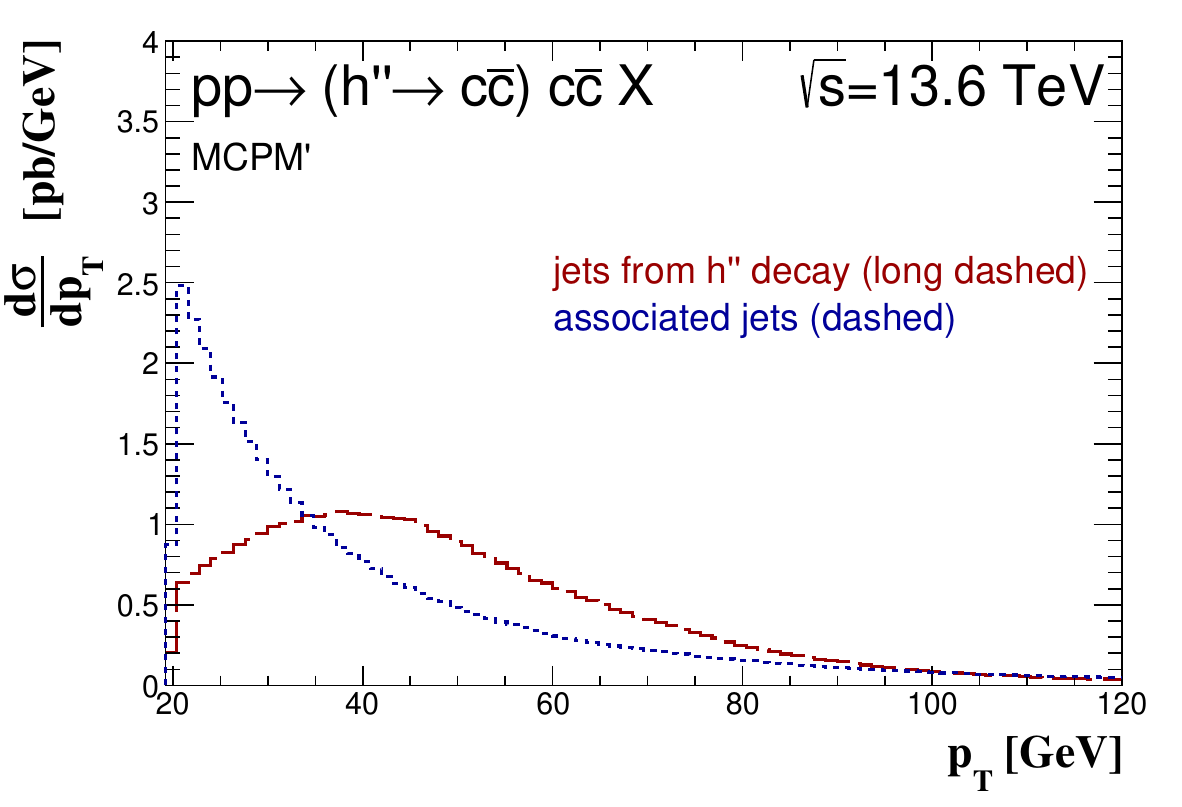}
   \caption{}
\label{fig:pt-signal-a}
 \end{subfigure}
  \begin{subfigure}{0.5\textwidth}
\centering
\includegraphics[width=1. \textwidth]{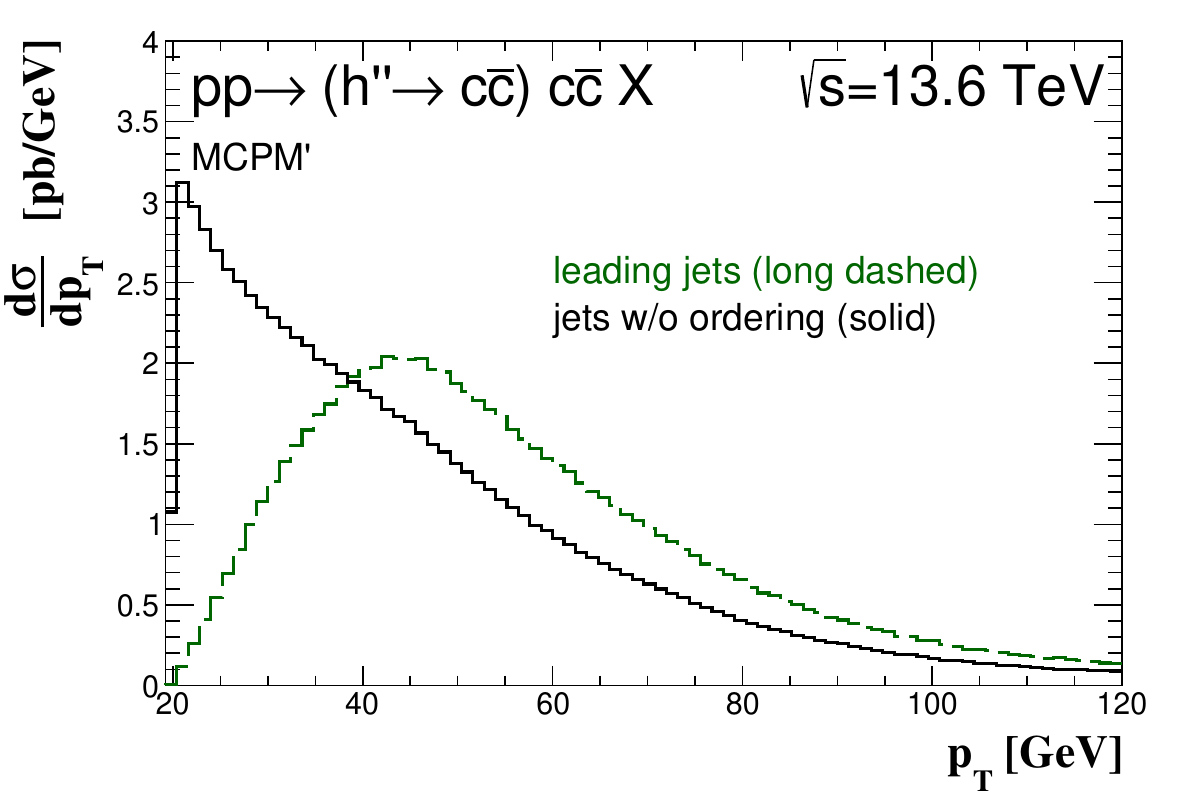}
   \caption{}
\label{fig:pt-signal-b}
 \end{subfigure}

\caption{\label{fig:pt-signal} Distribution of the transverse momentum of $c$ quark (or $\bar{c}$ antiquark) in the reaction \eqref{eq:h2cc}.
(a): The contributions from charm quarks originating from the decay of the $h''$ boson (long dashed) and those associated with the accompanying charm quarks (dashed). (b): The distributions without distinguishing the origin of the charm quarks. The solid histogram corresponds to the case where the $c$ quarks and $\bar c$ antiquarks are not sorted by their transverse momentum, while the long dashed histogram represents the distribution of the leading $p_{T}$ charm quark (or anti-charm quark).}
\end{figure}

Next we study various differential cross sections of this process. Figure \ref{fig:pt-signal} presents the differential distributions as a function of the transverse momentum of the $c$ quark (or $\bar c$ antiquark). In Fig.~\ref{fig:pt-signal-a} we show separately the contributions from charm quarks originating from the decay of the $h''$ boson (long dashed histogram) and those associated with the accompanying charm quarks (dashed histogram). In Fig.~\ref{fig:pt-signal-b} we display the distributions without distinguishing the origin of the charm quarks. Two histograms are shown: one corresponds to the case where the $c$ quarks and $\bar c$ antiquarks are not sorted by their transverse momentum (solid histogram), while the other represents the distribution of the leading charm quark (or leading anti-charm quark), i.e., the one with the highest $p_{T}$. The transverse momentum distributions of the jets from $h''$ decay in Fig.~\ref{fig:pt-signal-a} and of the leading jets in Fig.~\ref{fig:pt-signal-b} exhibit a maximum around $45 \;\mathrm{GeV} \approx m_{h''}/2$. This is expected from the $h''$ production and its decay to $c\bar c$ and corresponds to the Jacobian peak. 

%We denote in this context the {\em leading} $c$--$\bar{c}$ pair by the $c$ quark with the highest $p_T$ of the two $c$ quarks paired with the $\bar{c}$ quark with the %highest $p_T$ of the two $\bar{c}$ quarks.

%In Fig.~\ref{plot_h2_cccc_pt_avg-versus-lead_MCPM} we show the distribution of the transverse momentum~$p_T$ of the $c$ quark and $\bar{c}$ antiquark originating form %the decay of the boson~$h''$ in~\eqref{eq:h2cc} including cuts~\eqref{eq:2.1}. For comparison, we show in the same figure also the $p_T$ distribution of the largest-%$p_T$ $c$ and $\bar{c}$ quarks including cuts~\eqref{eq:2.1}. 

%
%In Fig.~\ref{plot_h2_cccc_delta_r_MCPM} we show the $\Delta R$ distribution for the leading $c$--$\bar{c}$ pair, that is, the pair formed with the highest-$p_T$ $c$ %and $\bar{c}$. \textbf{Note that the cut in \eqref{eq:2.1} is applied to the soft charm pair in the signal process \eqref{eq:2.3a} where the decay of the $h''$ into %another charm pair is not taken into account. Therefore we can also get contributions to the cross section for values of $\Delta R < 0.4$ [Comment: Unclear])}. 

\begin{figure}
  \begin{subfigure}{0.5\textwidth}
\begin{center}
\includegraphics[width=1. \textwidth]{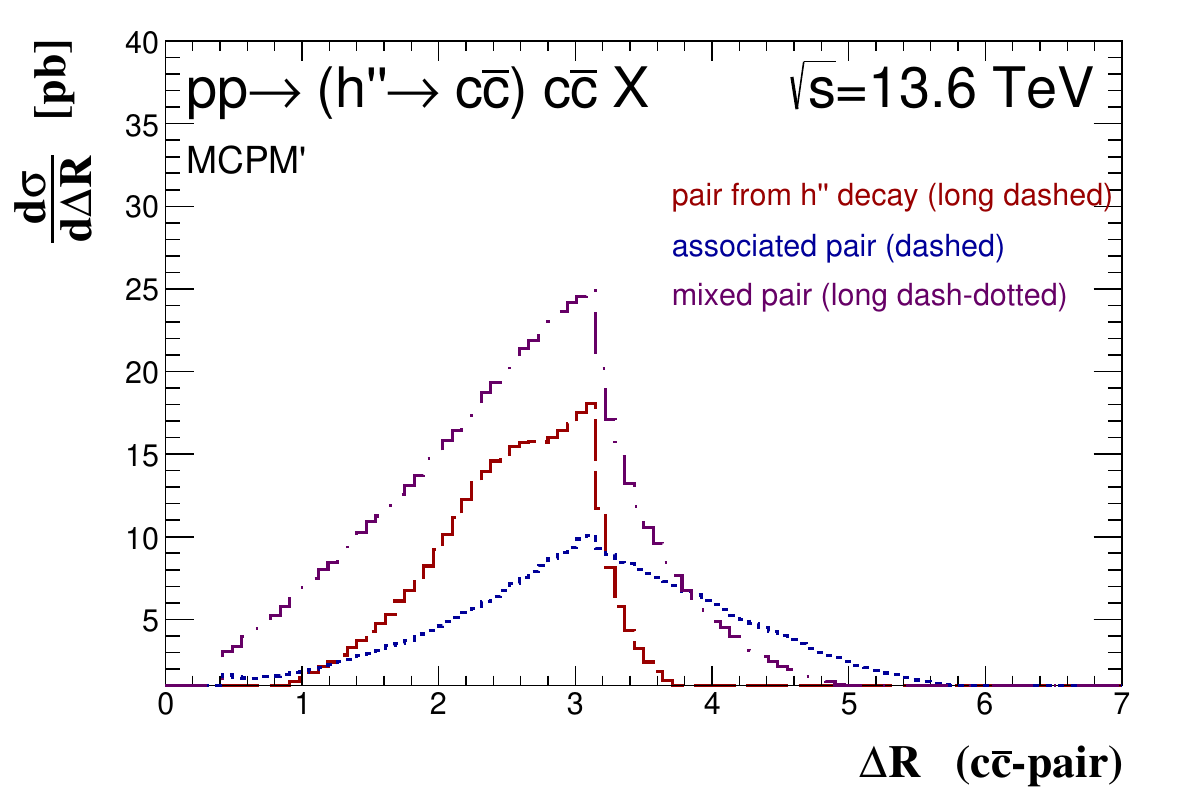}
\end{center}
   \caption{}
\label{fig:R-signal-a}
 \end{subfigure}
  \begin{subfigure}{0.5\textwidth}
\begin{center}
\includegraphics[width=1. \textwidth]{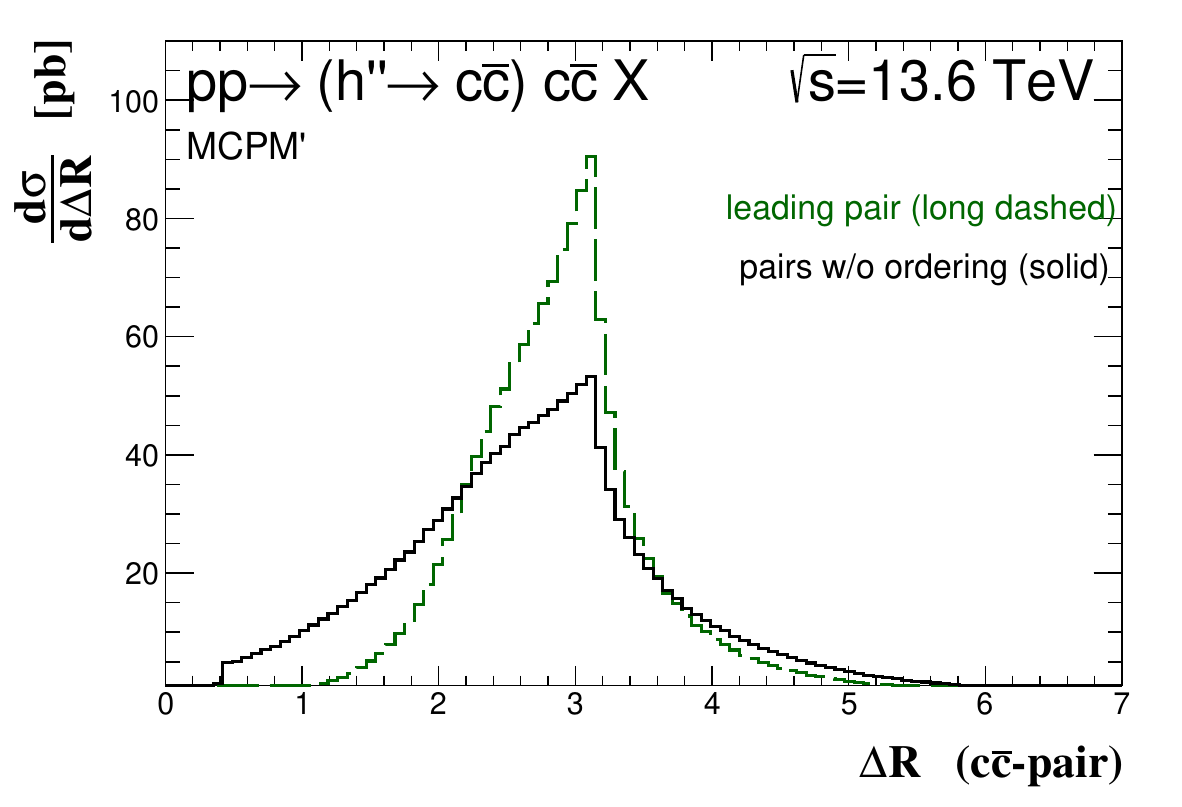}
\end{center}  
   \caption{}
\label{fig:R-signal-b}
 \end{subfigure}
   
\caption{\label{fig:R-signal} $\Delta R$ distribution of the $c\bar c$ pair in the reaction \eqref{eq:h2cc}. (a): The contributions corresponding to $c\bar c$ pairs formed by quarks and antiquarks originating from the decay of the $h''$ (long dashed), by associated charm quarks (dashed), or by combinations where one quark comes from the decay and the other is an associated quark (long dash-dotted). (b): The distributions without distinguishing the origin of the charm quarks. The solid histogram corresponds to the case where the $c\bar c$ pair is formed by quarks and antiquarks not sorted by their transverse momentum, while the long dashed histogram represents the distribution of the leading $c\bar c$ pair.}
\end{figure}

\begin{figure}
  \begin{subfigure}{0.5\textwidth}
\begin{center}
\includegraphics[width=1.0 \textwidth]{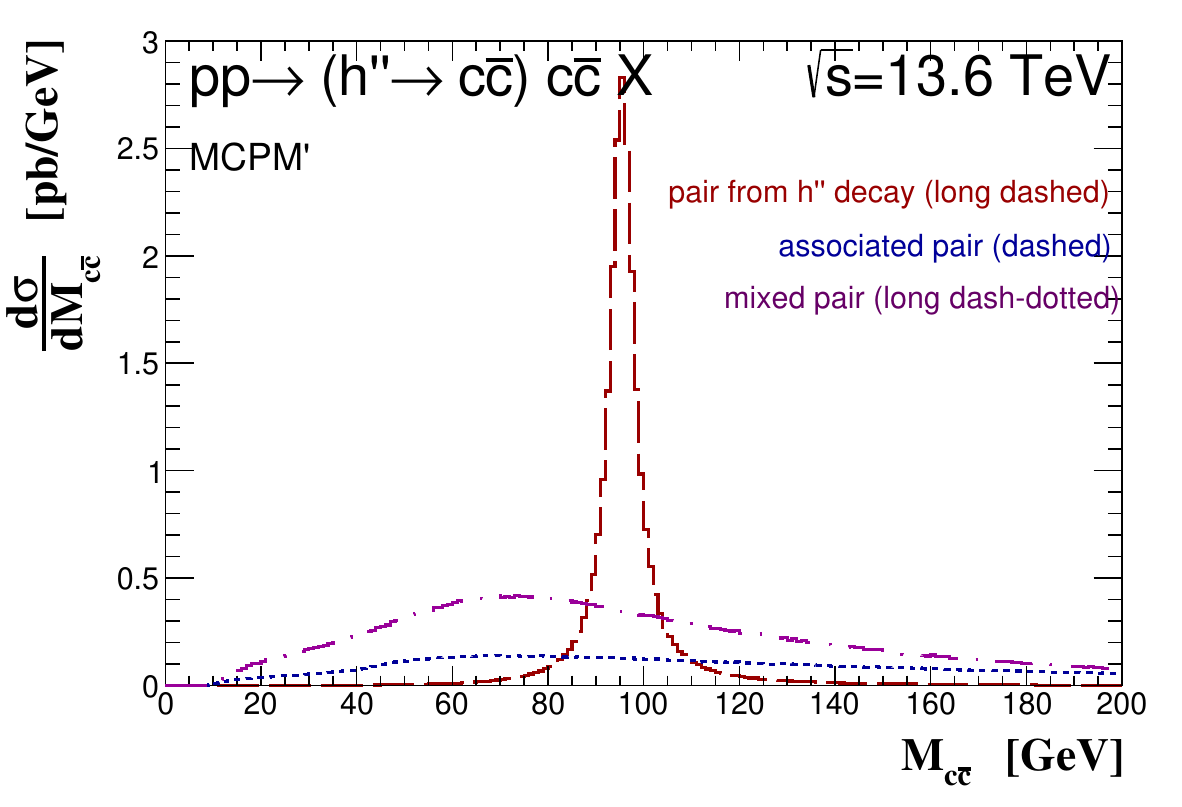}
\end{center}
   \caption{}
\label{fig:M-signal-a}
 \end{subfigure}
  \begin{subfigure}{0.5\textwidth}
\begin{center}
\includegraphics[width=1.0 \textwidth]{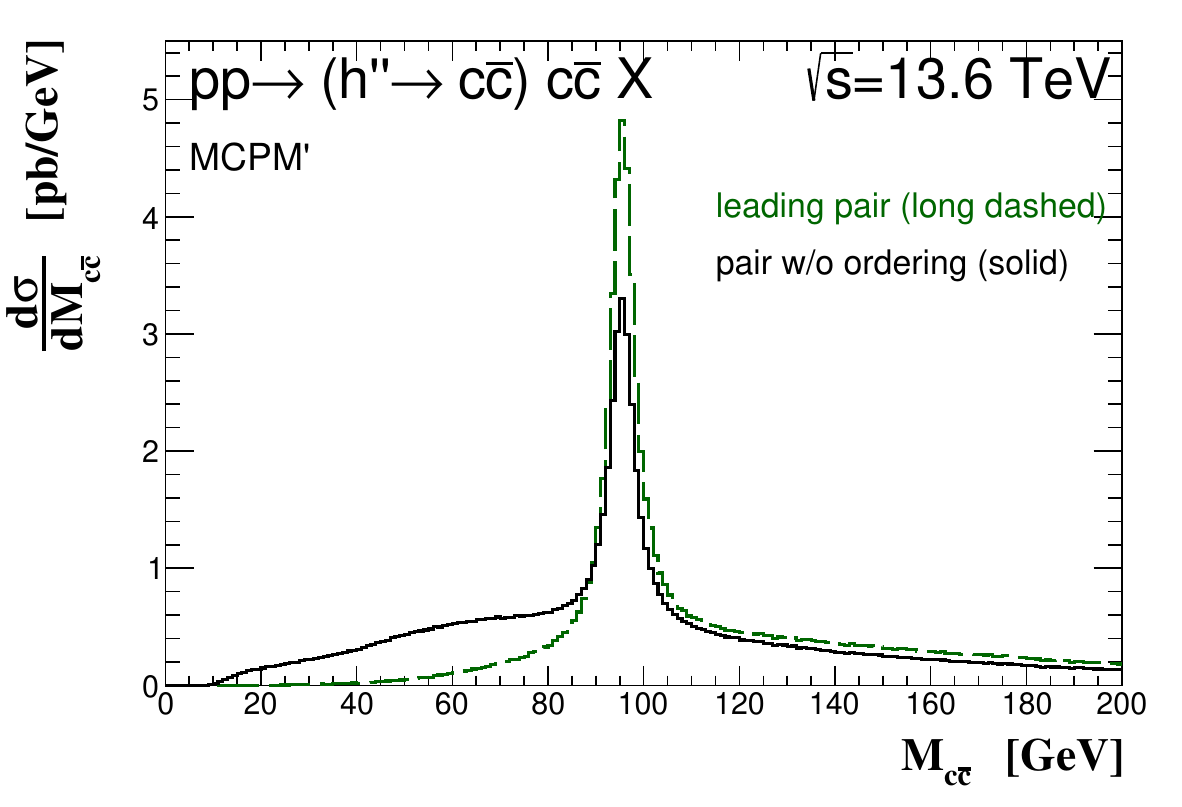}
\end{center}
   \caption{}
\label{fig:M-signal-b}
 \end{subfigure} 
\caption{\label{fig:M-signal} The same as in Fig.~\ref{fig:R-signal} but for the invariant mass $M_{c\bar{c}}$ distribution.}
\end{figure}

In a similar manner, we also present the differential distributions as functions of $\Delta R$ and the invariant mass $M_{c\bar c}$, as shown in Figures~\ref{fig:R-signal} and ~\ref{fig:M-signal}, respectively. For the production of the final state $c\bar c c\bar c$ in \eqref{eq:h2cc}, the possible $c\bar c$ pairs can be formed by quarks and antiquarks originating from the decay of the $h''$, by associated charm quarks, or by combinations where one quark comes from the decay and the other is an associated quark. In Figs.~\ref{fig:R-signal-a} and ~\ref{fig:M-signal-a} we show these three different contributions separately. On the other hand, in Figs.~\ref{fig:R-signal-b} and ~\ref{fig:M-signal-b} we present combined distributions for $c\bar c$ pairs without (solid histograms) and with $p_{T}$ ordering (long dashed histograms).
The so-called leading pair is defined as the pair formed by the charm
quark with the highest transverse momentum (leading $c$-jet) and the
anti-charm quark with the highest transverse momentum (leading $\bar
c$-jet). The $\Delta R$ distribution for the leading pair in
Fig.~\ref{fig:R-signal-b} is less smeared and more concentrated in the
rapidity range between 2 and 3.5, compared to the distribution without $p_{T}$ ordering. This may be advantageous in the context of maximizing the signal-to-background ratio. Similarly, the invariant mass distribution for the leading pair in Fig.~\ref{fig:M-signal-b} also appears more favorable in this regard. In this case, a higher peak is observed compared to the distribution obtained without sorting the pairs by transverse momentum.


%\clearpage
%%%%%%%%%%%%%%%%%%%%%%%%%%%%%%%%%%%%%%

\subsection{Cross sections for $p + p \to c + \bar{c}+ c + \bar{c}+X$ }
\label{sec:3.3}

So far we have given in~Figs.~\ref{fig:pt-signal}--\ref{fig:M-signal} distributions generated from the signal process~\eqref{eq:h2cc}. In reality, experimentalists can only select the process, where two $c$ quarks and two $\bar{c}$ antiquarks are produced:
\begin{equation} \label{eq:2.4}
p + p \to c + \bar{c}+ c + \bar{c}+X\;.
\end{equation}
This reaction gets contributions from the signal process~\eqref{eq:h2cc} but also from the SM; see Fig.~\ref{fig:feyn1b} for an example. For the SM contribution
we take only the QCD processes into account. We study the process~\eqref{eq:2.4} in the MCPM' and compare to the Standard Model result. 
It should be noted that the QCD diagrams appearing in the Standard Model are
likewise present in the MCPM'.  

We find with the standard cuts~\eqref{eq:2.1} from the single parton scattering (SPS) calculation a total cross section
in the SM of
\begin{equation} \label{eq:2.4a}
\left. \sigma_{\text{SM, tot}}(p + p \to c + \bar{c}+ c + \bar{c}+X) \right|_{\text{SPS}} = 1327.3~\text{pb.}
\end{equation}
For the MCPM' we find
\begin{equation} \label{eq:2.4b}
\left. \sigma_{\text{MCPM', tot}}(p + p \to c + \bar{c}+ c + \bar{c}+X) \right|_{\text{SPS}} = 1421.7~\text{pb.}
\end{equation}
Thus, we find an enhanced cross section in the MCPM' compared to the SM result.
However, in our opinion, the difference is not large enough to be used
to judge about the presence or abscence of the $h''$ contribution.
In the following we will therefore explore the $M_{c \bar c}$-distribution.

%commented by Rafal
%In Fig.~\ref{plot_cccc_inv_mass_MCPM_SM} we compare the invariant mass 
%of the $c$--$\bar{c}$ pair with the highest transverse momentum $c$ and $\bar{c}$. 
%\begin{figure}
%\begin{center}
%\includegraphics[width=0.8 \textwidth]{Plots/plot_cccc_inv_mass_MCPM_SM}
%\end{center}
%\caption{\label{plot_cccc_inv_mass_MCPM_SM} Invariant mass $m_{c\bar{c}}$ of the $c$--$\bar{c}$ pair with highest transverse momentum $c$ quark and $\bar{c}$ antiquark. }
%\end{figure}
%We see a clear resonance enhancement corresponding to the $h''$ in the MCPM' distribution.

\begin{figure}[h]
  \begin{subfigure}{0.5\textwidth}
    \centering
    \includegraphics[width=1.0\textwidth]{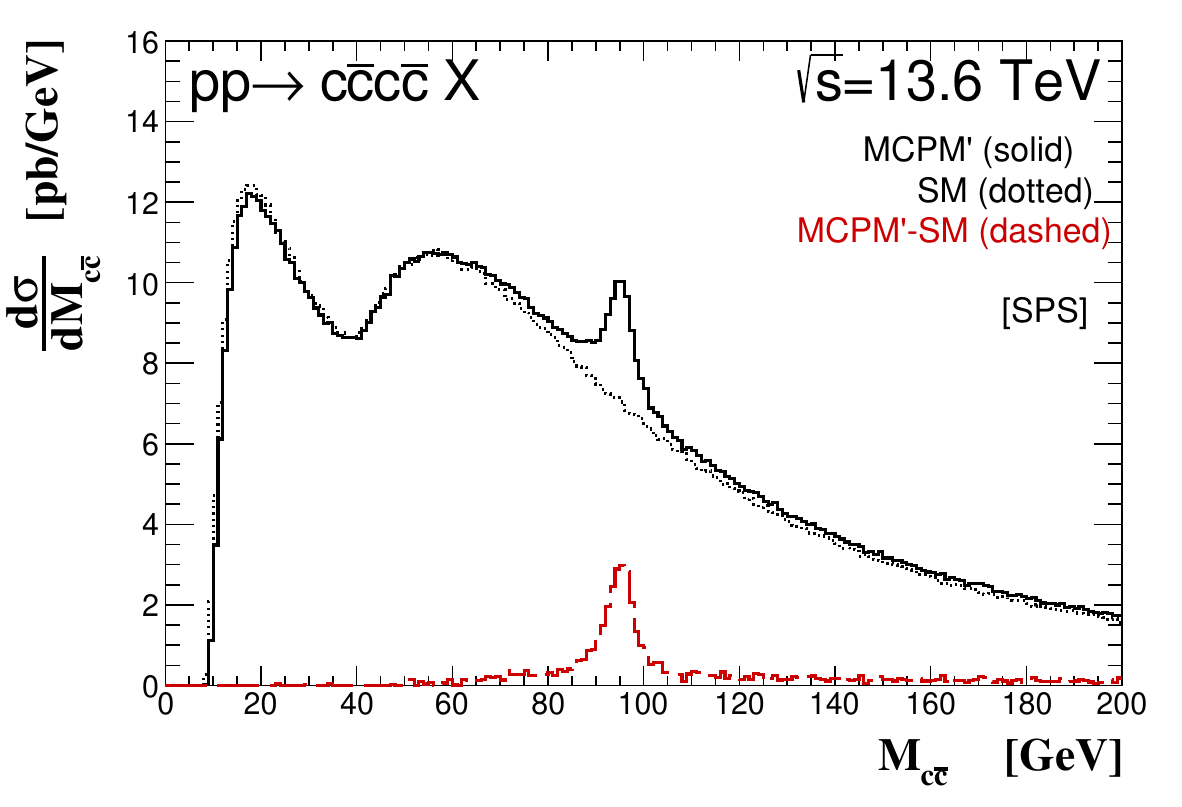}
       \caption{}
      \label{fig:1-rafal-a}
      \end{subfigure}
      \begin{subfigure}{0.5\textwidth}
      \centering
    \includegraphics[width=1.0\textwidth]{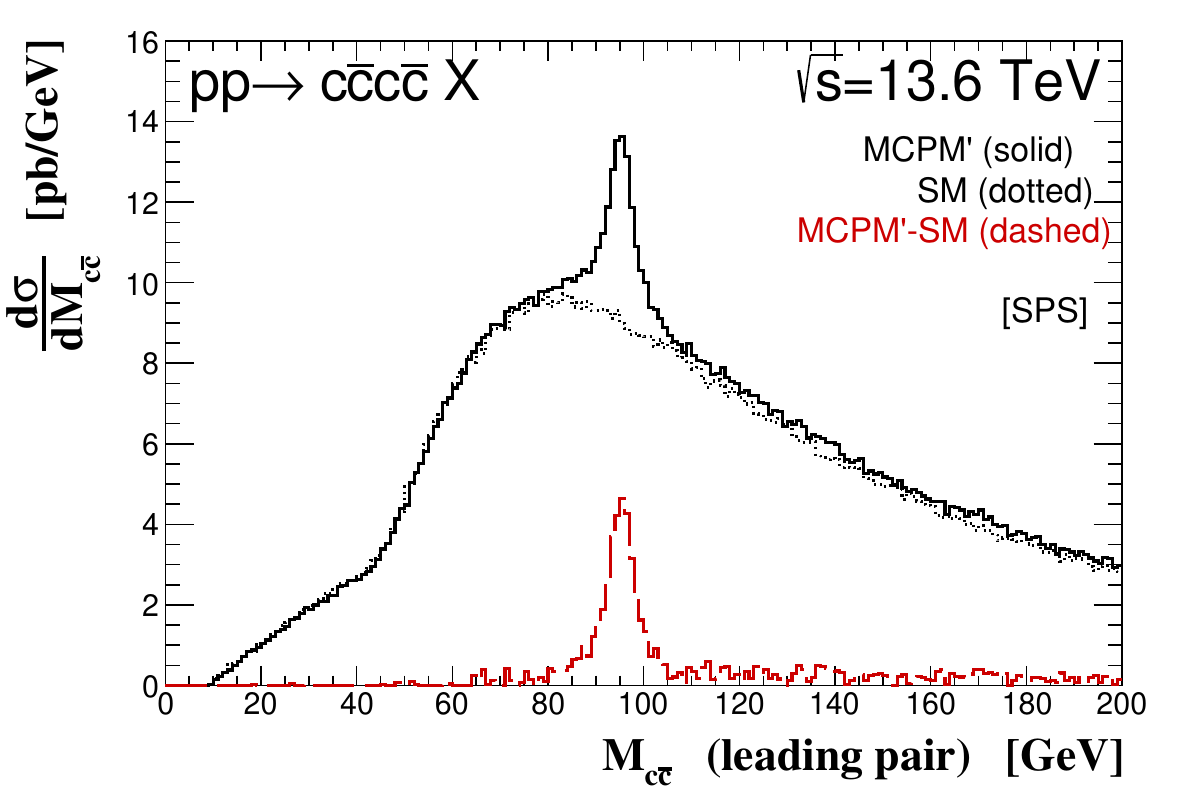}
       \caption{}
      \label{fig:1-rafal-b}
      \end{subfigure}
    \caption{The differential cross sections as a function of the $c\bar c$-pair invariant mass $M_{c\bar c}$. (a) The invariant mass distribution calculated for the $c\bar c$ pair without $p_{T}$-ordering. (b) The distribution for the leading $c\bar c$ pair.  Here the cuts \eqref{eq:2.1} are applied, including $p_T > 20$ GeV.}
    \label{fig:1-rafal}
\end{figure}

In Fig. \ref{fig:1-rafal} we present the differential cross sections as
a function of the $c\bar c$-pair invariant mass $M_{c\bar c}$. Specifically, Fig.~\ref{fig:1-rafal-a} shows the invariant mass distribution calculated for the $c\bar c$ pair without $p_{T}$-ordering, whereas Fig.~\ref{fig:1-rafal-b} displays the distribution for the leading $c\bar c$ pair. In both panels we see a clear resonance enhancement corresponding to the $h''$ in the MCPM' distribution. But a comparison of both plots clearly demonstrates that the variable defined for studying the leading pair is more advantageous, yielding larger cross sections in the signal region.

\begin{figure}[h]
  \begin{subfigure}{0.5\textwidth}
    \centering
        \includegraphics[width=0.7\textwidth]{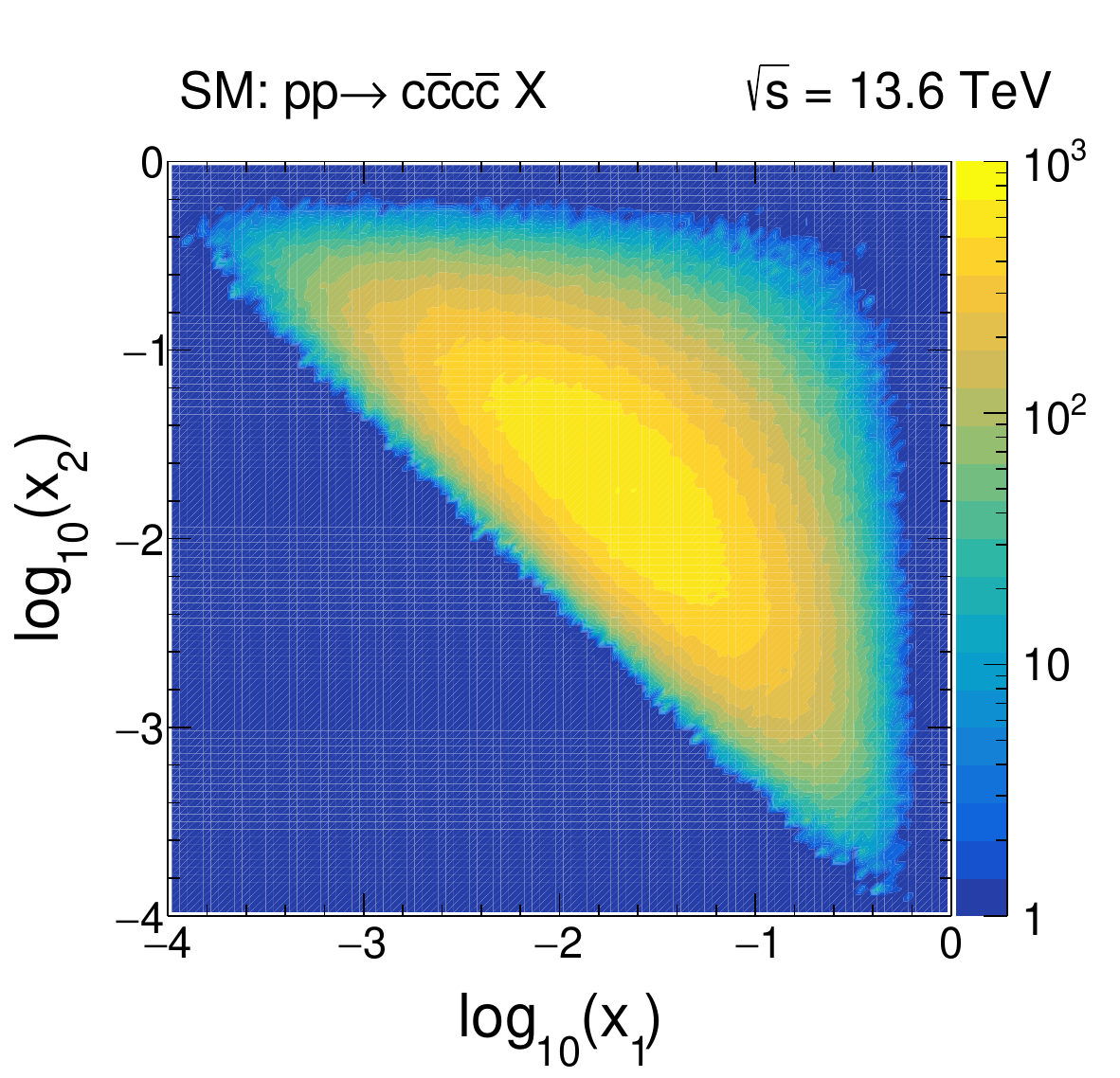}
       \caption{}
      \label{fig:2D-x1-x2}
      \end{subfigure}
  \begin{subfigure}{0.5\textwidth}
    \centering
        \includegraphics[width=0.7\textwidth]{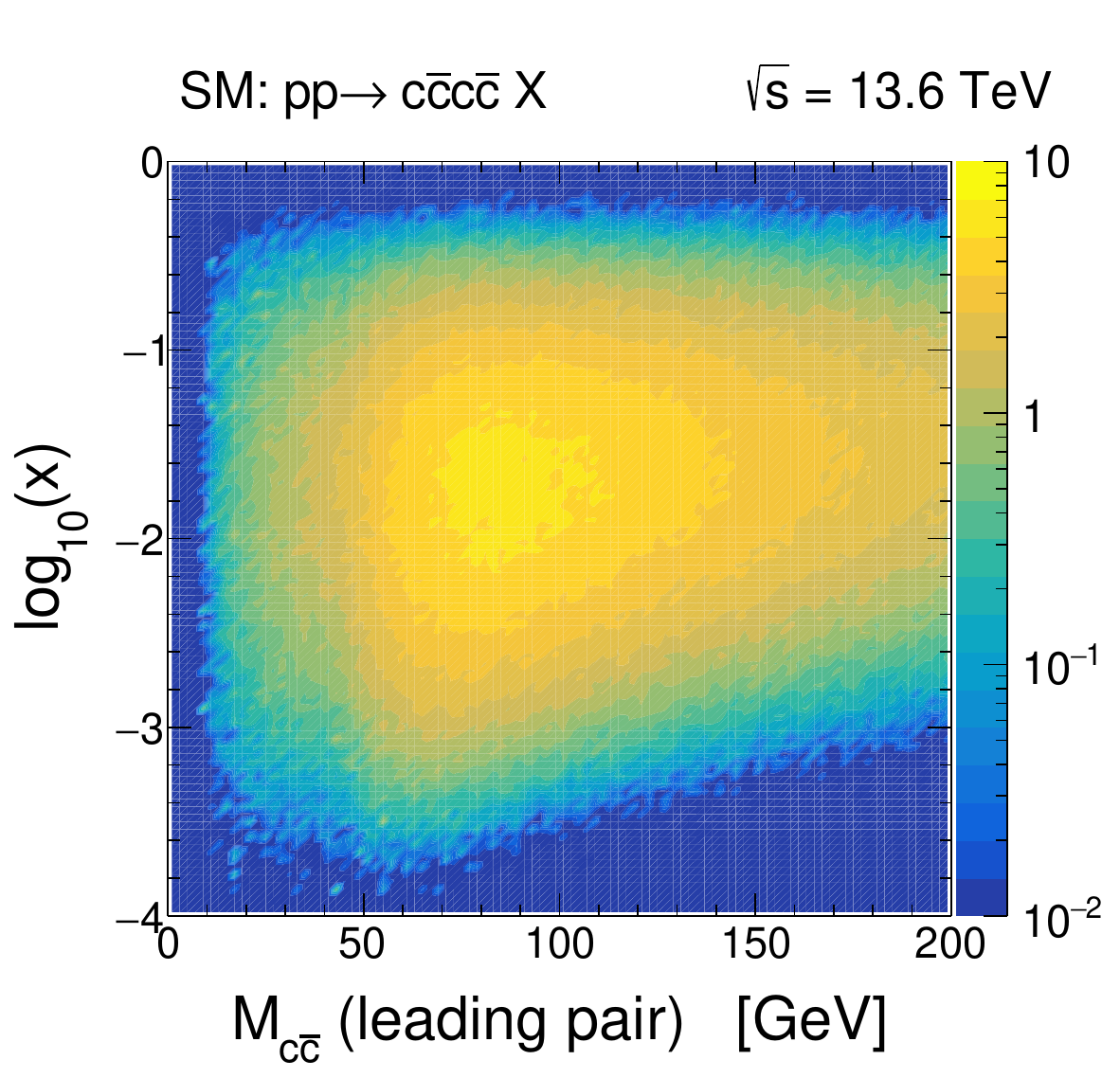}
       \caption{}
      \label{fig:2D-x-Mccbar}
      \end{subfigure}

    \caption{Double differential cross sections for the SM background (SPS LO collinear) as a function of: (a) logarithms of longitudinal momentum fractions $x_1$ and $x_2$ (b) logarithm of longitudinal momentum fraction $x = x_1$ or $x_2$ and the leading pair invariant mass $M_{c\bar c}$. Here the cuts \eqref{eq:2.1} are applied.}
    \label{fig:2D-SPS}
\end{figure}

Figure~\ref{fig:2D-SPS} presents the double-differential cross sections for the collinear SM SPS calculations as a function of logarithm of the longitudinal momentum fractions $x_1, x_2$ of the protons carried by initial partons (Fig.~\ref{fig:2D-x1-x2}), and as a function of logarithm of the longitudinal momentum fraction $x = x_1$ or $x_2$ carried by one initial state parton and the invariant mass of the leading pair $M_{c\bar c}$ (Fig.~\ref{fig:2D-x-Mccbar}). These distributions illustrate the kinematical characteristics of the studied process \eqref{eq:2.4} at the center-of-mass energy of 13.6 TeV, including the applied selection cuts \eqref{eq:2.1}. We observe that the maximum of the cross section for this process is concentrated in the region $10^{-2} \lesssim x \lesssim 10^{-1}$, which indicates that the probed longitudinal momentum fractions are relatively large. In the region of an invariant mass close to the signal, such values also dominate.

\begin{figure}[h]
 \begin{subfigure}{0.5\textwidth}
\centering
        \includegraphics[width=1.0\textwidth]{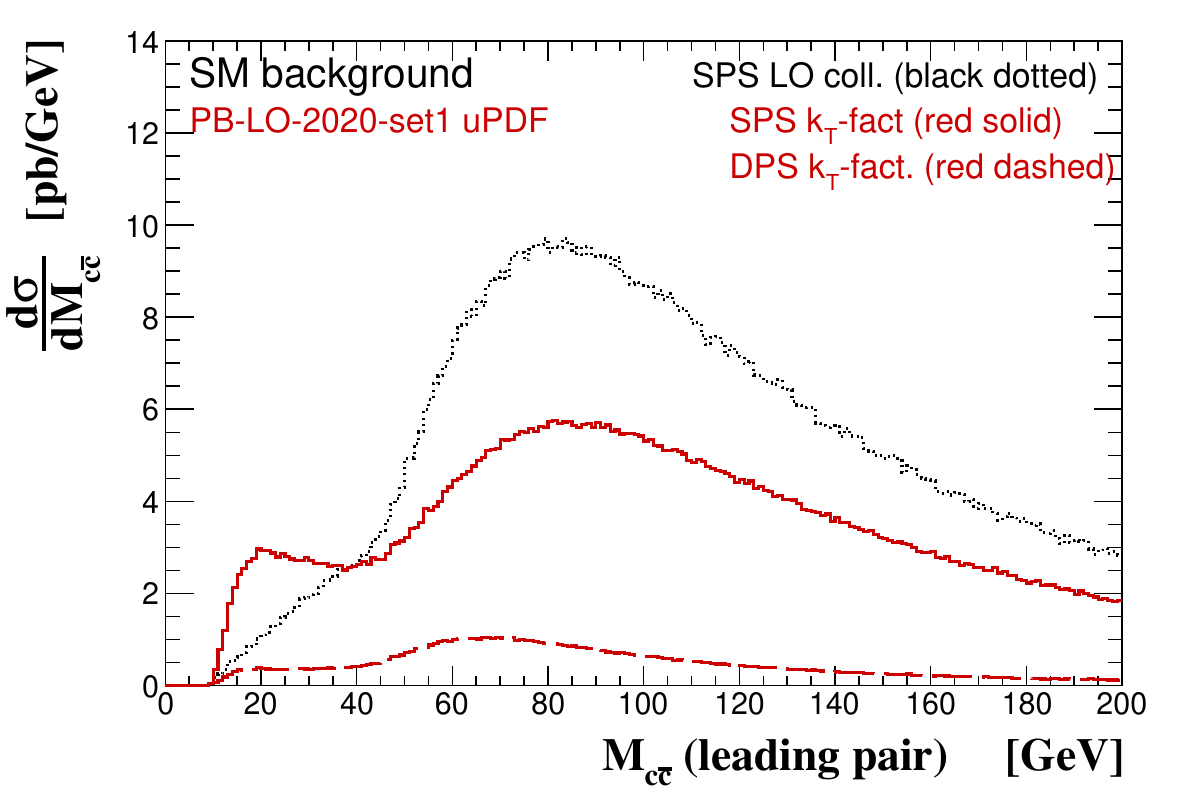}
       \caption{}
      \label{fig:background-a}
      \end{subfigure}        
 \begin{subfigure}{0.5\textwidth}
\centering       
        \includegraphics[width=1.0\textwidth]{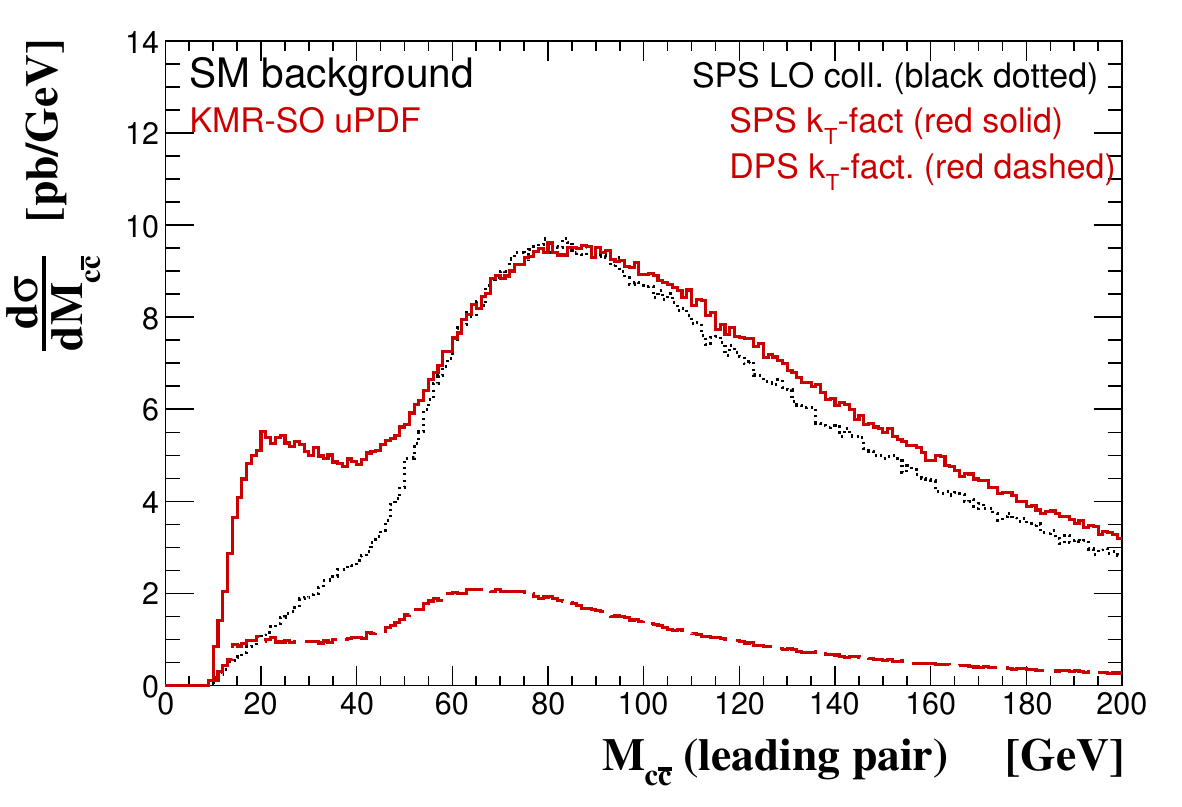}
       \caption{}
      \label{fig:background-b}
      \end{subfigure}   
    \caption{A comparison of SM calculations for the $d\sigma/dM_{c\bar c}$ from the reaction \eqref{eq:2.4}. We show the results obtained using LO collinear (dotted) and $k_{T}$-factorization approaches, including both, the SPS (solid) and the DPS (dashed) contributions. The $k_{T}$-factorization calculations are performed by using two different models for unintegrated parton densities: (a) PB-LO-2020-set1, and (b) KMR-SO. }
    \label{fig:background}
\end{figure}

Figure \ref{fig:background} presents a comparison of SM calculations for $d\sigma/dM_{c\bar c}$  obtained using different theoretical approaches. The production of the $c\bar{c}c\bar{c}$ final state within the Standard Model has been thoroughly analyzed in~\cite{Maciula:2013kd,vanHameren:2014ava,vanHameren:2015wva}, where both single-parton scattering (SPS) and double-parton scattering (DPS) mechanisms were considered. In the present study, we adopt the same theoretical frameworks and computational setup. The corresponding numerical results here are obtained using the \texttt{KaTie} Monte Carlo event generator~\cite{vanHameren:2016kkz}, employing both the collinear and $k_T$-factorization approaches\footnote{We have checked numerically, that the results from \texttt{MadGraph} are consistent with those obtained using the \texttt{KaTie} generator in the collinear mode.}. For the unintegrated (transverse momentum dependent) parton distribution functions (uPDFs), we use the PB-LO-2020-set1 model \cite{Jung:2025mtd}, obtained within the parton-branching method \cite{Hautmann:2017fcj}, as well as the widely adopted Kimber-Martin-Ryskin (KMR) model~\cite{Kimber:2001sc, Watt:2003mx}, constructed with the strong-ordering condition and by using the MSHT20lo$\!\_\!$as130 collinear PDFs as input. These two models can be reliably used in the probed kinematical domain, since they are based on the DGLAP evolution equations. The use of distributions specifically designed for small values of $x$ (as is the case in most of the literature) may not be fully justified in this region.
In the case of DPS, we follow a commonly used uncorrelated factorized ansatz \cite{Gaunt:2009re,Diehl:2011yj,Blok:2011bu,Blok:2013bpa}, taking as effective cross section $\sigma_{\mathrm{eff}} = 15$~mb. This is a value extracted from experimental analyses of various processes at different kinematic configurations (for recent results see e.g. \cite{ATLAS:2016rnd,LHCb:2016wuo,CMS:2013huw,CMS:2017han}). The $k_{T}$-factorization approach with the PB-LO-2020-set1 uPDFs yields smaller cross sections in the signal region compared to the collinear calculations (see Fig.~\ref{fig:background-a}). In contrast, the calculations performed within the KMR uPDFs yield higher cross sections and result in distributions that are very close to those obtained in the collinear approach (see Fig.~\ref{fig:background-b}). Moreover, as evident from the figure, the contribution from the double parton scattering (DPS) mechanism is significantly lower than that of the standard single parton scattering (SPS) process. In the signal region the DPS effect is around 15\% of the SPS cross section.

For the final state \( c + \bar{c} + c + \bar{c} + X \), the DPS cross section is largely driven by the production of two independent \( c\bar{c} \) pairs through Standard Model (QCD) mechanisms. This dominance reflects the large SPS cross section for single \( c\bar{c} \)-pair production. The DPS contributions specific to the MCPM', such as those involving the double production of \( h'' \) bosons or the mixed production of an \( h'' \) and a \( c\bar{c} \) pair, are found to be strongly suppressed. Using the KMR unintegrated parton distributions and applying the standard kinematic cuts given in Eq.~\eqref{eq:2.1}, the total DPS cross section for the production of two \( c\bar{c} \) pairs within the Standard Model is
\begin{equation} \label{eq:2.4a}
\left. \sigma_{\text{SM, tot}}(p + p \to c + \bar{c}+ c + \bar{c}+X) \right|_{\text{DPS}} = 193.80~\text{pb.}
\end{equation}
In comparison, the DPS contributions calculated within the MCPM' give the following difference to the SM result:
\begin{equation} \label{eq:2.4b}
\left. \sigma_{\text{MCPM'-SM, tot}}(p + p \to c + \bar{c}+ c + \bar{c}+X) \right|_{\text{DPS}} = 2.74~\text{pb.}
\end{equation}
Thus, the MCPM'-specific DPS effects contribute at the level of only about 1--2\% to the total MCPM' DPS cross section.
However, in the region of the signal peak, the MCPM'-SM contributions reach up to about 15\% of the dominant SM DPS component.

\begin{figure}[h]
    \centering
    \includegraphics[width=0.5\textwidth]{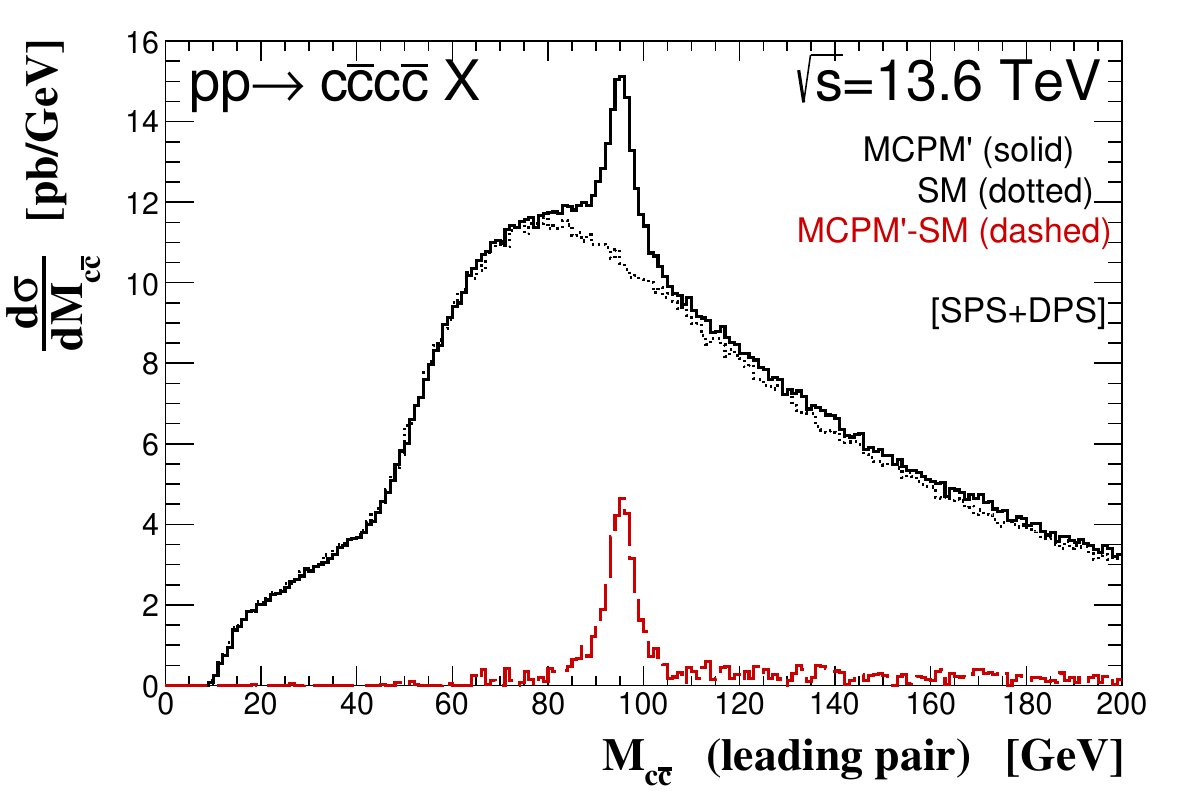}

    \caption{The differential cross sections as a function of the leading $c\bar c$-pair invariant mass $M_{c\bar c}$. Here the MCPM' as well as the SM background results contain the SPS plus the DPS contributions. The cuts \eqref{eq:2.1} are applied, including $p_{T} > 20$ GeV. The figure is the analog of Fig.~\ref{fig:1-rafal-b} where only the SPS contributions are shown.}
    \label{fig:background-dps}
\end{figure}

In Fig.~\ref{fig:background-dps} we show the corresponding effect of including the DPS contribution on the shape and the normalization of the predicted distributions.
The DPS contribution slightly reduces the signal-to-background (S/B) ratio. However, as will be shown in the following, imposing extra cuts, in particular on the leading jets transverse momenta, makes the relative DPS contribution even smaller -- but not negligible.

\begin{figure}[h]
  \begin{subfigure}{0.5\textwidth}
    \centering
        \includegraphics[width=0.7\textwidth]{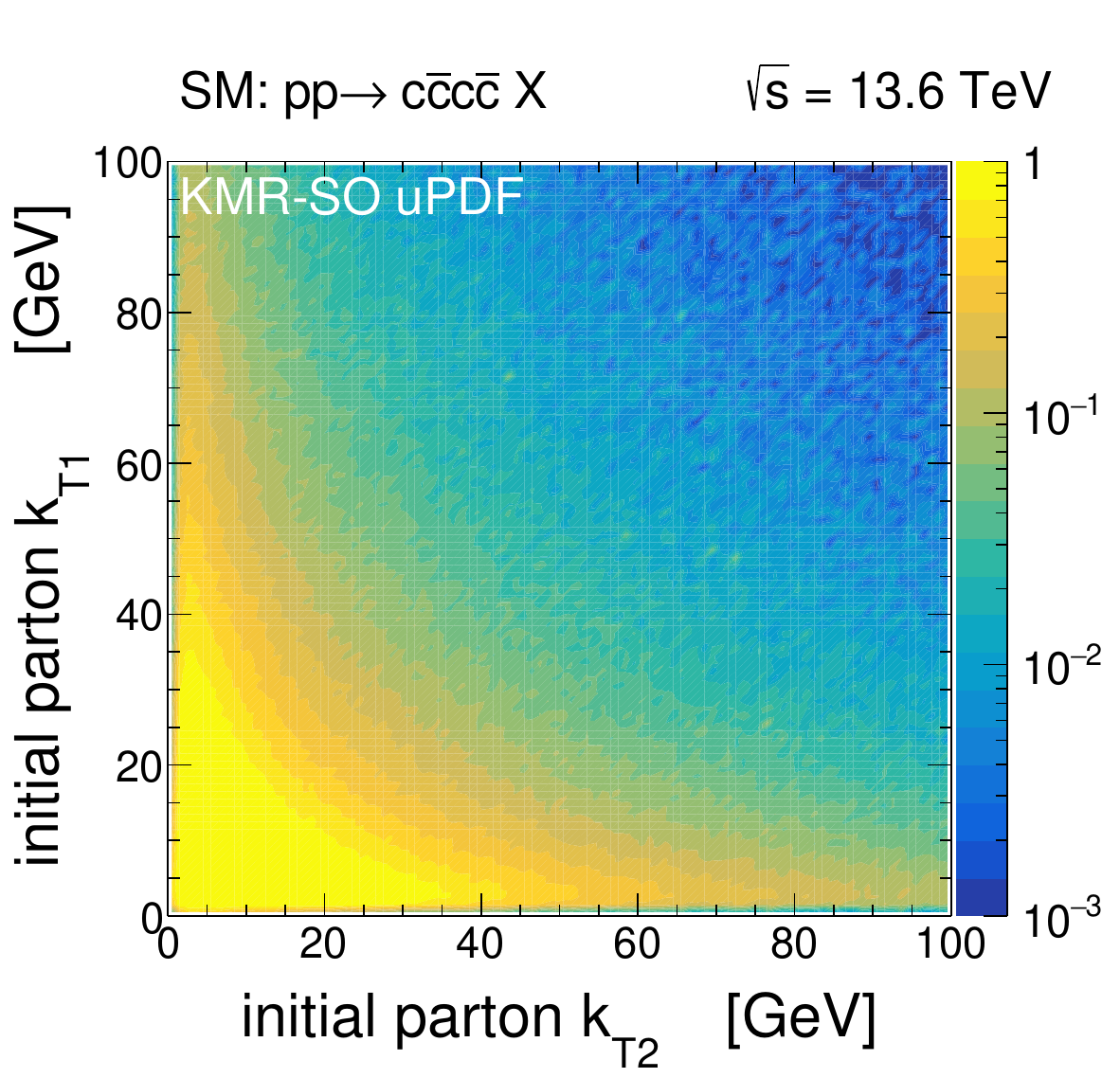}
       \caption{}
      \label{fig:2D-kt1-kt2}
      \end{subfigure}
  \begin{subfigure}{0.5\textwidth}
    \centering
        \includegraphics[width=0.7\textwidth]{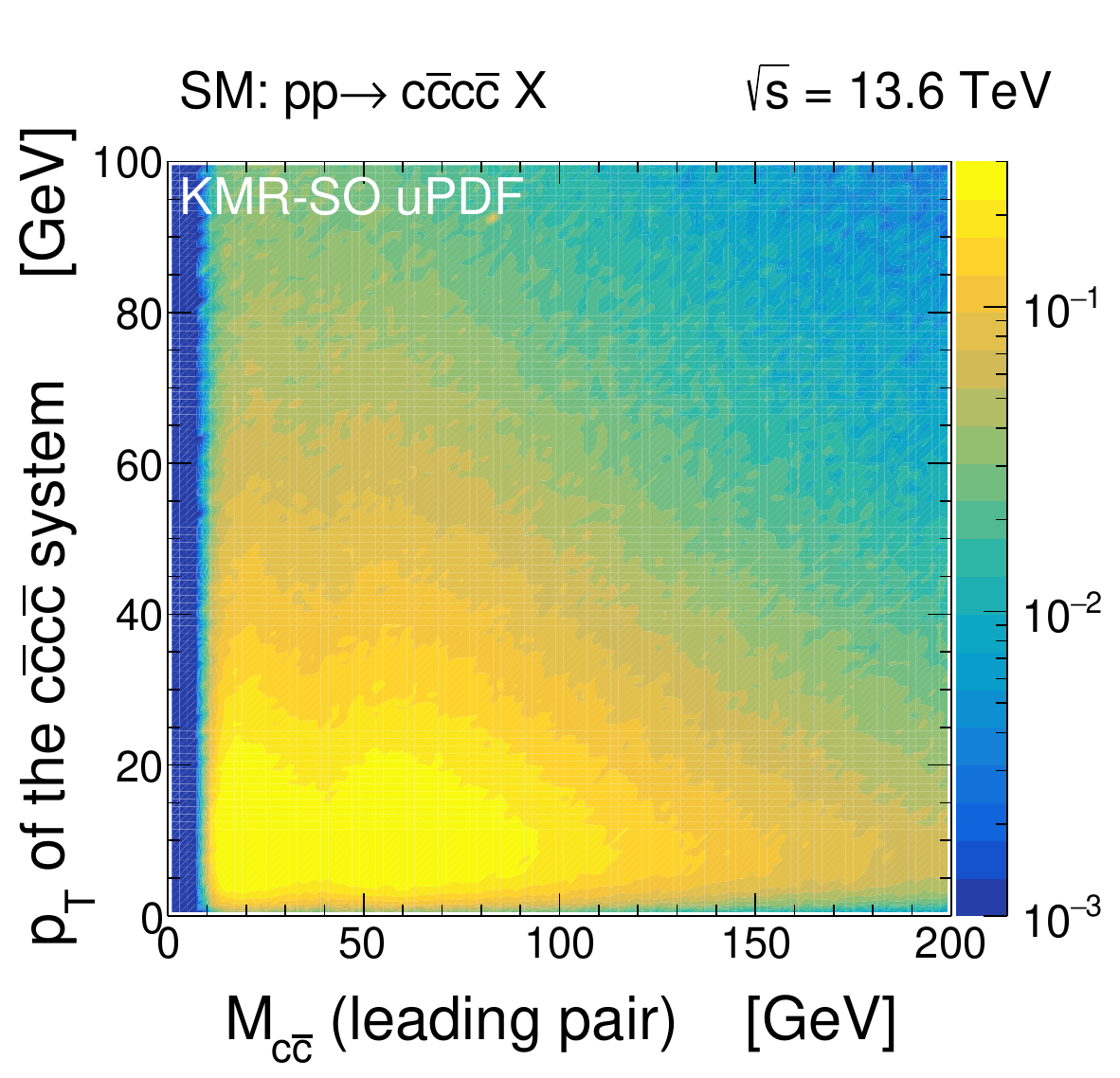}
       \caption{}
      \label{fig:2D-kt-Mccbar}
      \end{subfigure}

    \caption{Double differential cross sections for the SM background (SPS $k_{T}$-factorization with the KMR uPDF) as a function of: (a) transverse momenta of incident partons $k_{T1}$ and $k_{T2}$, and (b) transverse momentum of the $c\bar c c \bar c$-system and the leading pair invariant mass $M_{c\bar c}$. Here the cuts \eqref{eq:2.1} are applied.}
    \label{fig:2D-SPS-kT}
\end{figure}

As can be seen in Figs.~\ref{fig:background-a} and \ref{fig:background-b} it is remarkable that for $M_{c\bar c} \lesssim 50$ GeV the SM SPS results using the $k_{T}$-factorization approach give a much higher cross section than the leading order collinear calculations. The origin is kinematical: off-shell initial partons carry finite transverse momentum (see Fig.~\ref{fig:2D-kt1-kt2}), so the four-quark system acquires non-zero total 
transverse momentum already at leading-order, which enlarges the available phase space and preferentially populates the low-mass region (see Fig.~\ref{fig:2D-kt-Mccbar}). However, in the present paper we are mainly interested in the region $80 < M_{c\bar c} < 110$ GeV. There, the SM SPS calculations shown in Fig.~\ref{fig:background-b} give compatible results.

In the following we discuss two-dimensional correlations and variations of the cuts of \eqref{eq:2.1} with the aim of finding suitable regimes where the $h''$ signal is larger than for the standard cut values \eqref{eq:2.1}. 
Note that here and in the following our \textit{$h''$ signal} should be understood as the \textit{MCPM' signal produced by the $h''$ boson}.
The peaks which we show are for the leading $c\bar c$ pairs which do not come exclusively from the $h''$ decay. Also contributions of $c (\bar c)$ from $h''$ decays together with $\bar c (c)$ associated quark can contribute. Thus, for example, one should not try to estimate the width of the $h''$  
boson directly from these peaks.

\subsubsection{Two dimensional correlations}

\begin{figure}[h]
  \begin{subfigure}{0.32\textwidth}
    \centering
        \includegraphics[width=1.0\textwidth]{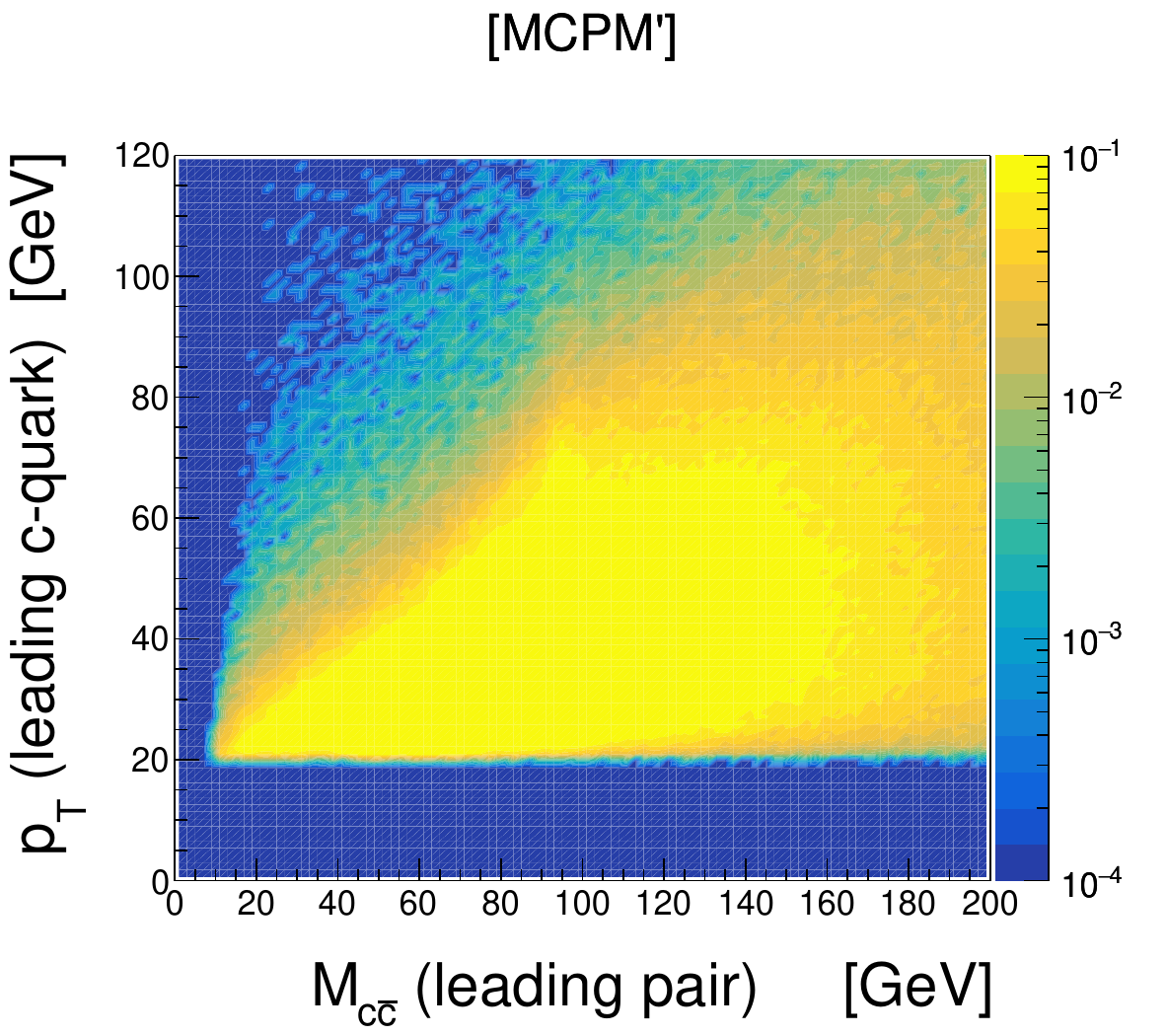}
       \caption{}
      \label{fig:2D-Mccbar-pTc-a}
      \end{subfigure}
  \begin{subfigure}{0.32\textwidth}
    \centering
        \includegraphics[width=1.0\textwidth]{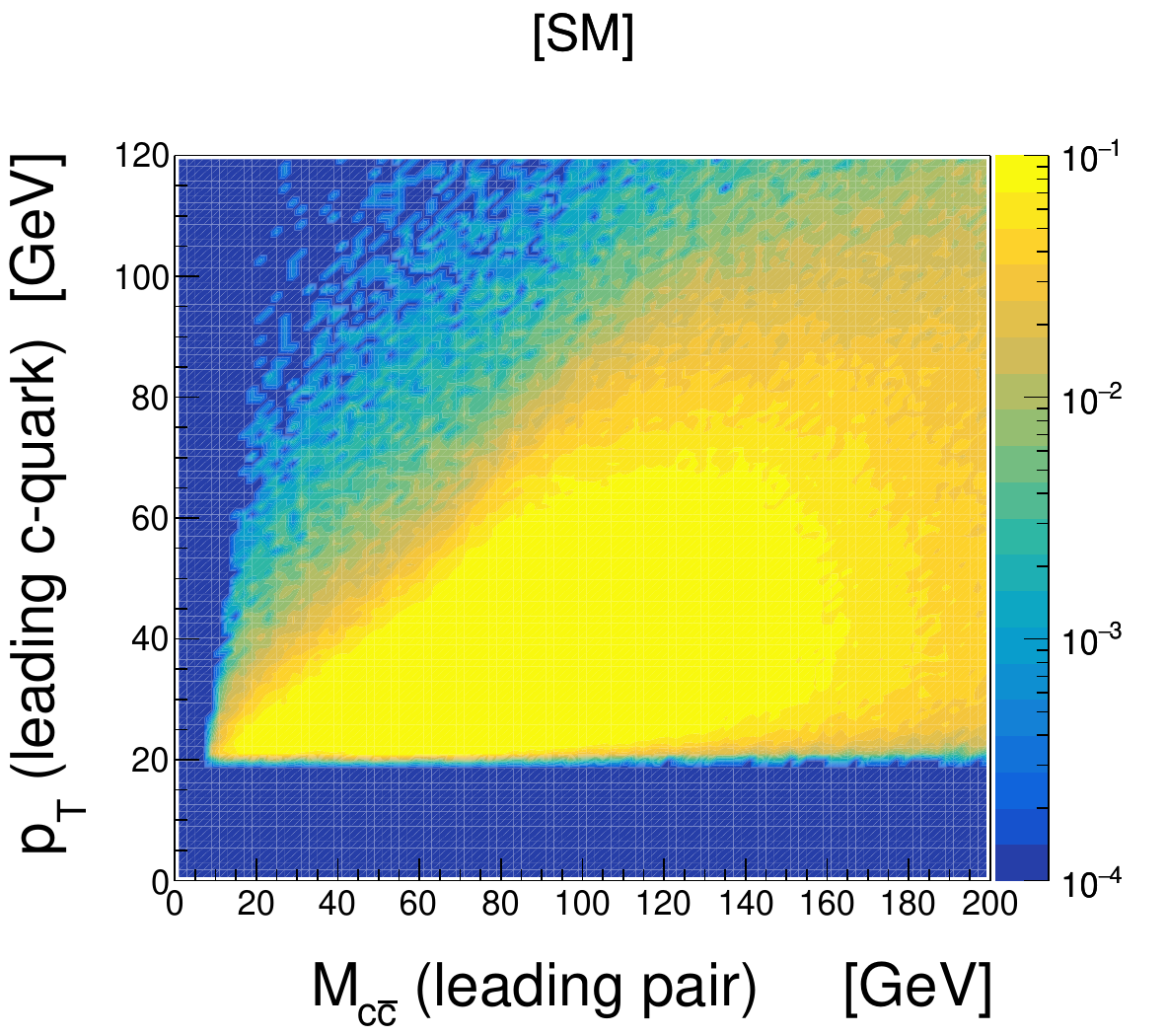}
       \caption{}
      \label{fig:2D-Mccbar-pTc-b}
      \end{subfigure}
  \begin{subfigure}{0.32\textwidth}
    \centering
        \includegraphics[width=1.0\textwidth]{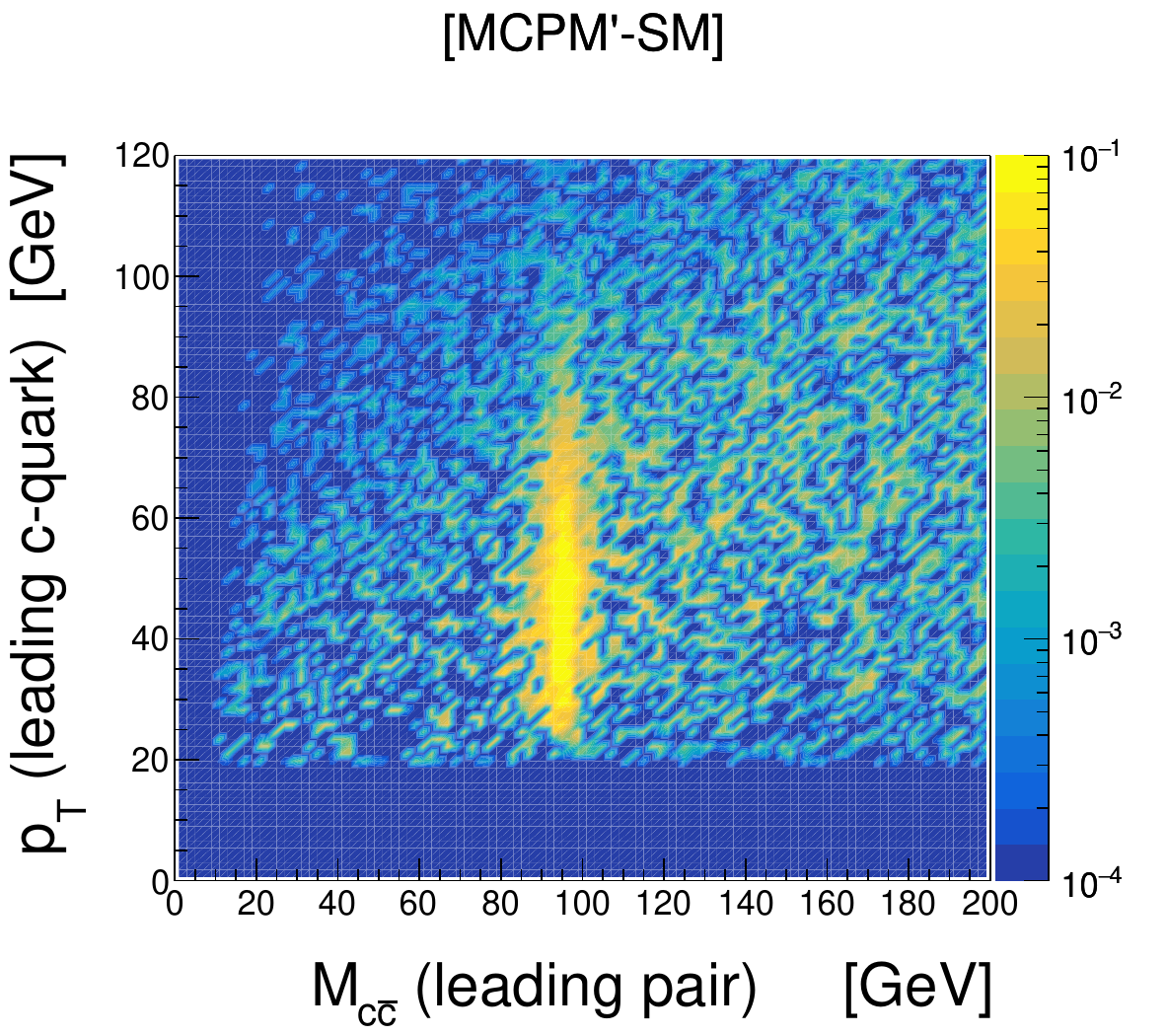}
       \caption{}
      \label{fig:2D-Mccbar-pTc-c}
      \end{subfigure}
    \caption{Double differential cross sections: $\frac{d^{2}\sigma}{dM_{c\bar c}dp_{T}}$ as a function of the leading pair invariant mass and leading quark (or antiquark) transverse momentum for the MCPM' (a), the SM (b), and for the MCPM'$\!-$SM (c). The calculations are done with the SPS method and cuts \eqref{eq:2.1} are applied which require $p_{T} > 20$ GeV. }
    \label{fig:2D-Mccbar-pTc}
\end{figure}

\begin{figure}[h]
  \begin{subfigure}{0.5\textwidth}
    \centering
        \includegraphics[width=1.\textwidth]{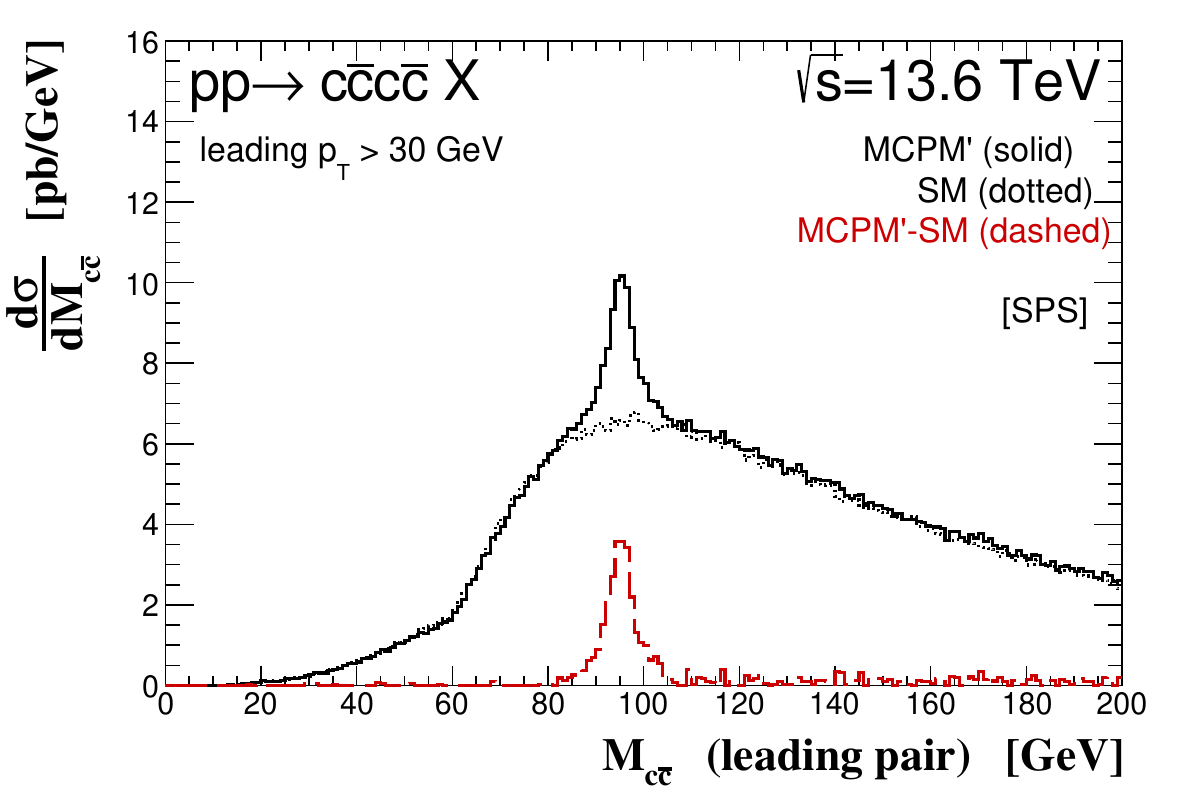}
  \caption{}
  \label{fig:M_extrapTcut-a}
  \end{subfigure}
  \begin{subfigure}{0.5\textwidth}
    \centering
        \includegraphics[width=1.\textwidth]{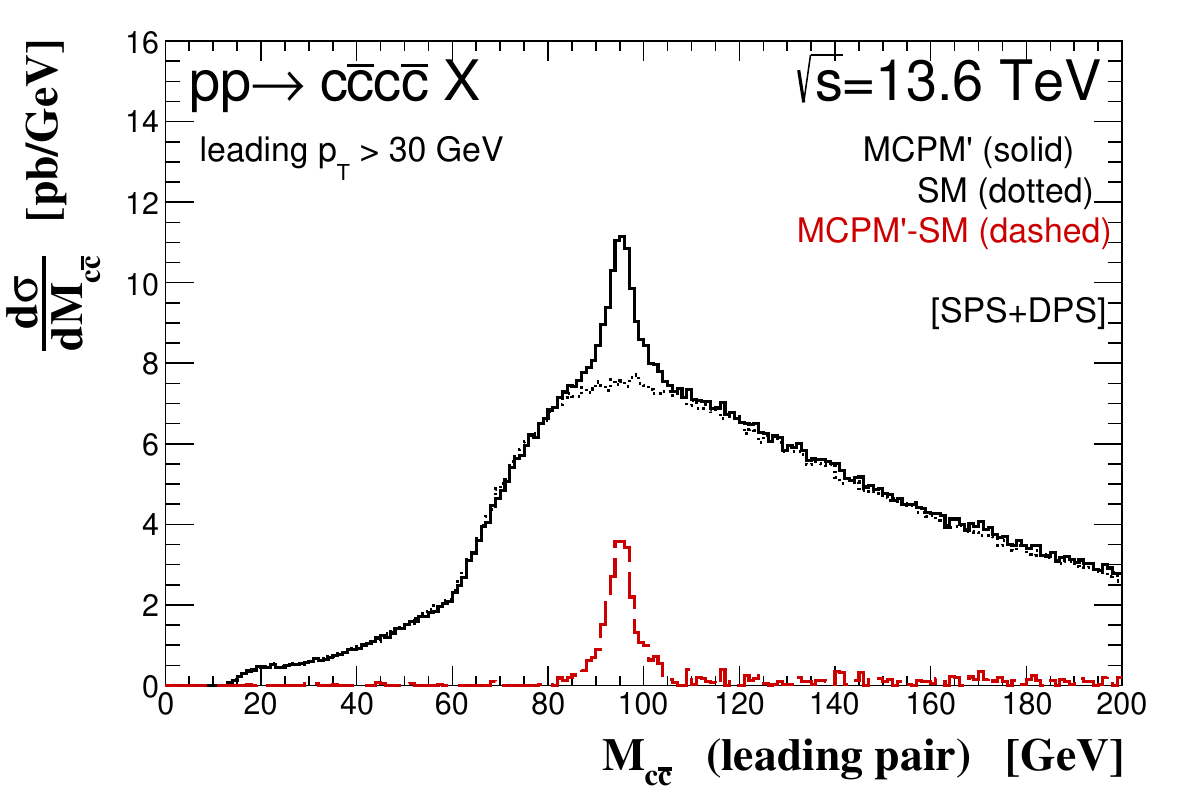}
  \caption{}
  \label{fig:M_extrapTcut-b}
  \end{subfigure}
    \caption{The differential cross sections as a function of the leading $c\bar c$-pair invariant mass $M_{c\bar c}$. The calculations for MCPM', SM, and MCPM'-SM are done with the following methods: in (a) with SPS, in (b) with SPS+DPS. Here the extra cut on the leading jet transverse momenta $p_{T} > 30$ GeV is applied.}
    \label{fig:M_extrapTcut}
\end{figure}

In this section, we investigate two-dimensional correlation distributions for the SM, the MCPM', and the signal MCPM'$\!-$SM. These are subsequently used to propose additional kinematic cuts aimed at enhancing the S/B ratio. We can observe from Fig.~\ref{fig:2D-Mccbar-pTc-c} a clear enhancement at 
$M_{c\bar c} \approx 95$ GeV in the signal MCPM'$\!-$SM distribution. We also see that in the signal region $M_{c\bar c} \approx 95$ GeV the MCPM'-SM distribution is mostly concentrated in the region of leading jets $30 < p_{T} < 60$ GeV. In contrast, from Figs.~\ref{fig:2D-Mccbar-pTc-a} and ~\ref{fig:2D-Mccbar-pTc-b} we observe that the background distributions give a significant contribution in the region of $20 < p_{T} < 30$ GeV. Therefore, it is reasonable to consider applying an extra cut of $p_{T} > 30$ GeV (for both, leading $c$-quark and leading $\bar c$-antiquark), which should enhance the signal-to-background ratio without a significant loss of signal. On the other hand, imposing an upper limit on the transverse momentum does not appear to be useful. The effect of applying such an additional lower cut on leading charm and anticharm jet transverse momenta is presented in Figure \ref{fig:M_extrapTcut}. There we show separately predictions for two cases.
In Fig.~\ref{fig:M_extrapTcut-a} the distributions for MCPM', SM, and MCPM'-SM are calculated using only the SPS mechanisms. In Fig.~\ref{fig:M_extrapTcut-b} the DPS contributions are also included.
The corresponding relative S/B ratios at the signal peak are found to be $55\%$ and $47\%$, respectively. The DPS contribution reduces the ratio, however, in both cases the numbers are slightly enhanced compared to the ones obtained  with the basic cuts \eqref{eq:2.1}, where the counterpart numbers are $52\%$ and $45\%$.

\begin{figure}[h]
  \begin{subfigure}{0.32\textwidth}
    \centering
        \includegraphics[width=1.0\textwidth]{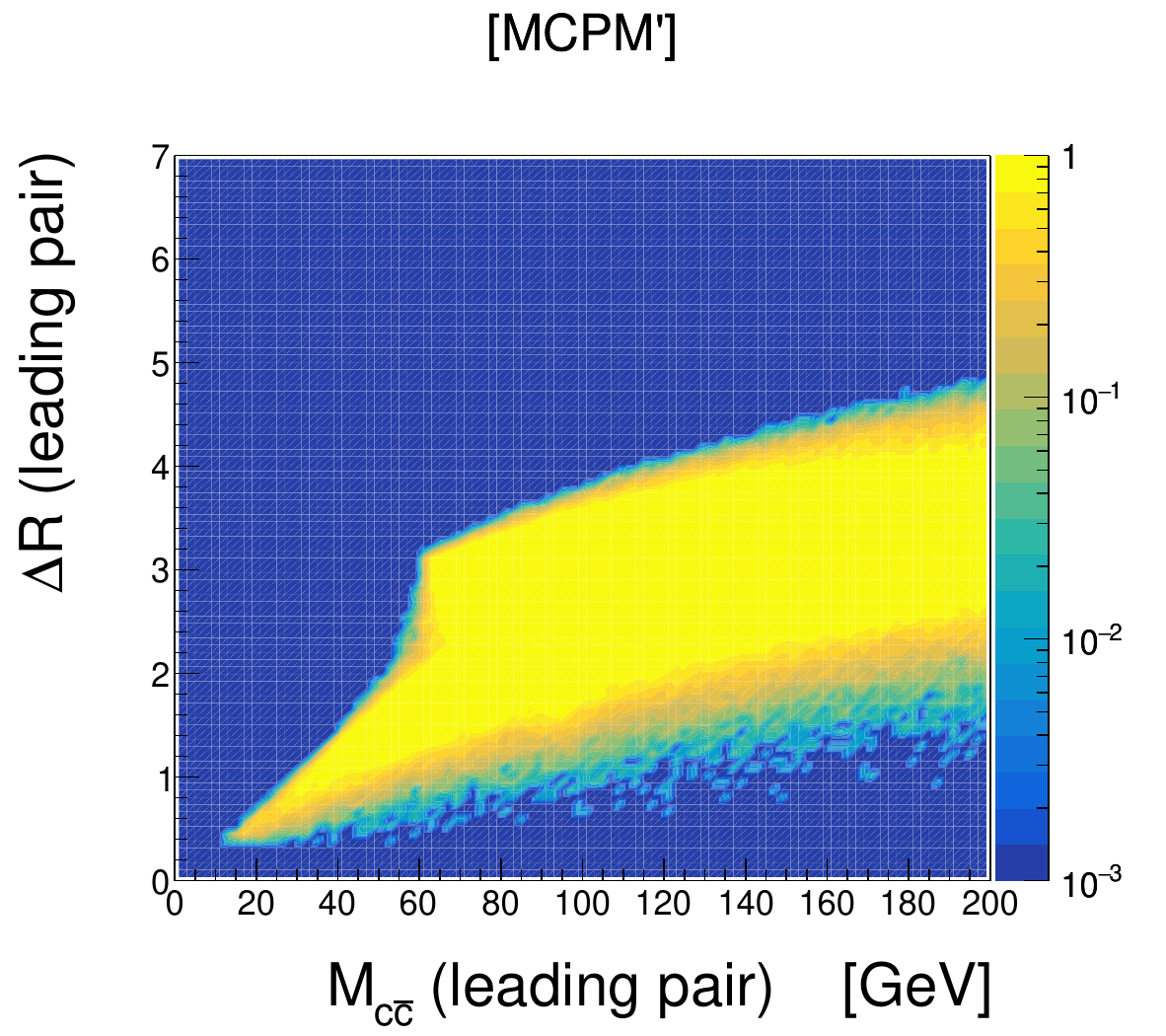}
       \caption{}
      \label{fig:2D-Mccbar-R-a}
      \end{subfigure}
  \begin{subfigure}{0.32\textwidth}
    \centering
        \includegraphics[width=1.0\textwidth]{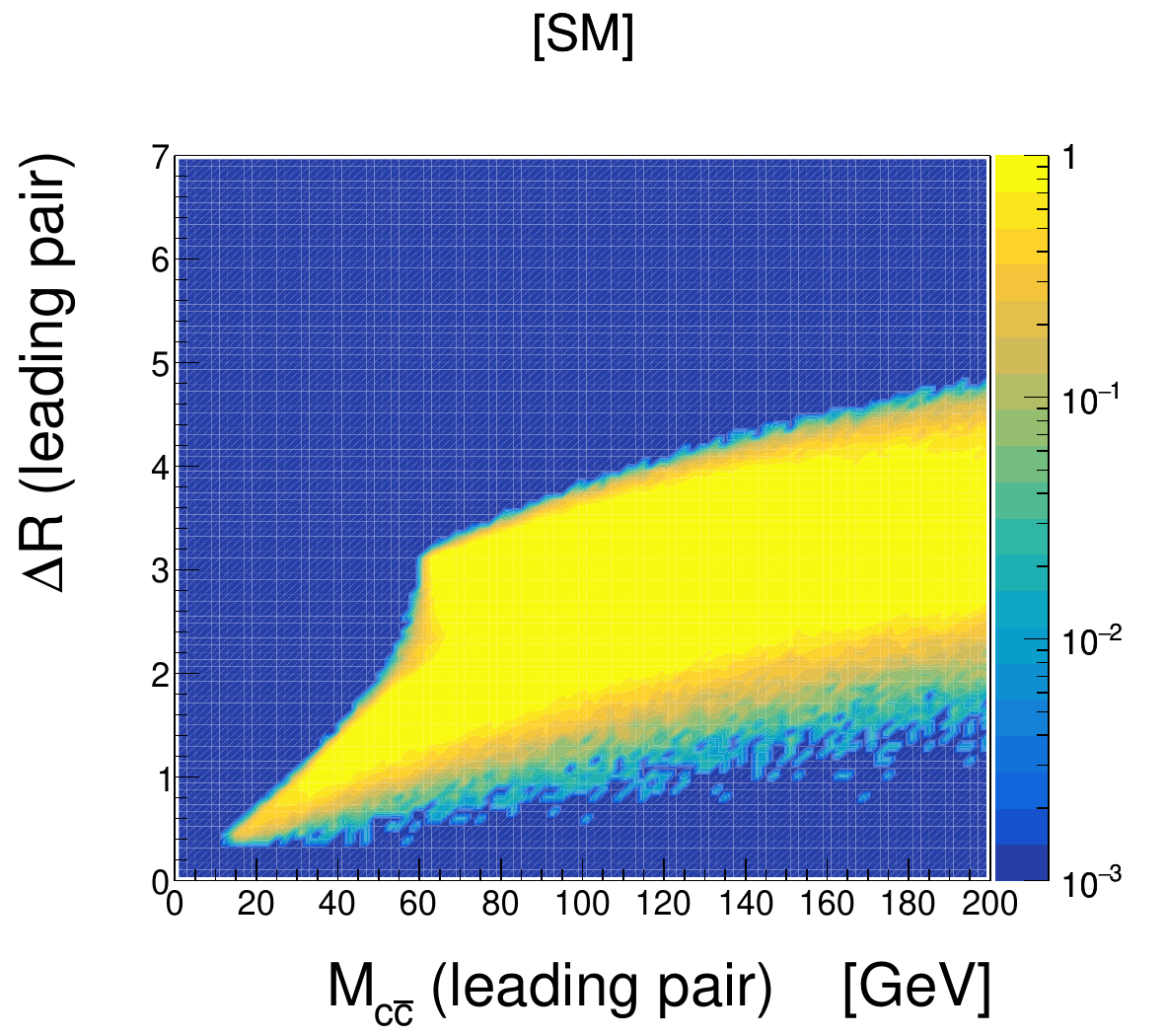}
       \caption{}
      \label{fig:2D-Mccbar-R-b}
      \end{subfigure}
  \begin{subfigure}{0.32\textwidth}
    \centering
        \includegraphics[width=1.0\textwidth]{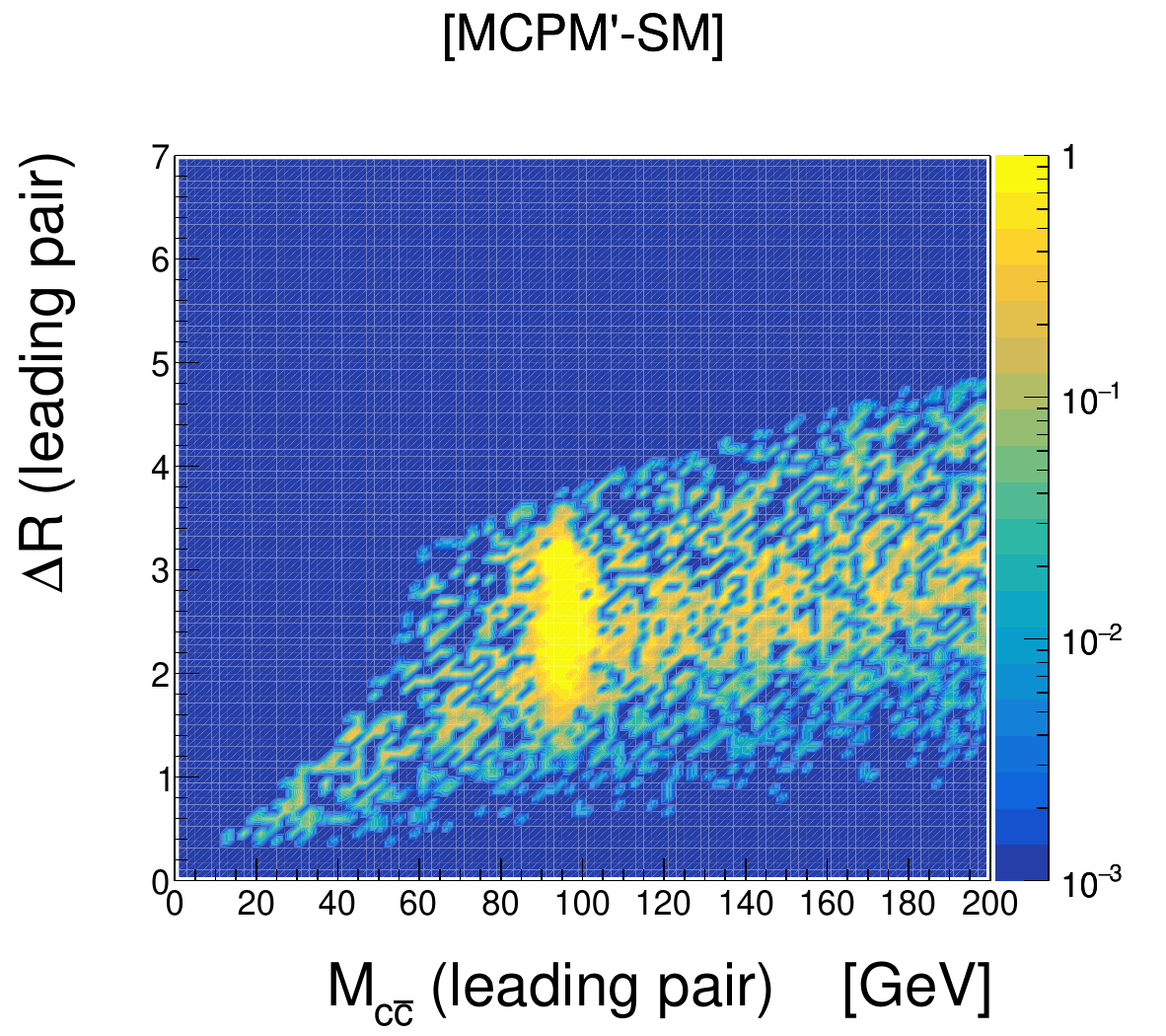}
       \caption{}
      \label{fig:2D-Mccbar-R-c}
      \end{subfigure}
    \caption{Double differential cross sections: $\frac{d^{2}\sigma}{dM_{c\bar c}d\Delta R}$ as a function of the leading pair invariant mass and leading pair $\Delta R$ for the MCPM' (a), the SM (b), and for the MCPM'$\!-$SM (c). These distributions were computed using the SPS method and applying cuts on the transverse momenta of the jets from the leading $c\bar c$ pair of $p_{T} > 30$ GeV.}
    \label{fig:2D-Mccbar-R}
\end{figure}

\begin{figure}[h]
  \begin{subfigure}{0.5\textwidth}
    \centering
        \includegraphics[width=1.\textwidth]{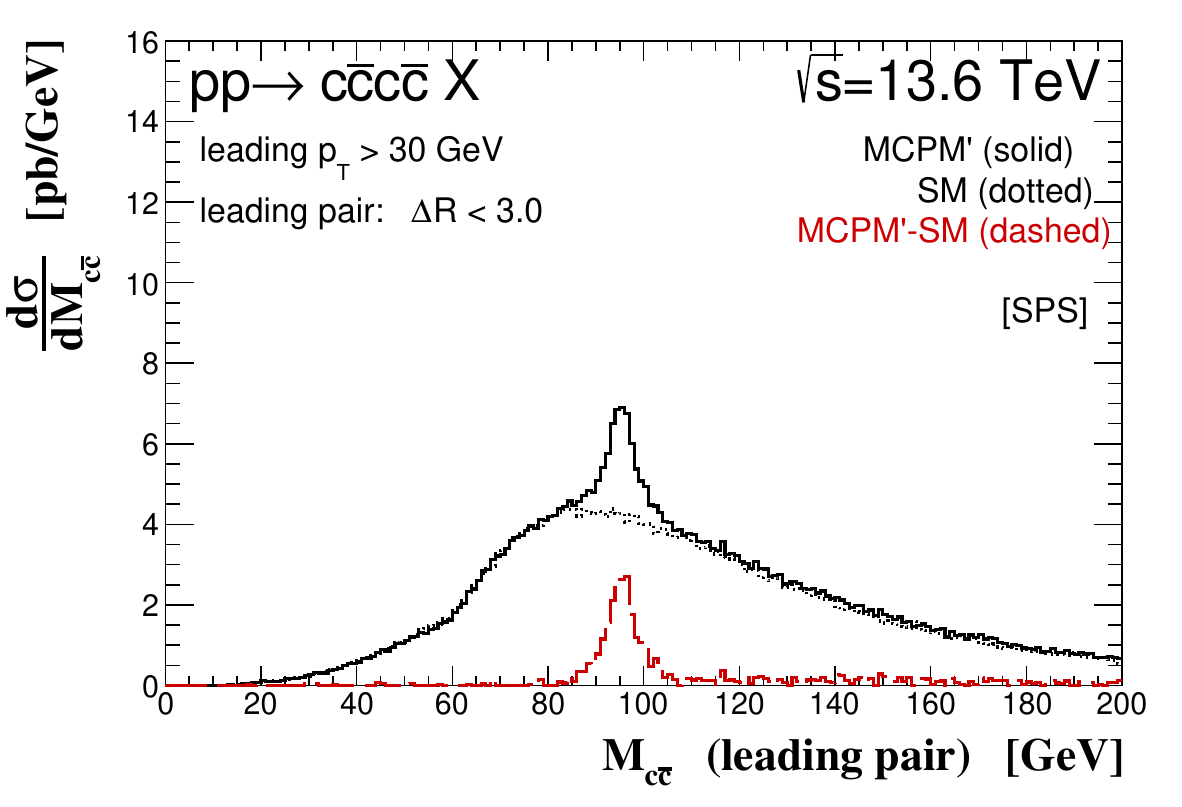}
  \caption{}
  \label{fig:M_extrapTcut_extraRcut-a}
  \end{subfigure}        
  \begin{subfigure}{0.5\textwidth}
    \centering
        \includegraphics[width=1.\textwidth]{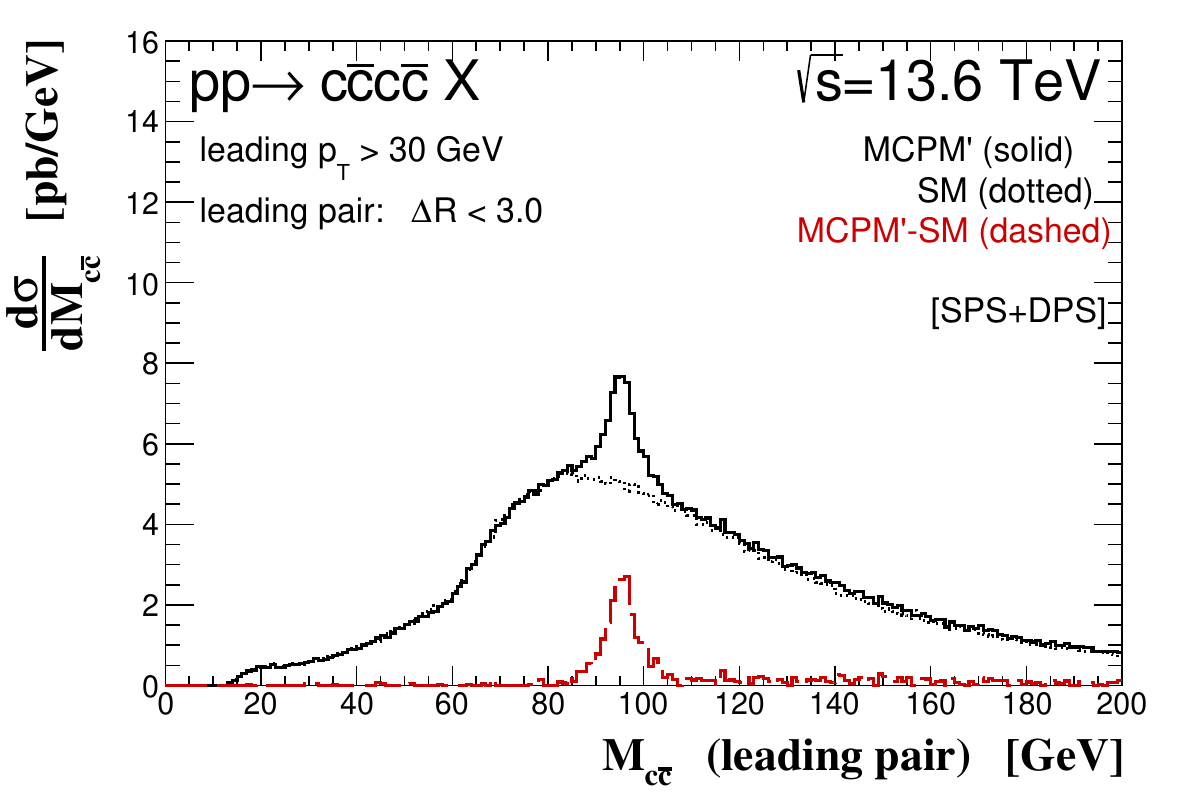}
  \caption{}
  \label{fig:M_extrapTcut_extraRcut-b}
  \end{subfigure}        
    \caption{The differential cross sections as a function of the leading $c\bar c$-pair invariant mass $M_{c\bar c}$. The calculations for MCPM', SM, and MCPM'-SM are done with the following methods: in (a) with SPS, in (b) with SPS+DPS. Here the extra cut on the leading jet transverse momenta $p_{T} > 30$ GeV is applied, together with the cut $\Delta R < 3.0$.}
    \label{fig:M_extrapTcut_extraRcut}
\end{figure}

Figure \ref{fig:2D-Mccbar-R} presents kinematical correlations between the invariant mass of the leading $c\bar c$-pair and the leading pair $\Delta R$. Here again Figs.~\ref{fig:2D-Mccbar-R-a}, ~\ref{fig:2D-Mccbar-R-b}, and \ref{fig:2D-Mccbar-R-c} correspond to the MCPM', to the SM, and to MCPM'$\!-$SM, respectively. These distributions were computed after applying cuts on the leading transverse momenta of the jets. The analysis of these distributions indicates a potential benefit in enhancing the signal-to-background ratio by applying an upper limit on $\Delta  R$ \eqref{eq:2.11}, specifically $\Delta R < 3.0$. Rejection of $\Delta R > 3.0$ excludes back-to-back configurations in the azimuthal angle between leading $c$ and $\bar c$ jets. The corresponding effect of the $\Delta R < 3.0$ limitation on the invariant mass distribution of the leading $c\bar c$-pair for the MCPM', the SM, and the difference MCPM'$\!-$SM, is shown in Figure \ref{fig:M_extrapTcut_extraRcut}. Here the S/B ratio values are only slightly larger than above and reach values of $63\%$ (SPS only) and $55\%$ (DPS inlcuded). 

\begin{figure}[h]
  \begin{subfigure}{0.32\textwidth}
    \centering
        \includegraphics[width=1.0\textwidth]{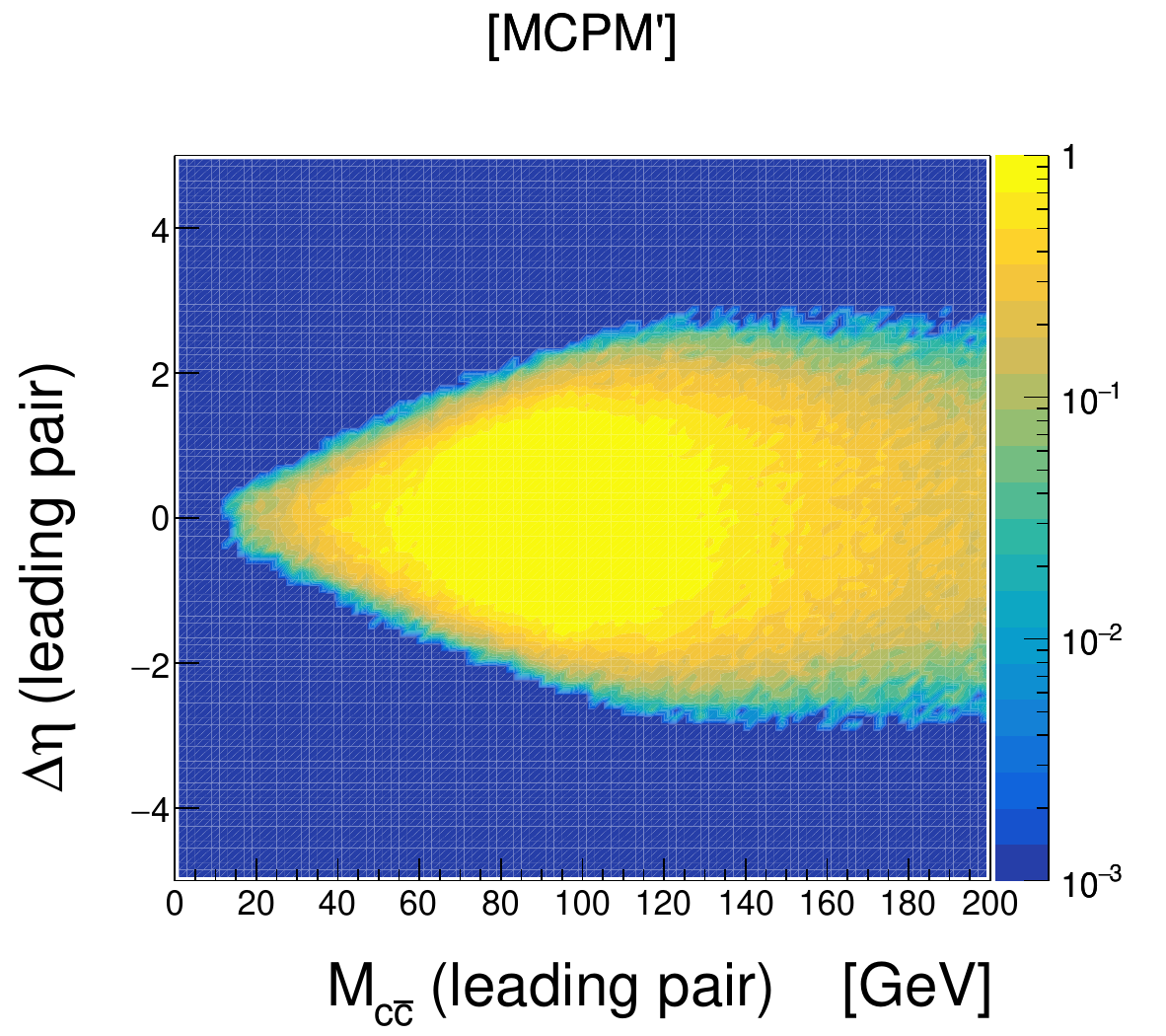}
       \caption{}
      \label{fig:2D-Mccbar-deta-a}
      \end{subfigure}
  \begin{subfigure}{0.32\textwidth}
    \centering
        \includegraphics[width=1.0\textwidth]{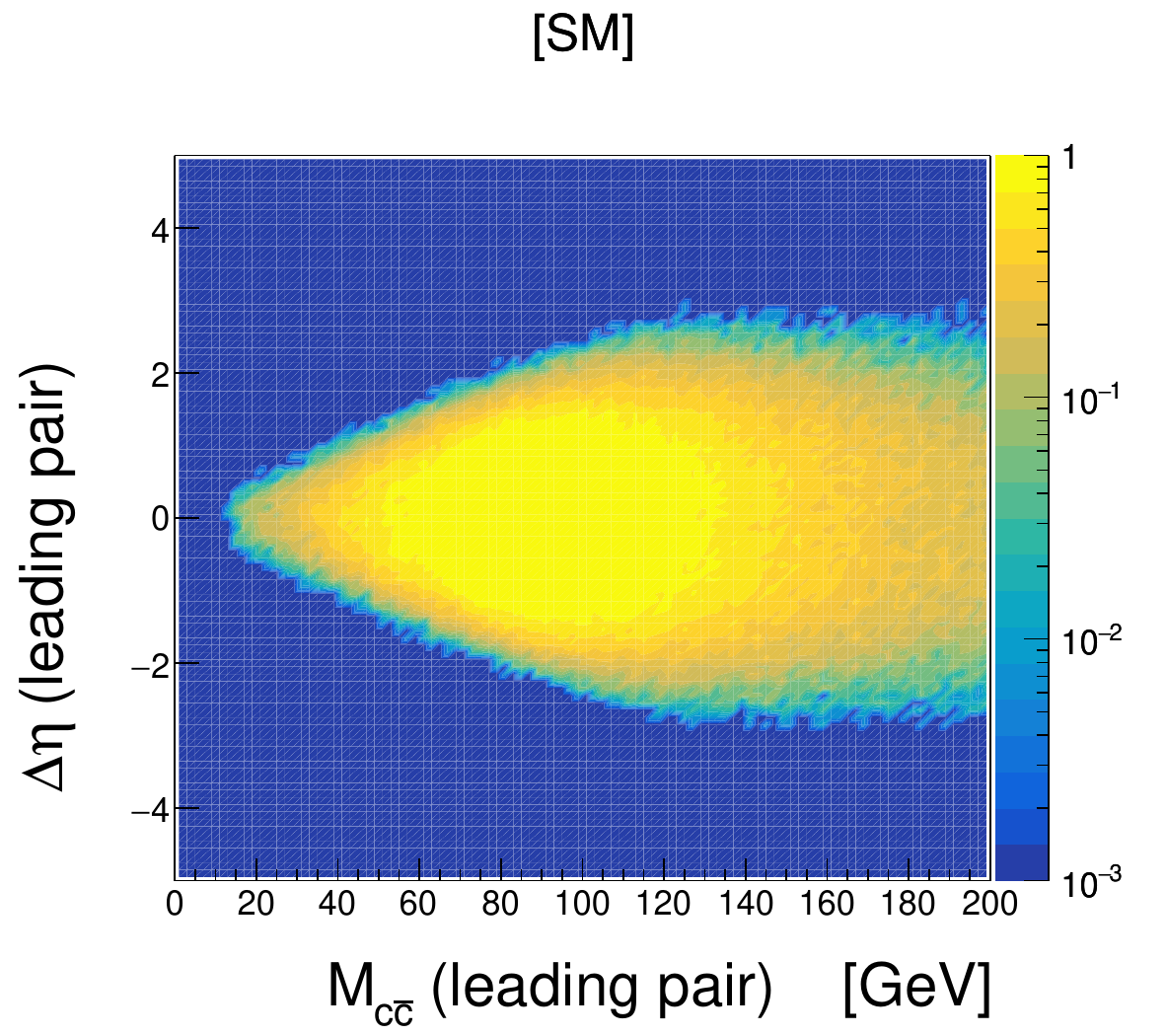}
       \caption{}
      \label{fig:2D-Mccbar-deta-b}
      \end{subfigure}
  \begin{subfigure}{0.32\textwidth}
    \centering
        \includegraphics[width=1.0\textwidth]{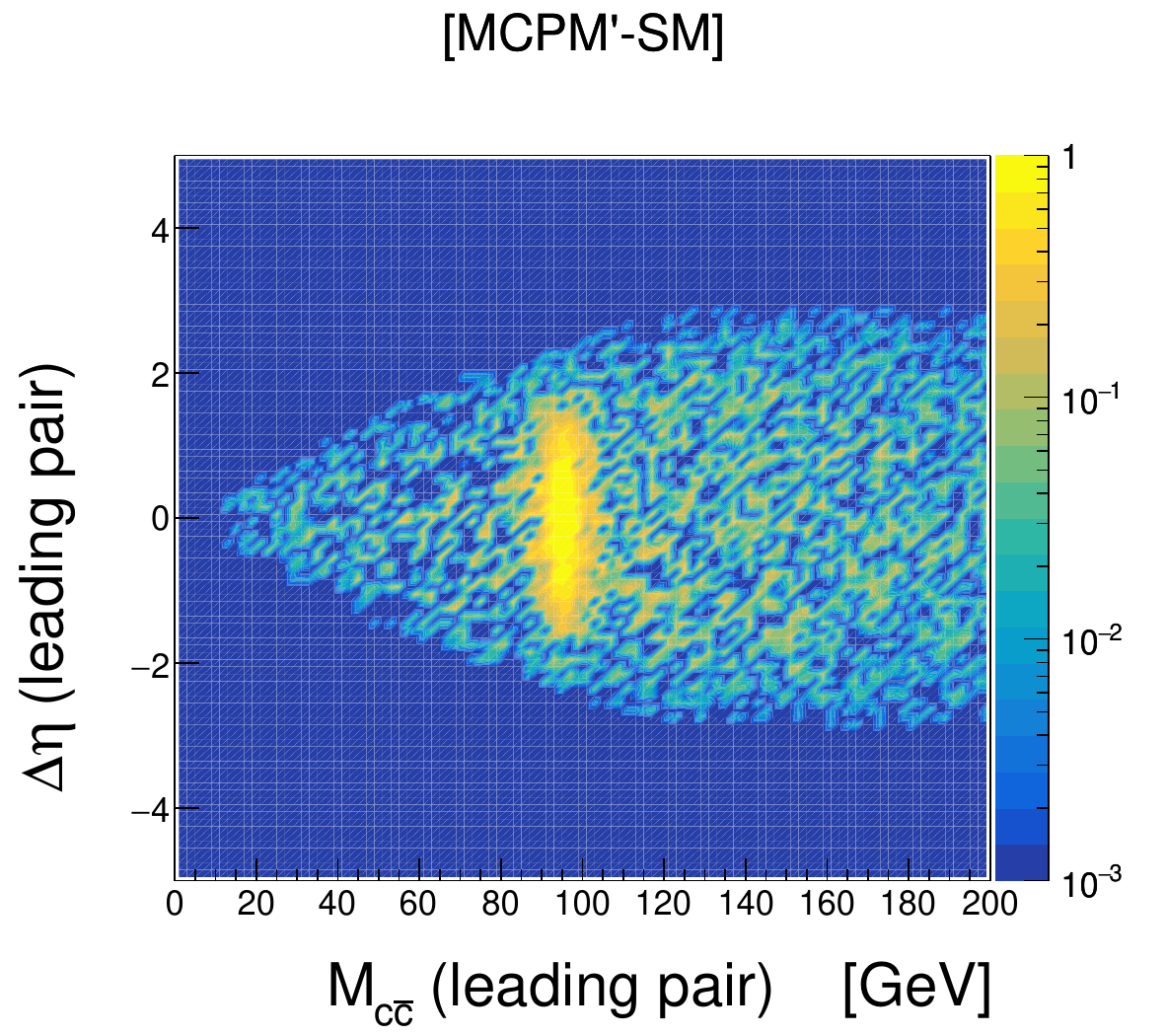}
       \caption{}
      \label{fig:2D-Mccbar-deta-c}
      \end{subfigure}
    \caption{Double differential cross sections: $\frac{d^{2}\sigma}{dM_{c\bar c}d\Delta \eta}$ as a function of the leading pair invariant mass and pseudorapidity distance between jets from the leading pair $\Delta \eta$ for the MCPM' (a), the SM (b), and for the MCPM'$\!-$SM (c). These distributions were computed using the SPS method and applying cuts on the transverse momenta of the jets from the leading $c\bar c$ pair of $p_{T} > 30$ GeV and $\Delta R < 3.0$.}
    \label{fig:2D-Mccbar-deta}
\end{figure}

\begin{figure}[h]
  \begin{subfigure}{0.5\textwidth}
    \centering
        \includegraphics[width=1.\textwidth]{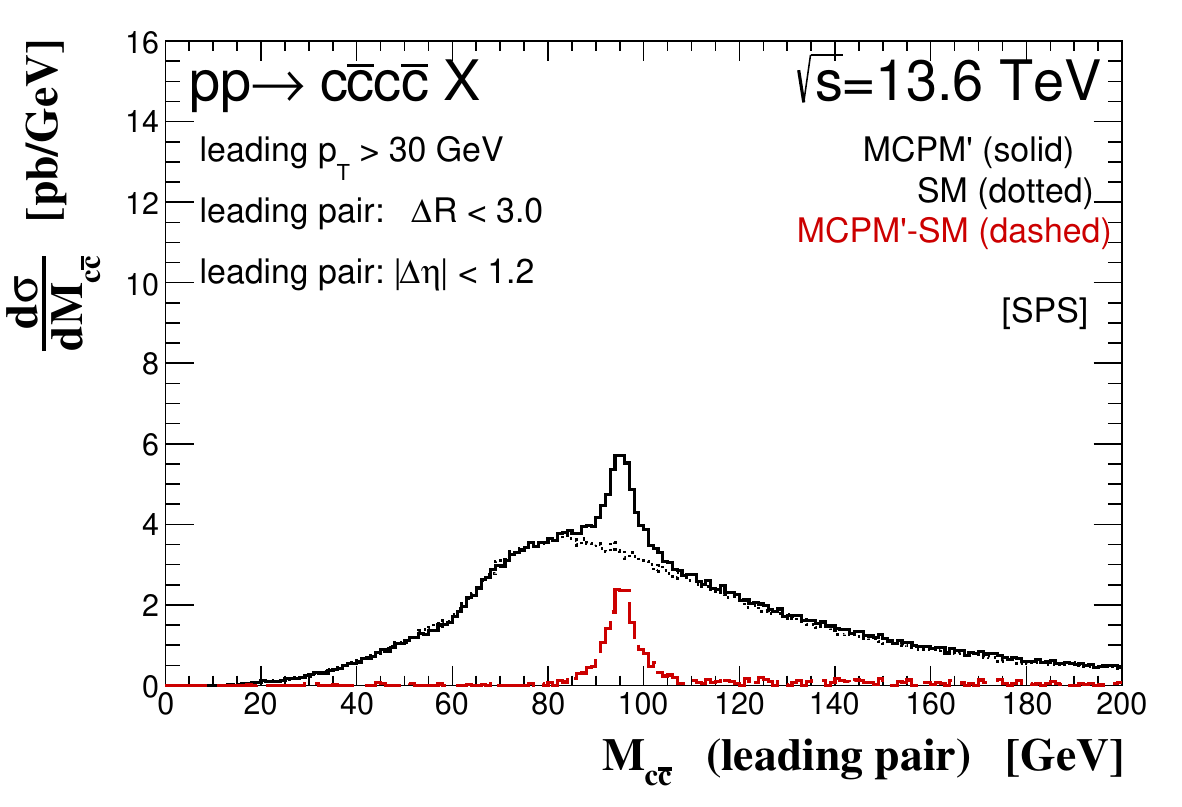}
  \caption{}
  \label{fig:M_extrapTcut_extraRcut_extraDetacut-a}
  \end{subfigure}                
   \begin{subfigure}{0.5\textwidth}
    \centering
        \includegraphics[width=1.\textwidth]{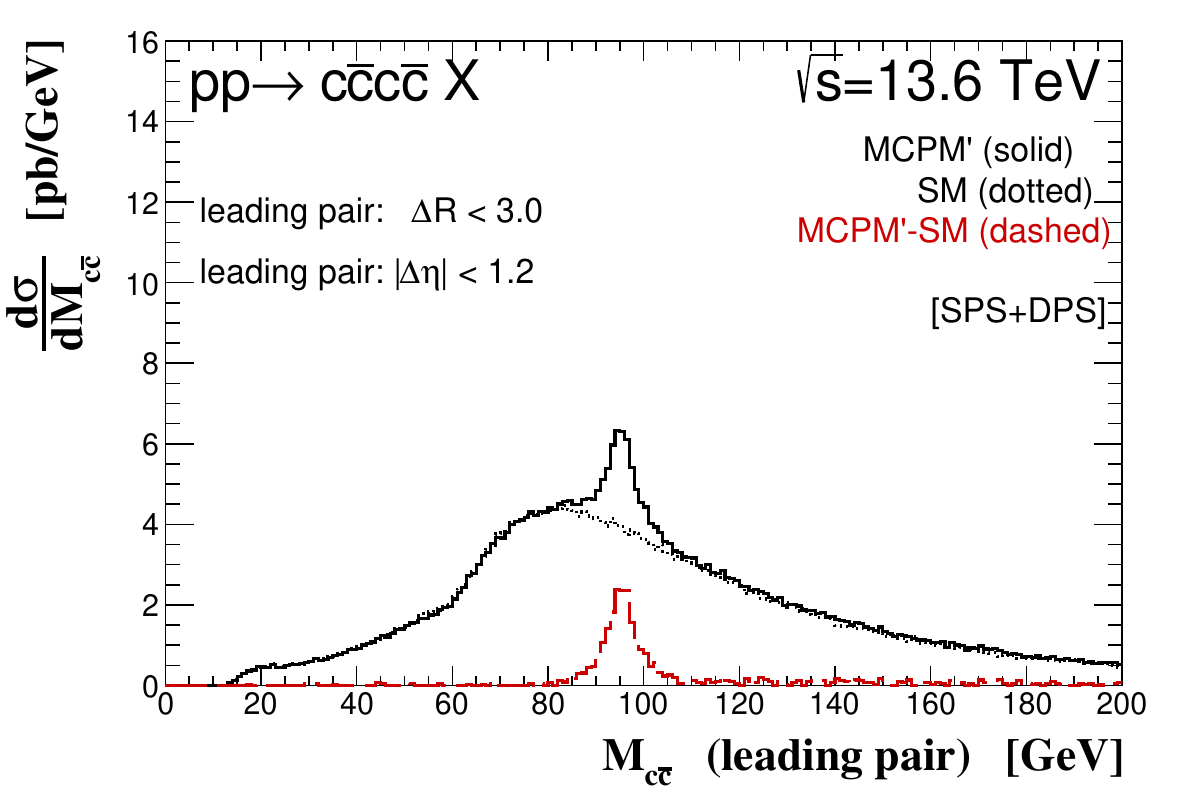}
  \caption{}
  \label{fig:M_extrapTcut_extraRcut_extraDetacut-b}
  \end{subfigure}        
    \caption{The differential cross sections as a function of the leading $c\bar c$-pair invariant mass $M_{c\bar c}$. The calculations for MCPM', SM, and MCPM'-SM are done with the following methods: in (a) with SPS, in (b) with SPS+DPS. Here the extra cut on the leading jet transverse momenta $p_{T} > 30$ GeV is applied, together with the cuts $\Delta R < 3.0$, and $|\Delta \eta| < 1.2$.}
    \label{fig:M_extrapTcut_extraRcut_extraDetacut}
\end{figure}

Figure \ref{fig:2D-Mccbar-deta} shows kinematical correlations between the invariant mass of the leading $c\bar c$-pair and the leading pair $\Delta \eta$.
The distributions were calculated under the condition that the jets forming the leading $c\bar c$ pair satisfy $p_{T}$ > 30 GeV and $\Delta R < 3.0$.
We observe that the MCPM'$-$SM signal (Fig.~\ref{fig:2D-Mccbar-deta-c}) is more concentrated at lower values of $\Delta \eta$ compared to the SM background (Fig.~\ref{fig:2D-Mccbar-deta-b}). Therefore, imposing a constraint of $|\Delta \eta| < 1.2$ may lead to a further improvement in the signal-to-background ratio. The effect of the cuts requiring for the leading $c\bar c$-pair jets $p_{T} > 30$ GeV, $\Delta R < 3.0$, and $|\Delta \eta| < 1.2$ is shown in Fig.~\ref{fig:M_extrapTcut_extraRcut_extraDetacut}. Here we see that these cuts increase the S/B ratios significantly to the level of $73\%$ (SPS only) and $63\%$ (DPS included).

%%%%%%%%%%%%%%%%%%%%%%%%%%%%%%%%%%%%%%
\subsubsection{Optimizing cuts}

We are interested in optimized cuts which give the highest significance of the process~\eqref{eq:h2cc}, that is, $p + p \to h'' c \bar{c}X$ with subsequent decay $h'' \to c \bar{c}$, with respect to the Standard Model background.  We vary various cuts and compute the significance in order to detect the optimal cut values.

\begin{figure}[h]
  \begin{subfigure}{0.5\textwidth}
    \centering
        \includegraphics[width=1.\textwidth]{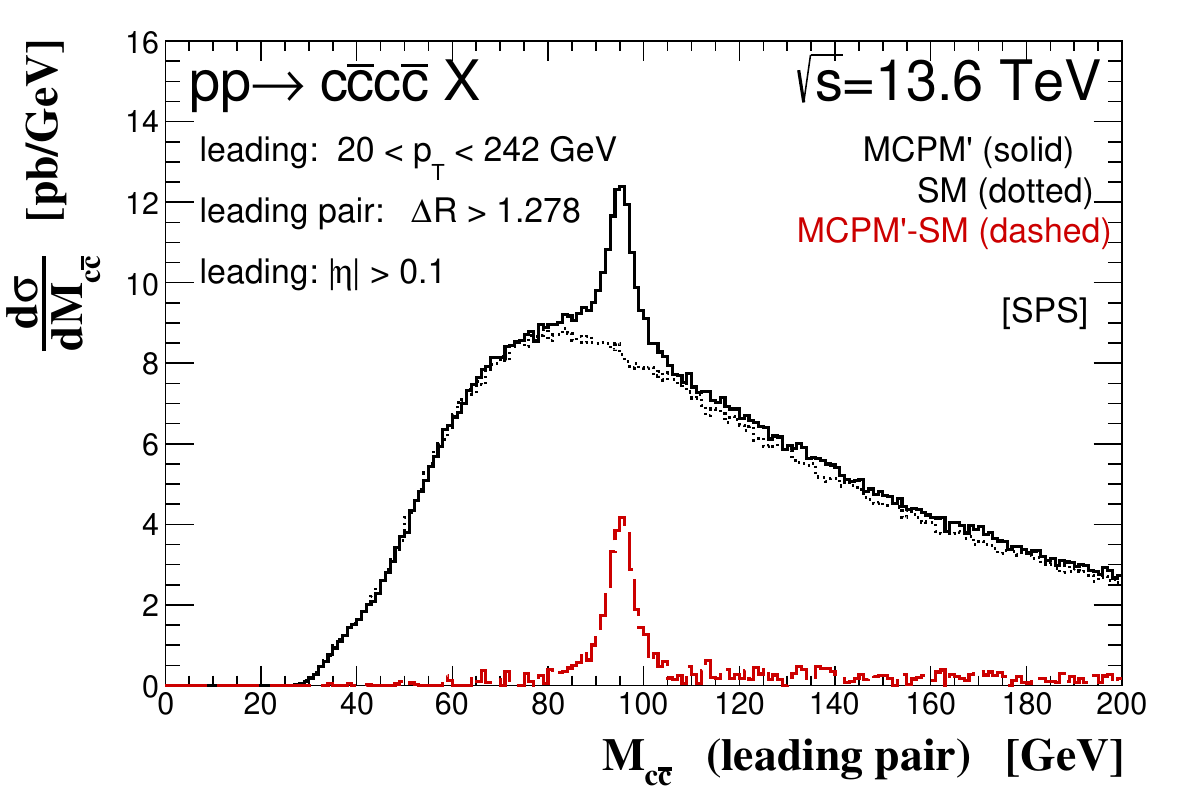}
  \caption{}
  \label{plot_inv_mass_MCPM_SM_cuts-a}
  \end{subfigure}                
  \begin{subfigure}{0.5\textwidth}
    \centering
        \includegraphics[width=1.\textwidth]{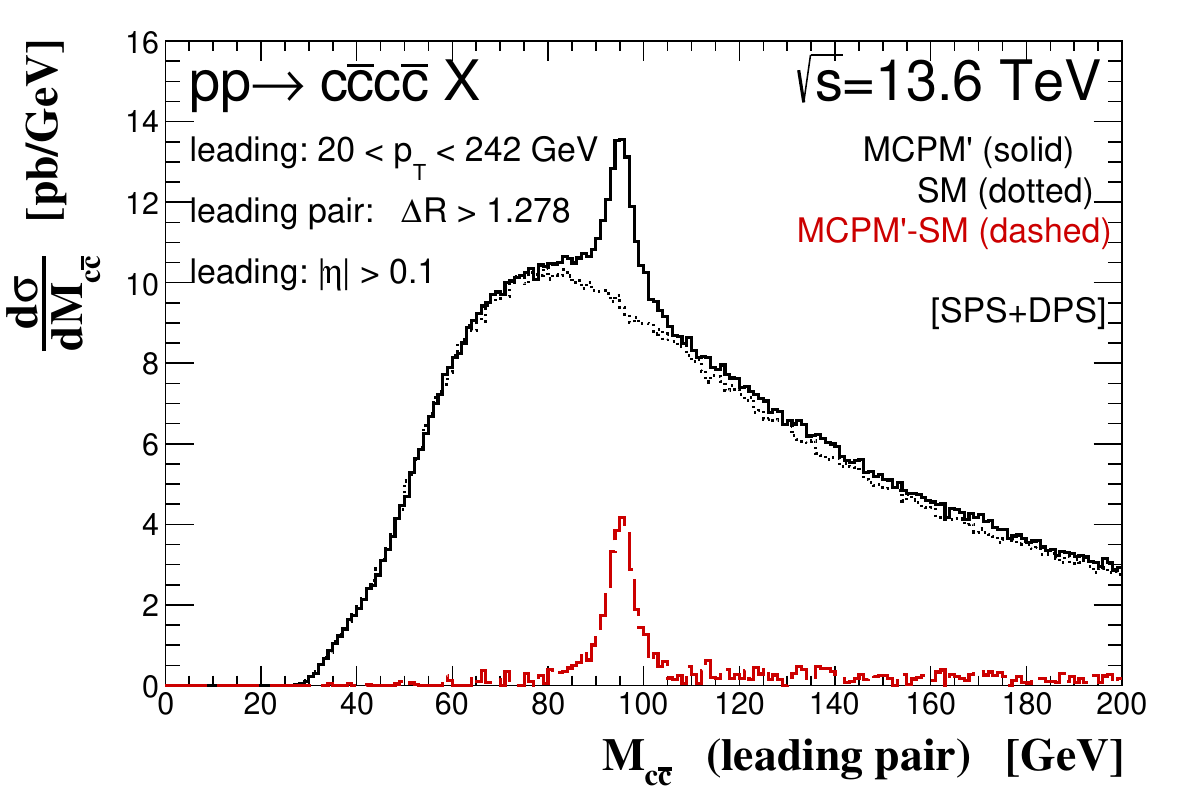}
  \caption{}
  \label{plot_inv_mass_MCPM_SM_cuts-a}
  \end{subfigure} 
\caption{\label{plot_inv_mass_MCPM_SM_cuts} The differential cross sections as a function of the leading $c\bar c$-pair invariant mass $M_{c\bar c}$.The calculations for MCPM', SM, and MCPM'-SM are done with the following methods: in (a) with SPS, in (b) with SPS+DPS. Here we apply the optimized cuts~\eqref{optimize_cuts}.}
\end{figure}

 For this optimization we vary (1) the minimal $p_T$ required for each quark of the leading pair, (2) the maximal $p_T$ of each quark of the leading pair,  (3) the minimal $\Delta R$ of the leading pair and (4) the minimal pseudorapidity of each quark forming the leading pair.  
This optimization is done varying these four parameters within a large range. For each chosen set of the four parameters, the significance is computed.
The significance is here the number of signal events given by events originating from $p + p \to h'' + c + \bar{c} + X$ with $h'' \to c \bar{c}$ divided by the square root of the sum of the number of signal events and the number of SM background events $p + p \to c + \bar{c} + c + \bar{c} + X$.
We find numerically in this four-dimensional search for the best significance the following optimized cut values:
\begin{equation} \label{optimize_cuts}
20~\text{GeV} < p_T < 242~\text{GeV}, \qquad
\Delta R > 1.278, \qquad
|\eta| > 0.1
\end{equation}

Applying these optimized cuts~\eqref{optimize_cuts} we compute the differential cross section with respect to the invariant mass of the leading charm pair. This differential cross section is shown in Fig.~\ref{plot_inv_mass_MCPM_SM_cuts}.
Comparing Figs.~\ref{plot_inv_mass_MCPM_SM_cuts} and 
\ref{fig:1-rafal-b} we see almost the same results with very similar S/B ratios.

\section{Conclusions}

In the present paper we have worked out further consequences of the model MCPM' for experiments at the LHC. The MCPM' is a two-Higgs-doublet model (2HDM) which has been introduced in \cite{Maniatis:2023aww}. The Higgs sector of the MCPM'contains, as in any 2HDM, three neutral Higgs bosons ($\rho', h', h'' $) and a charged pair ($H^{\pm}$).
The CP-symmetry relations forming the basis of the MCPM' fix the couplings of these Higgs bosons to the fermions to a large extent. The $\rho'$ couples to the fermions like the SM Higgs particle. The bosons
$h'$ and $h''$ couple mainly to charm quarks in a scalar and pseudoscalar way, respectively. The
corresponding coupling constants are, however, proportional to the top-quark mass.

In \cite{Maniatis:2023aww} the MCPM' was compared to the results for $\gamma \gamma$ production from \cite{CMS:2023yay} where a possible enhancement at $M_{\gamma \gamma} = 95.4$ GeV was reported. Choosing the mass of the pseudoscalar boson $h''$
of the MCPM' at this value compatibility of the MCPM' with the results of \cite{CMS:2023yay} was found.

This gave us the motivation to look for further experimental consequences at the LHC of a possible $h''$
pseudoscalar Higgs boson with the properties dictated by the MCPM'. We have studied here in detail
the production of the $h''$ associated with a charm-anticharm pair,
\begin{equation} \label{4.1}
p + p \to h'' + c + \bar{c} + X.
\end{equation} 
According to theory the $h''$ boson of mass $95.4$ GeV decays to nearly 100\% to $h'' \to c + \bar{c}$. Thus the final state at the LHC is  
\begin{equation} \label{4.2}
p + p \to c + \bar{c} + c + \bar{c} + X.
\end{equation}
Of course, this final state can also be populated by SM processes.

Our main findings are as follows. The total cross section for \eqref{4.1} is large, of the order of $12 ~\text{nb}$ (see \eqref{eq:2.1f_dps}). This has to be compared 
to the Drell-Yan cross section (see \eqref{eq:2.1e}) 
\begin{equation} \label{4.3}
\left. \sigma_{\mathrm{tot}}(p + p \to h'' + X) \right|_{\text{DY}} \approx 15 ~\text{nb.}
\end{equation}

We assumed that the $c$ and $\bar c$ jets can be tagged. We introduced suitable cuts which
should make measurement of the charm and anticharm jets in the reaction \eqref{4.2} possible. A main requirement was to consider the leading charm and anticharm jets, that is,
those with the largest $p_{T}$. The probability of these jets to come from the $h''$ decay is high. With suitable cuts we found that the distribution in the invariant mass of the leading $c$ and $\bar c$ jets, $M_{c\bar c}$, shows a resonance-like structure at the $h''$ mass.
With very stringent cuts we get with the SPS+DPS calculation from Fig.~\ref{fig:M_extrapTcut_extraRcut_extraDetacut-b}
a signal cross section at the peak of the resonance of
\begin{equation} \label{4.4}
\left.  \frac{d\sigma}{dM_{c\bar c}} \right|_{\text{MCPM'-SM}} \approx 2.4 ~\text{pb/GeV}
\end{equation}
over a background from the SM of
\begin{equation} \label{4.5}
\left. \frac{d\sigma}{dM_{c\bar c}} \right|_{\text{SM}} \approx 3.8 ~\text{pb/GeV}.
\end{equation}

We have taken into account contributions driven by the effective $h''GG$ coupling. However, the corresponding  
cross sections have been found to be much smaller than those obtained within the standard tree-level diagrams.
We have also discussed the reaction $p + p \to h'' + h'' + X$ with both $h''$ decaying to $c + \bar c$. However, the corresponding cross section is found to be negligible.

To summarize: we have carefully analyzed the cross section for the $p + p \to c + \bar c + c + \bar c + X$ reaction.
In addition to the signal process $p+p \to (h'' \to c + \bar c) + c + \bar c + X$, there are sizeable nonresonant backgrounds, corresponding to the single-parton scattering and double-parton scattering mechanisms giving the same $c + \bar c + c + \bar c + X$ final state.
In our paper we have assumed 100\% charm tagging.
We have calculated the distributions of the invariant mass of the leading $c\bar c$ quark pair in the MCPM' applying appropriate kinematical cuts. 
Our analysis has shown relatively large cross sections and promising resonance-like enhancements within the MCPM' with respect to the SM results. We hope that our present study will be helpful to verify or reject the hypothesis that the 
enhancement observed by the CMS collaboration at $M_{\gamma \gamma}$ = 95.4 GeV is due to the $h''$ boson of the MCPM'. 

On the experimental side the identification of $c$ and $\bar c$ jets is improving (see \cite{LHCb:2015tna,Vertesi:2020gqr,CMS:2017wtu}).
For example, the LHCb collaboration has measured the cross section for $c \bar c$ dijets \cite{LHCb:2020frr} (see also \cite{Maciula:2022bfv}).
The CMS and ATLAS would need to perform their own analyses including 
jet missidentification probabilities.
In our opinion the ALICE-3 project \cite{ALICE:2022wwr} would be ideal 
in the future to perform experimental studies corresponding to our 
theoretical studies discussed here.

\acknowledgments

The authors would like to thank the participants of the EMMI-RRTF meeting "Next Generation Facility for Forward Physics at the LHC"
held in February 2025 in Heidelberg for interesting and stimulating discussions concerning the topics of the present paper. This work is supported, in part, by the Chilean project Fondecyt 1250132.

\appendix
\section{Partial decay width of $h''$ into a gluon pair and the effective  $h''$-gluon-gluon interaction}
\label{appA}

We now determine the effective $h''$--gluon--gluon interaction. This interaction has no 
tree-level contribution but arises from a triangle quark loop. Based on this interaction, 
we get in particular contributions to the process 
$p + p \to h'' + c + \bar{c} + X$ via gluon--gluon fusion; see Fig.~\ref{fig:100}. 
We estimate that for a mass $m_{h''} \sim 95.4~\mathrm{GeV}$ in proton--proton collisions, 
with a center-of-mass energy of $13.6~\mathrm{TeV}$, the corresponding Bjorken-$x$ for the 
gluons is about $0.007$. Gluon densities in the protons are known to be very high at 
low Bjorken-$x$. On the other hand, the coupling of a gluon pair to the $h''$ boson is 
loop suppressed, with the dominant contribution arising from the charm-quark loop.
In the following, we use the $\gamma$-matrix conventions of~\cite{Bjorken:1965zz} and the 
Levi-Civita symbol with the normalization $\varepsilon_{0123} = 1$.

We shall first calculate the matrix elements for the decay 
of $h'' \to GG$ and the production $GG \to h''$. We have:
\begin{equation}\label{eq:A1}
    h''(k) \to G(p_1, \varepsilon_1, a) + G(p_2, \varepsilon_2, b), 
    \tag{A.1}
\end{equation}
\noindent and:
\begin{equation}
    G(p_1, \varepsilon_1, a) + G(p_2, \varepsilon_2, b) \to h''(k), 
    \tag{A.2}
\end{equation}
\noindent where $k, p_1, p_2$ with $k = p_1 + p_2$ are the momenta of the particles, and 
$\varepsilon_{1,2}$, $a$, $b$ are the polarization vectors and color indices of 
the gluons, respectively. From CPT invariance we have:
\begin{align}
    \bra{G(p_1, \varepsilon_1, a), G(p_2, \varepsilon_2, b)} T \ket{h''(k)} 
    &= \bra{h''(k)} T \ket{G(p_1, -\varepsilon_1^*, a), G(p_2, -\varepsilon_2^*, b)}.
    \tag{A.3}
\end{align}

\begin{figure}[h!]
\centering
\begin{tikzpicture}
  \begin{feynman}
    % Wierzchołki (czubki trójkąta pętli)
    \vertex (h) {\( h''(k) \)};
    \vertex [right=1.5cm of h] (v1);
    \vertex [above right=1.8cm of v1] (v2);
    \vertex [below right=1.8cm of v1] (v3);
    \vertex [right=2.0cm of v2] (g1) {\( G(p_1,\varepsilon_1,a) \)};
    \vertex [right=2.0cm of v3] (g2) {\( G(p_2,\varepsilon_2,b) \)};

    % Diagram
    \diagram* {
      (h) -- [scalar, thick,] (v1),
      (v1) -- [fermion, thick, edge label=\( q \)] (v2) -- [fermion, thick, edge label=\( q \)] (v3) -- [fermion, thick, edge label=\( q \)] (v1),
      (v2) -- [gluon, thick,] (g1),
      (v3) -- [gluon, thick,] (g2),
    };
          \draw [fill, black] (v1) circle (.05);
          \draw [fill, black] (v2) circle (.05);
          \draw [fill, black] (v3) circle (.05);                    
  \end{feynman}

  % Tekst: + (perm)
  \node at ($(g1)!0.5!(g2)+(4.2cm,0)$) {\normalsize \( + \left( G(p_1,\varepsilon_1,a) \leftrightarrow G(p_2,\varepsilon_2,b) \right) \)};
\end{tikzpicture}
\caption{Diagrams at one-loop level for the decay \( h'' \to GG \) \eqref{eq:A1}, with the quarks $q=c$ and $q=s$ in the loop.}
\label{fig:A1}
\end{figure}
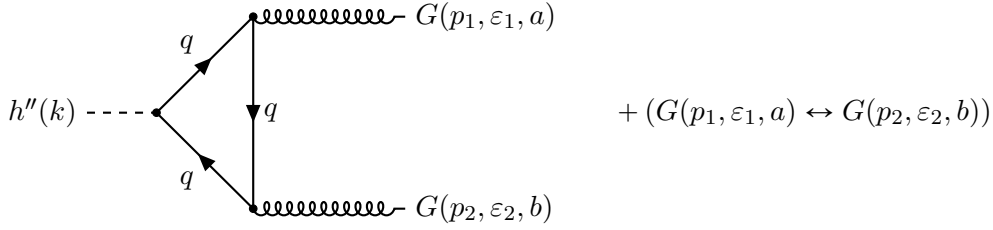
At one loop level we have in the MCPM' for \eqref{eq:A1} the diagrams shown in Fig.~\ref{fig:A1}. The evaluation of the diagrams of Fig.~\ref{fig:A1} is straightforward using the $h'' q \bar{q}$ couplings from (A.81) of Appendix~A of~\cite{Maniatis:2023aww}:
\begin{align}
    i\,\Gamma^{(h'' c c)}(k_1, k_2) &= \frac{m_t}{v_0} \gamma_5, \nonumber \\
    i\,\Gamma^{(h'' s s)}(k_1, k_2) &= -\frac{m_b}{v_0} \gamma_5, \nonumber \\
    v_0 = 246~\mathrm{GeV}. 
    \tag{A.4}
\end{align}
\noindent Here $v_0$ is the standard Higgs vacuum-expectation value.
We get:
\begin{align}\label{eq:A5}
    \bra{G(p_1, \varepsilon_1, a),\, G(p_2, \varepsilon_2, b)}\, T \ket{h''(k)} 
    &= \delta_{ab} \cdot \frac{2\alpha_s}{\pi} \cdot \frac{1}{m_{h''}^2} \,
    \varepsilon^{\mu \nu \rho \sigma} \, 
    \varepsilon_{1\mu}^* \, \varepsilon_{2\nu}^* \, p_{1\rho} \, 
p_{2\sigma} \nonumber \\
    &\quad \times \left\{
        \frac{m_t\, m_c}{v_0} \, f\left( \frac{4m_c^2}{m_{h''}^2} \right)
        - \frac{m_b\, m_s}{v_0} \, f\left( \frac{4m_s^2}{m_{h''}^2} \right)
    \right\},
    \tag{A.5}
\end{align}
where
\begin{equation}\label{eq:A6}
f(z) =
\begin{cases}
-\dfrac{1}{4} \left[ \ln \left( \dfrac{1 + \sqrt{1 - z}}{1 - \sqrt{1 - z}} \right) - i\pi \right]^2,
& \text{for } 0 < z < 1, \\[10pt]
\left[ \arcsin \left( \sqrt{1/z} \right) \right]^2,
& \text{for } z \geq 1.
\end{cases}
\tag{A.6}
\end{equation}
\noindent
In \eqref{eq:A6}, the positive values for the square roots have to be taken.

Now we shall construct an effective \( h'' \)--gluon--gluon coupling:
\begin{equation}
\label{eq:A7}
\mathcal{L}^{(h''GG)}_{\text{eff}}(x) = \mathcal{C}_{h''GG} \, h''(x) \, G^a_{\mu\nu}(x) \, \widetilde{G}^{a\mu\nu}(x),
\tag{A.7}
\end{equation}
\noindent where
\begin{equation}
\widetilde{G}^{a\mu\nu}(x) = \frac{1}{2} \, \varepsilon^{\mu\nu\rho\sigma} \, G^a_{\rho\sigma}(x), \nonumber
\end{equation}

\noindent which we require to reproduce the result \eqref{eq:A5}. In \eqref{eq:A7}, \( \widetilde{G}^{a\mu\nu}(x) \) is the gluonic dual field-strength tensor, and 
\( \mathcal{C}_{h''GG} \) is a complex constant which we determine below. From \eqref{eq:A7} we get:
\begin{align}
    \bra{G(p_1, \varepsilon_1, a),\, G(p_2, \varepsilon_2, b)}\, T \ket{h''(k)}_{\text{eff}} 
    = \delta_{ab} \cdot 4 \mathcal{C}_{h''GG} \cdot 
    \varepsilon^{\mu \nu \rho \sigma} \,
    \varepsilon_{1\mu}^* \, \varepsilon_{2\nu}^* \, p_{1\rho} \, p_{2\sigma}.
    \tag{A.8}
\end{align}
\noindent Comparing with \eqref{eq:A5}, we find:
\begin{equation}
\mathcal{C}_{h''GG} = 
\frac{\alpha_s}{2\pi} \cdot \frac{1}{m_{h''}^2} \left\{
\frac{m_t m_c}{v_0} \, f\left( \frac{4m_c^2}{m_{h''}^2} \right)
- \frac{m_b m_s}{v_0} \, f\left( \frac{4m_s^2}{m_{h''}^2} \right)
\right\}.
\tag{A.9}
\end{equation}
\noindent With the mass values for the quarks listed in~\cite{ParticleDataGroup:2024cfk} and 
\( \alpha_s = 0.12 \), we find numerically:
\begin{equation}
\mathcal{C}_{h''GG} = (-3.01 + i2.53) \times 10^{-5} ~\text{GeV}^{-1}.
\tag{A.10}
\end{equation}

\section{Digression on total cross sections and comparison of $h''$ and $Z$ distributions}
\label{appCS}

Now we want to calculate the total cross section for $h''$ production, 
composed of the cross sections for the Drell--Yan process with $c\bar{c}$ fusion and the $GG$ fusion process discussed in (37) and (38) of \cite{Maniatis:2023aww}
and the process \eqref{eq:1.5} here. For this we shall first make some general considerations.

\begin{figure}[!ht]
\centering
\begin{tikzpicture}
  \begin{feynman}
  %place first vertices
    \vertex (i1) at (0,1.5);
    \vertex (i2) at (0,-1.5);
     \vertex (a) at (2.0, 1.5) ;
     \vertex (b) at (2.0, -1.5) ;

     \vertex (g) at (3.2,-1.2);
     \vertex (g1) at (6.7,1.75) {\( c \)};
     \vertex (g2) at (6.7,-1.75) {\( \bar{c} \)};
  
       \vertex (c1) at (4.5, 1.2); 
     \vertex (c) at (5.0, 0);     
            \vertex (c2) at (4.5, -1.2); 

\vertex (d1) at (6.5, 1.7);
\vertex (d2) at (6.5, -1.7);

     \vertex[red] (z1) at (6.49,0.62) {\( hard \)};

     \vertex[teal] (z2) at (2.2,2.0) {\( soft \)};
     \vertex[teal] (z3) at (2.2,-2.0) {\( soft \)};

     \vertex[gray] (z4) at (6.3,1.2) { \scriptsize \( p_{T} \)};
     \vertex[gray] (z5) at (6.3,-1.2) { \scriptsize \( p'_{T} \)};

     \vertex (h) at (6.5, 0.0) {\( h'' \)};

     \vertex (X1) at (4, 1.8) ;
     \vertex (X1a) at (4, 2.0) ;
     \vertex (X1b) at (4, 1.6) ;
     \vertex (X2) at (4, -1.8) ;
     \vertex (X2a) at (4, -1.6) ;
     \vertex (X2b) at (4, -2.0) ;
    \diagram* {
      (i1) -- [plain, very thick, edge label=\(p\)] (a),
      (i2) -- [plain, very thick,edge label'=\(p\)] (b),
      (a) -- [gluon, thick, edge label'=\(G\)] (c1),
      (b) -- [gluon, thick, edge label=\(G\)] (c2),

      %(b) -- [gluon, thick] (g),

      (c2) -- [fermion, thick] (c),
      (c) -- [fermion, thick] (c1), 
      (c1) -- [fermion, thick] (d1),
      (d2) -- [fermion, thick] (c2),

      (h) -- [scalar, thick] (c),

      (a) -- [plain, thick] (X1),
      (a) -- [plain, thick] (X1a),
      (a) -- [plain, thick] (X1b),
      (b) -- [plain, thick] (X2),      
      (b) -- [plain, thick] (X2a),      
      (b) -- [plain, thick] (X2b),      
    };
    
\draw[red, thin] (4.5,-0.8) rectangle (7.,0.8);
   	\draw [fill, gray] (a) circle (.2);
 	\draw [fill, gray] (b) circle (.2);
 	\draw [fill] (c) circle (.055);
 	 	\draw [fill] (c1) circle (.035);
 	 	 	\draw [fill] (c2) circle (.035);
 	 	 	
\draw[gray, dotted, thick] (4.5,1.2) -- (6.5,1.0); 	 	 	
\draw[arrows = {-Stealth[length=5pt, inset=2pt]}, gray, thick] (6.,1.05) -- (6.01,1.6); 

\draw[gray, dotted, thick] (4.5,-1.2) -- (6.5,-1.0); 	 	 	
\draw[arrows = {-Stealth[length=5pt, inset=2pt]}, gray, thick] (6.,-1.05) -- (6.01,-1.6);

  \end{feynman}
\end{tikzpicture}
 
\caption{\label{fig:300}
A contribution to the Drell-Yan process $c\bar c \to h''$.}
\end{figure}
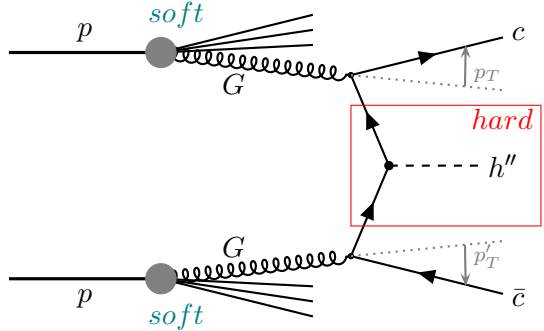

In the Drell--Yan (DY) process we consider $c$ and $\bar{c}$ from the wave 
functions of the two protons, respectively, which fuse to give $h''$. 
One type of diagrams contributing is shown in Fig.~\ref{fig:300}: two gluons split into $c$ and $\bar{c}$ 
which fuse to give $h''$.

The wave functions, respectively the distribution functions, of the protons are evolved up to a certain scale and then folded with the hard cross section for $c\bar{c} \to h''$. We take the relevant scales, the factorization scale $\mu_F$ and the renormalization scale $\mu_R$, to be
\begin{equation}
\mu_F = \mu_R = m_{h''} = 95.4 \; \text{GeV}. 
\end{equation}
This follows the recommendations of \cite{Ellis:1996mzs}. The evolution of the proton wave function can also be considered as follows in a simple physical picture; see \cite{Kogut:1974ni,Kogut:1973ub} and also sections 19.2 and 19.3 of \cite{Nachtmann:1990ta}. As we increase the energy of the $pp$ collision we increase the temporal resolution. Fluctuations with higher and higher transverse momenta have to be included in the effective proton wave functions. Thus, in Fig.~\ref{fig:300} $c$ and $\bar{c}$ quarks with transverse momenta
\begin{equation}
p_T \leq p_{T,\text{soft},\max}
\end{equation}
should be considered as part of the proton wave functions.

Now we come to the reaction \eqref{eq:1.5}, where the diagrams are essentially the same as in Fig.~\ref{fig:300}, 
but now with a different partition of soft and hard parts.
\begin{figure}[!ht]
\centering
\begin{tikzpicture}
  \begin{feynman}
  %place first vertices
    \vertex (i1) at (0,1.5);
    \vertex (i2) at (0,-1.5);
     \vertex (a) at (2.0, 1.5) ;
     \vertex (b) at (2.0, -1.5) ;

     \vertex (g) at (3.2,-1.2);
     \vertex (g1) at (6.7,1.75) {\( c \)};
     \vertex (g2) at (6.7,-1.75) {\( \bar{c} \)};
  
       \vertex (c1) at (4.5, 1.2); 
     \vertex (c) at (5.0, 0);     
            \vertex (c2) at (4.5, -1.2); 

\vertex (d1) at (6.5, 1.7);
\vertex (d2) at (6.5, -1.7);

     \vertex[red] (z1) at (6.49,0.62) {\( hard \)};

     \vertex[teal] (z2) at (2.2,2.0) {\( soft \)};
     \vertex[teal] (z3) at (2.2,-2.0) {\( soft \)};

     \vertex[gray] (z4) at (6.3,1.2) { \scriptsize \( p_{T} \)};
     \vertex[gray] (z5) at (6.3,-1.2) { \scriptsize \( p'_{T} \)};

     \vertex (h) at (6.5, 0.0) {\( h'' \)};

     \vertex (X1) at (4, 1.8) ;
     \vertex (X1a) at (4, 2.0) ;
     \vertex (X1b) at (4, 1.6) ;
     \vertex (X2) at (4, -1.8) ;
     \vertex (X2a) at (4, -1.6) ;
     \vertex (X2b) at (4, -2.0) ;
    \diagram* {
      (i1) -- [plain, very thick, edge label=\(p\)] (a),
      (i2) -- [plain, very thick,edge label'=\(p\)] (b),
      (a) -- [gluon, thick, edge label'=\(G\)] (c1),
      (b) -- [gluon, thick, edge label=\(G\)] (c2),

      %(b) -- [gluon, thick] (g),

      (c2) -- [fermion, thick] (c),
      (c) -- [fermion, thick] (c1), 
      (c1) -- [fermion, thick] (d1),
      (d2) -- [fermion, thick] (c2),

      (h) -- [scalar, thick] (c),

      (a) -- [plain, thick] (X1),
      (a) -- [plain, thick] (X1a),
      (a) -- [plain, thick] (X1b),
      (b) -- [plain, thick] (X2),      
      (b) -- [plain, thick] (X2a),      
      (b) -- [plain, thick] (X2b),      
    };
    
\draw[red, thin] (4.1,-1.99) rectangle (7.,1.99);
   	\draw [fill, gray] (a) circle (.2);
 	\draw [fill, gray] (b) circle (.2);
 	\draw [fill] (c) circle (.055);
 	 	\draw [fill] (c1) circle (.035);
 	 	 	\draw [fill] (c2) circle (.035);
 	 	 	
\draw[gray, dotted, thick] (4.5,1.2) -- (6.5,1.0); 	 	 	
\draw[arrows = {-Stealth[length=5pt, inset=2pt]}, gray, thick] (6.,1.05) -- (6.01,1.6); 

\draw[gray, dotted, thick] (4.5,-1.2) -- (6.5,-1.0); 	 	 	
\draw[arrows = {-Stealth[length=5pt, inset=2pt]}, gray, thick] (6.,-1.05) -- (6.01,-1.6);

  \end{feynman}
\end{tikzpicture}
 
\caption{\label{fig:400}
Partition of soft and hard parts in a diagram contributiong to the reaction \eqref{eq:1.5}.}
\end{figure}
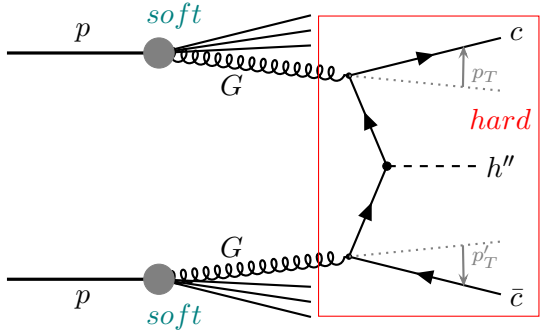
Here two gluons of the proton wave functions initiate the hard process
\begin{equation}
G + G \to c + \bar{c} + h''.
\end{equation}
Clearly, to avoid double counting, we must require
\begin{equation}
p_T, \; p_T' \; > \; p_{T,\text{hard},\min},
\end{equation}
with
\begin{equation}
\label{eq.255}
p_{T,\text{hard},\min} \;\geq\; p_{T,\text{soft},\max}.
\end{equation}
Ideally we should have the equality in \eqref{eq.255} and we should be able 
to shift the border value to some extent freely.

Below we compare the obtained integrated cross sections of \eqref{eq:2.1b}--\eqref{eq:2.1f_dps} for associated $h''+ c \bar c$ production to the corresponding total cross sections for $h''$ production (see e.g. numbers given in (37) and (38) of \cite{Maniatis:2023aww} for $\sqrt{s} = 13$ TeV). For the current LHC energy $\sqrt{s} = 13.6$ TeV and with the MSHT20lo$\!\_\!$as130 PDFs,  we find
\begin{equation} \label{eq:2.1e}
\left. \sigma_{\mathrm{tot}}(p + p \to h'' + X) \right|_{\text{DY}} = 14716.2 ~\text{pb.}
\end{equation} 
\begin{equation} \label{eq:2.1f}
\left. \sigma_{\mathrm{tot}}(p + p \to h'' + X) \right|_{\text{GG}} = 247.7 ~\text{pb.}
\end{equation} 
for the Drell-Yan process and $GG$ fusion, respectively. These numbers are obtained in the full phase space, i.e., without imposing any cuts on $h''$ transverse momentum. The cross sections \eqref{eq:2.1e} and \eqref{eq:2.1f} differ somewhat from those given in (37) and (38) of \cite{Maniatis:2023aww} for $\sqrt{s} = 13$ TeV:
\begin{equation} \label{eq:2.1ee}
\left. \sigma_{\mathrm{tot}}(p + p \to h'' + X) \right|_{\text{DY}} = 17874.5 ~\text{pb.}
\end{equation} 
\begin{equation} \label{eq:2.1ff}
\left. \sigma_{\mathrm{tot}}(p + p \to h'' + X) \right|_{\text{GG}} = 210.2 ~\text{pb.}
\end{equation}
We have checked that this difference is mainly due to the use of different PDFs, not primarily due to the slightly different c.m. energy. For our present paper 
the main observation is that the cross section for associated $h'' + c \bar c$ production \eqref{eq:2.1f_dps}
is large, around 78\% of the $h''$ cross section (\eqref{eq:2.1e} plus \eqref{eq:2.1f}).

Comparing \eqref{eq:2.1e} and \eqref{eq:2.1ee} we see that we have an uncertainty of this cross section of the order of 
20\% due to the PDF uncertainty. We note that these cross sections were calculated with the SPS collinear method.
Now we want to give a number for the total $h''$ production cross section, summing up the contributions from DY, GG and
associated $h''$ production; see \eqref{eq:2.1e}, \eqref{eq:2.1f} and \eqref{eq:2.1f_dps}. As we discussed above in this appendix 
there is the possible danger of the double counting. In order to avoid this one would have to calculate all cross sections
with some $k_{T}$-factorisation approach. But, as we see from Fig.~\ref{fig:background}, various of these approaches give rather different results. Also the SPS cross sections have an uncertainty of around 20\% as we have seen. Thus, at the moment we think that the best estimate which we can give for the total $h''$ cross section is to add the results of DY \eqref{eq:2.1e}, GG \eqref{eq:2.1f}, and associated $h''$ production \eqref{eq:2.1f_dps}. In this way we get a total cross section for $h''$ production and cross section times branching ratio for $h'' \to \gamma \gamma$ for $\sqrt{s}=13.6$ TeV as follows:
\begin{equation} \label{eq:2.1g}
\sigma_{\mathrm{tot}}(p + p \to h''+ X) = 26640.16 ~\text{pb.}
\end{equation}
The uncertainty of this number is estimated to be around 20\%. With the 
branching ratio $\text{B}(h''\to \gamma \gamma)$ as calculated in Eq.(28) of \cite{Maniatis:2023aww},
\begin{equation} \label{eq:2.1i}
\text{B}(h''\to \gamma \gamma) = 5.66 \times 10^{-7},
\end{equation} 
we get
\begin{equation} \label{eq:2.1h}
\sigma_{\mathrm{tot}}(p + p \to h''+ X) \times \text{B}(h''\to \gamma \gamma) = 1.51 \times 10^{-2} ~\text{pb.}
\end{equation} 

\begin{figure}[h]
  \begin{subfigure}{0.5\textwidth}
    \centering
    \includegraphics[width=1.\textwidth]{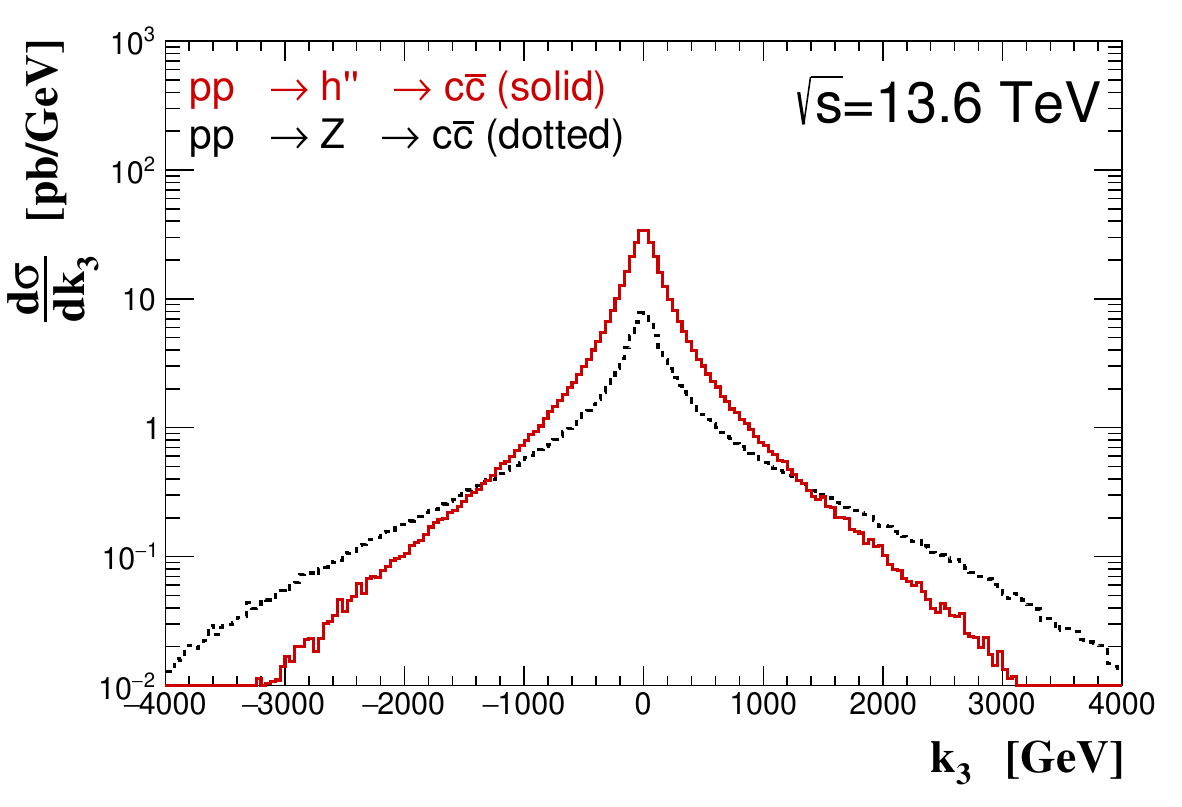}
  \caption{}
  \label{fig:k3-a}
  \end{subfigure}        
  \begin{subfigure}{0.5\textwidth}
    \centering
    \includegraphics[width=1.\textwidth]{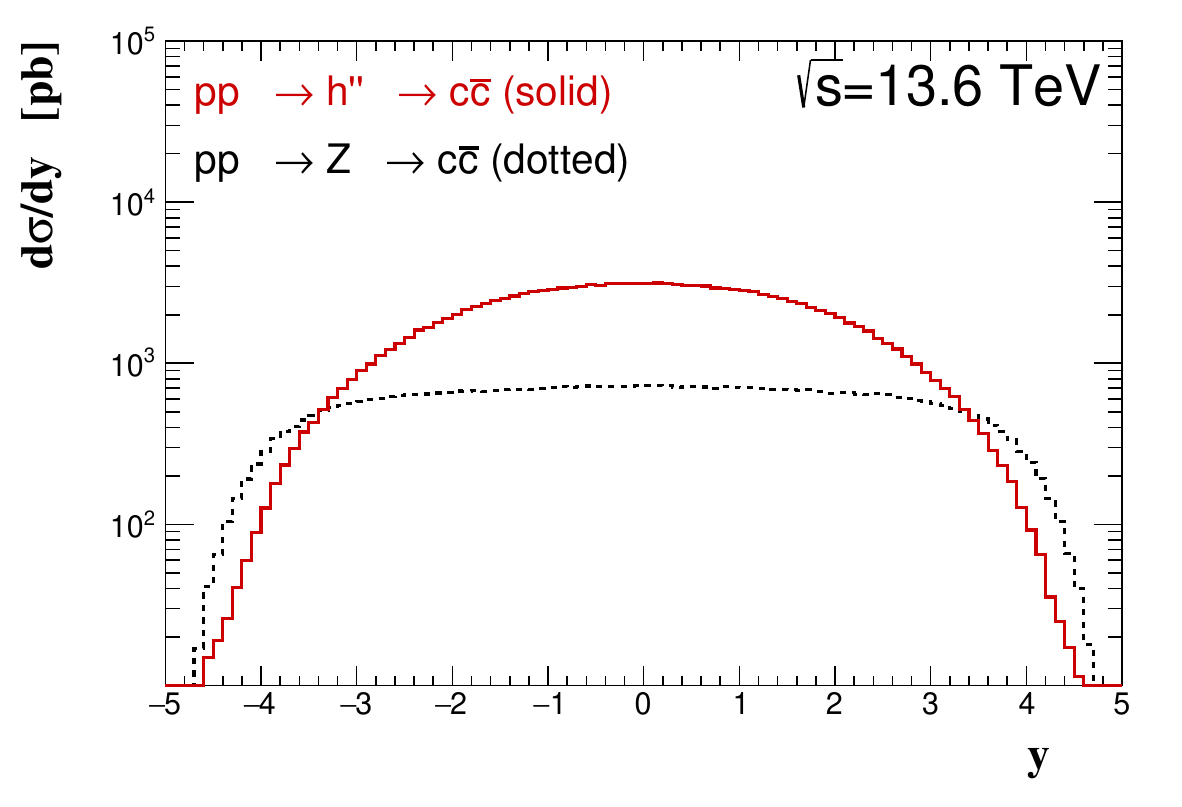}
  \caption{}
  \label{fig:k3-b}
  \end{subfigure}        
    \caption{Distribution of: (a) the longitudinal momentum $k_3$, and (b) the rapidity $y$ of $h''$ in \eqref{eq:2.1e} and of $Z$ in \eqref{eq:2.1i}.}
    \label{fig:k3}
\end{figure}

We find it interesting to compare the DY cross section \eqref{eq:2.1e} to the cross section for the production of the $Z$ boson with its subsequent decay to $c + \bar{c}$:
\begin{equation} \label{eq:2.1i}
p + p \to (Z \to c +\bar{c}) + X.
\end{equation}
From \cite{ParticleDataGroup:2024cfk} we have
\begin{align} \label{eq:2.1j}
m_{Z} &= 91.1880 \pm 0.0020 \; \text{GeV,} \nonumber \\
\Gamma_{Z} &= 2.4955 \pm 0.0023 \; \text{GeV,} \nonumber \\ 
B(Z \to c + \bar c) &= (12.03\pm 0.21) \times 10^{-2}.
\end{align}
The mass $m_Z$ is close to the $m_{h''}$ mass \eqref{eq:1.4}. With the same methods as we calculated \eqref{eq:2.1e} we find
\begin{equation} \label{eq:2.1k}
\left. \sigma_{\mathrm{tot}}(p + p \to Z + X) \right|_{\text{DY}} = 45752.7 ~\text{pb.} 
\end{equation} 
\begin{equation} \label{eq:2.1l}
\left. \sigma_{\mathrm{tot}}(p + p \to Z + X) \right|_{\text{DY}} \times B(Z \to c +\bar c) = 5490.3 ~\text{pb.} 
\end{equation} 
Thus, the cross section for the reaction \eqref{eq:2.1i} is much smaller than that of \eqref{eq:2.1e}.
Note that $h''$ boson of mass \eqref{eq:1.4} is predicted to decay to nearly 100\% to $c\bar c$; see Fig.~8 of \cite{Maniatis:2009vp}. Here we show the distributions in the longitudinal momentum $k_3$ and rapidity $y$ of the $h''$ in \eqref{eq:2.1e} and the $Z$ in \eqref{eq:2.1i}; see Figs.~\ref{fig:k3-a} and ~\ref{fig:k3-b}, respectively .

These distributions are not directly observable but have to be reconstructed
from the distribution of the $c\bar{c}$ pairs.
The invariant mass of the pairs $M_{c\bar{c}}$ is a superposition of two
Breit--Wigner distributions around the nominal values of the masses of
$h''$ and $Z$.

\begin{figure}[h]
  \begin{subfigure}{0.5\textwidth}
    \centering
    \includegraphics[width=1.\textwidth]{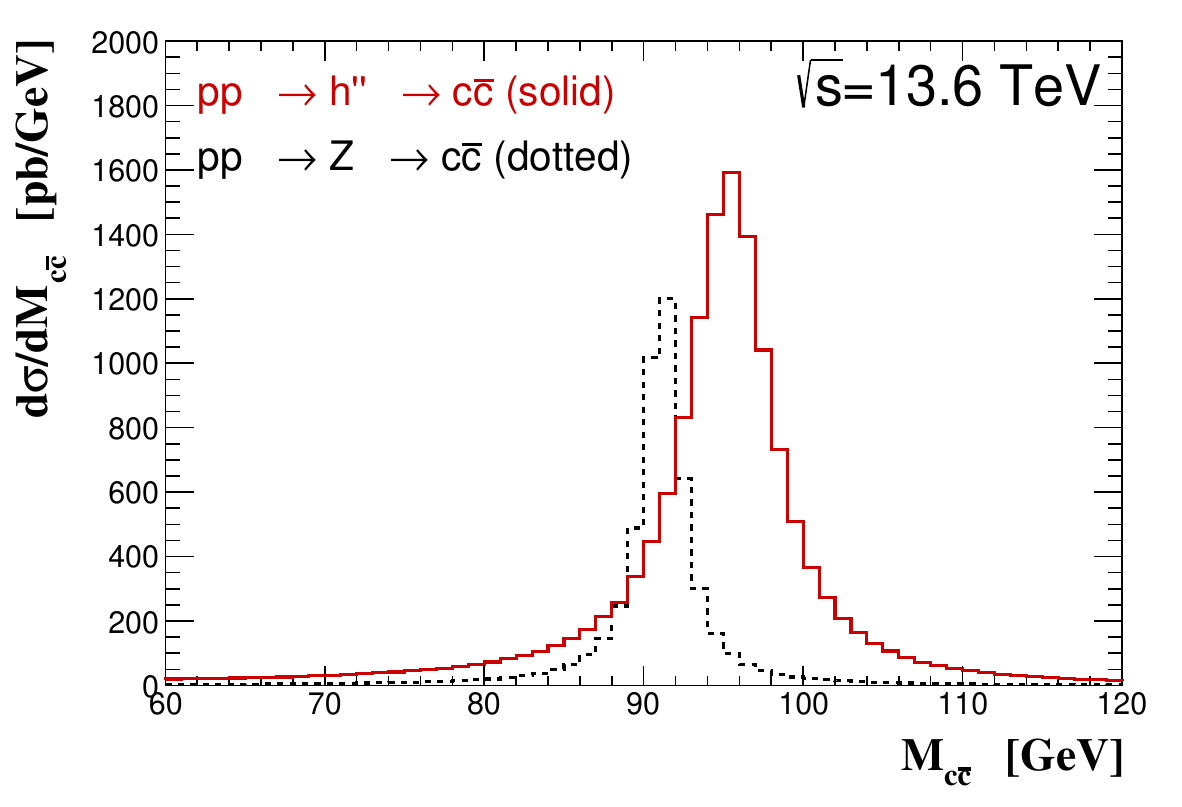}
  \caption{}
  \label{fig:Mccbar-a}
  \end{subfigure}        
  \begin{subfigure}{0.5\textwidth}
    \centering
    \includegraphics[width=1.\textwidth]{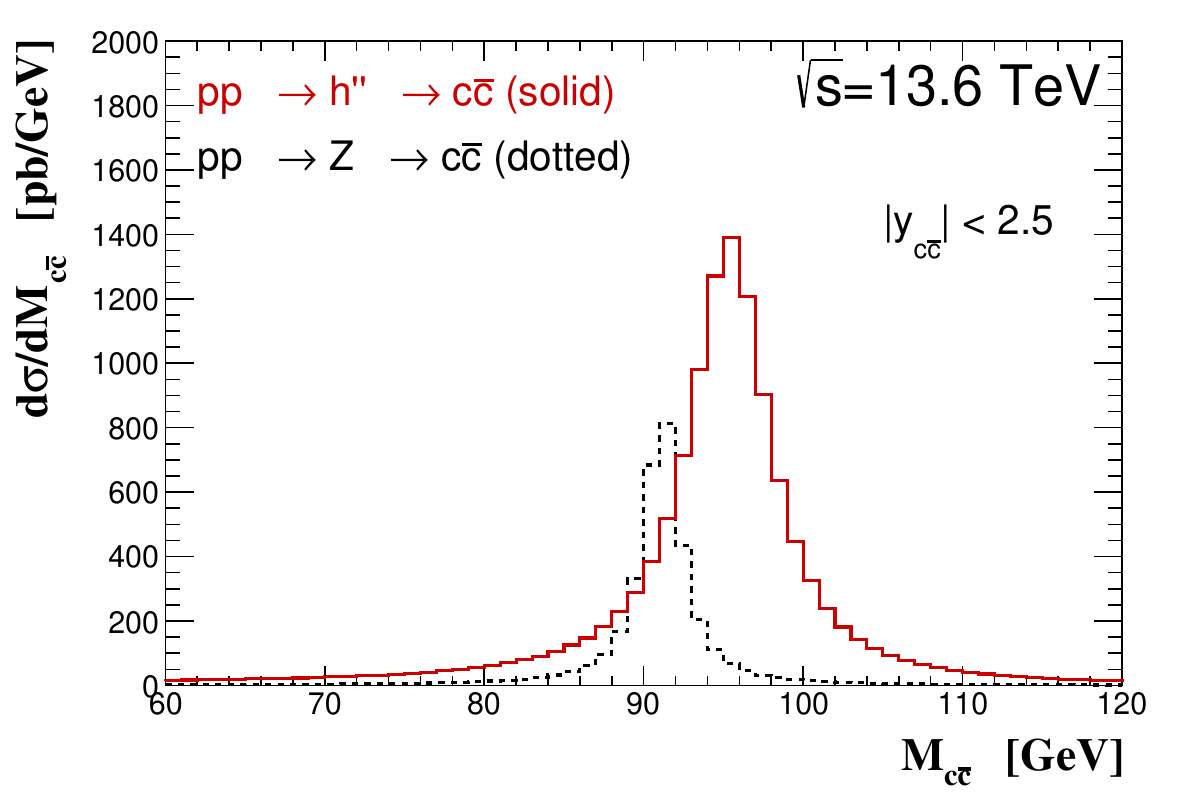}
  \caption{}
  \label{fig:Mccbar-b}
  \end{subfigure}        
    \caption{The distributions of \(M_{c\bar c}\) for \(h''\) production in Eq.~\eqref{eq:2.1e} and for \(Z\) production in Eq.~\eqref{eq:2.1i}, obtained (a) in the full phase space and (b) with the requirement \(|y| < 2.5\).}
    \label{fig:Mccbar}
\end{figure}

In Fig.~\ref{fig:Mccbar} we show the distribution of $M_{c\bar{c}}$. Two peaks for the $h''$ and the $Z$ are clearly visible.
Fig.~\ref{fig:Mccbar-a} corresponds to the calculations for the full phase-space, while Fig.~\ref{fig:Mccbar-b} represents results obtained within the $|y| < 2.5$ requirement (which is equivalent to $|k_{3}| \lesssim 650$ GeV). The corresponding QCD $c\bar c$-background in the invariant-mass region around the two signal peaks is approximately two orders of magnitude larger and therefore, it is not shown in the figure. Note that all these distributions are calculated at the parton level.
In the real world the momenta of $c$ and $\bar{c}$ have to be obtained
from the corresponding jet momenta.
This certainly will introduce a smearing of the distributions.

\begin{figure}[h]
  \begin{subfigure}{0.5\textwidth}
    \centering
    \includegraphics[width=1.\textwidth]{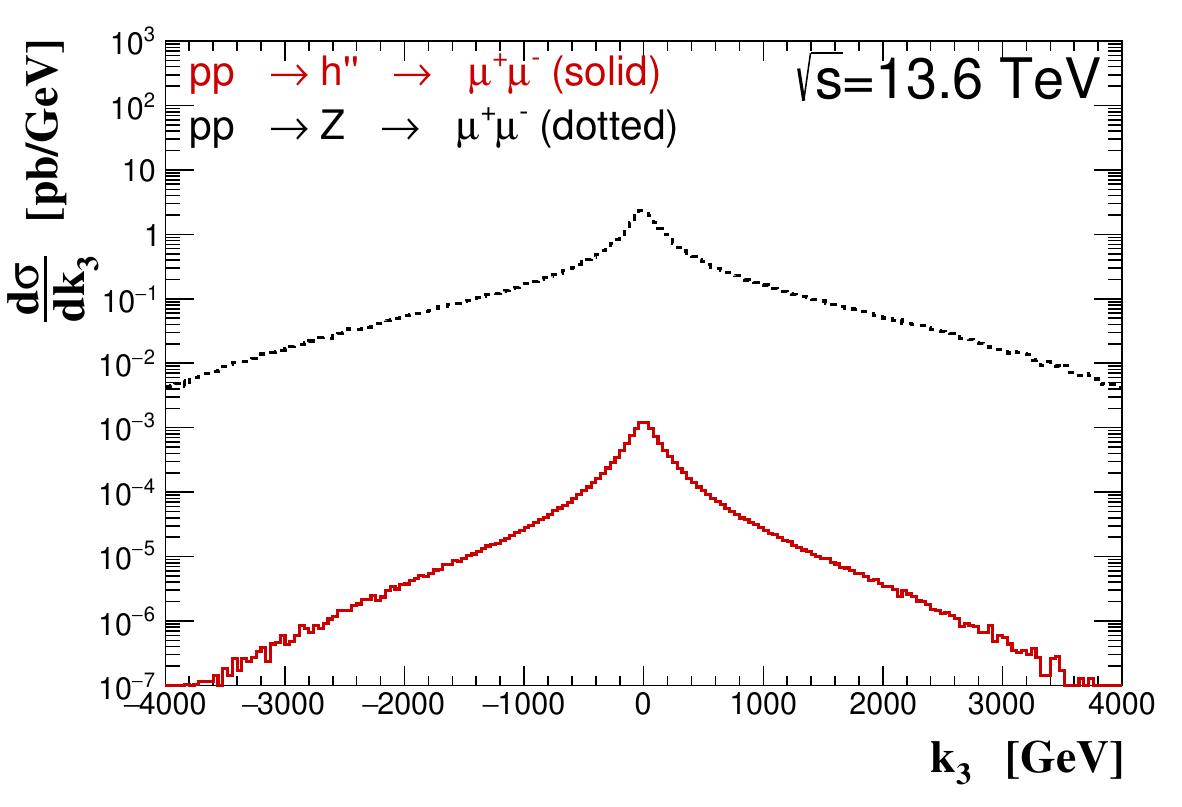}
  \caption{}
  \label{fig:k3-a-mupmum}
  \end{subfigure}        
  \begin{subfigure}{0.5\textwidth}
    \centering
    \includegraphics[width=1.\textwidth]{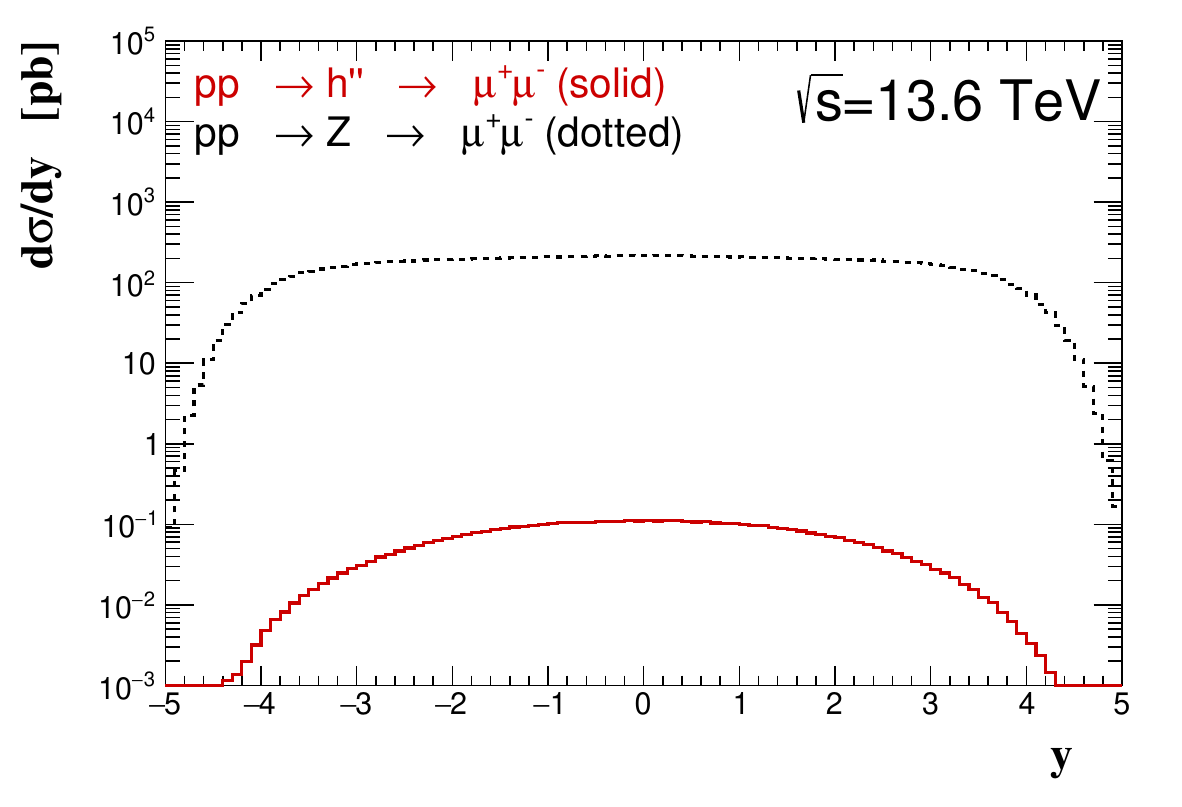}
  \caption{}
  \label{fig:k3-b-mupmum}
  \end{subfigure}        
    \caption{Distribution of: (a) the longitudinal momentum $k_3$, and (b) the rapidity $y$ of $h''$ in \eqref{B.17} and of $Z$ in \eqref{B.18}.}
    \label{fig:k3-mupmum}
\end{figure}

\begin{figure}[h]
    \centering
    \includegraphics[width=0.5\textwidth]{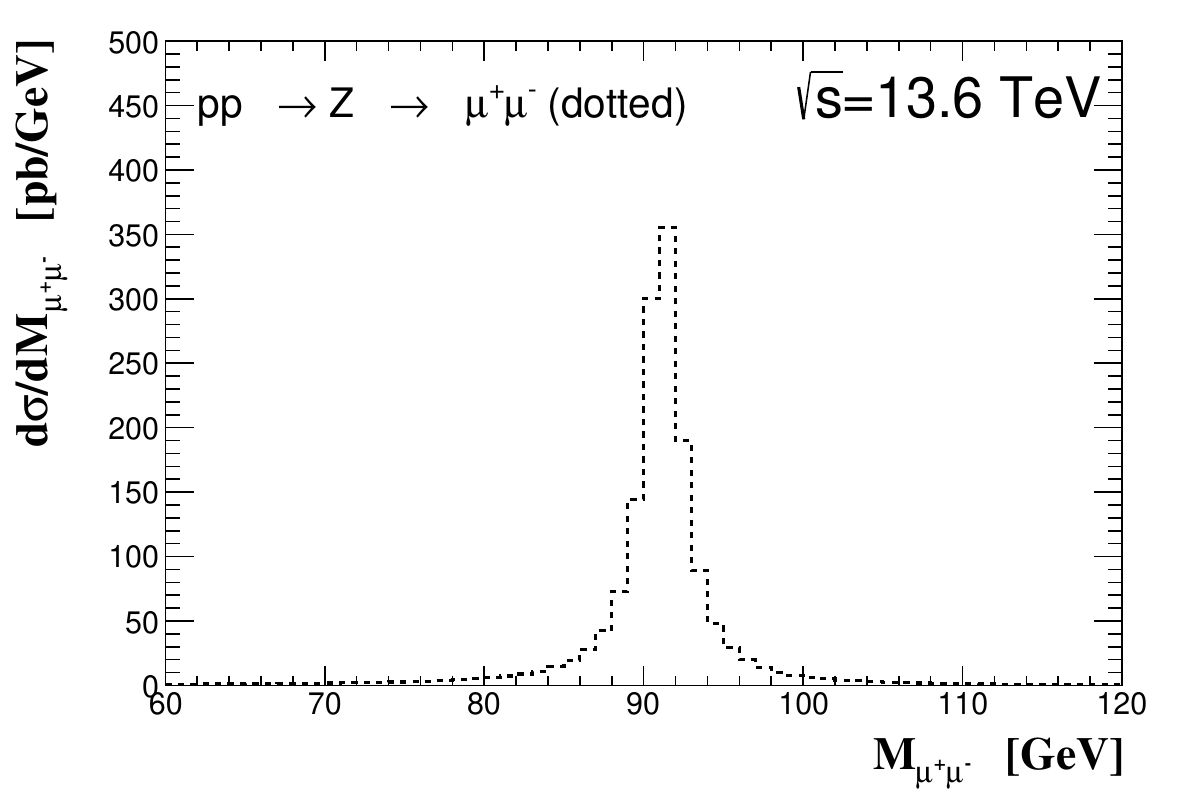}
    \caption{The distributions of \(M_{\mu^{+}\mu^{-}}\) for Z production in Eq.~\eqref{B.18} obtained in the full phase space.}
    \label{fig:Mmupmum}
\end{figure}

Finally we compare the distributions for the reactions
\begin{equation}
\label{B.17}
p + p \;\to\; (h'' \to \mu^+ \mu^-) + X ,
\tag{B.17}
\end{equation}
and
\begin{equation}
\label{B.18}
p + p \;\to\; (Z \to \mu^+ \mu^-) + X .
\tag{B.18}
\end{equation}
From Eq.~(3.2) and Table~2 of~\cite{Maniatis:2009vp} we get for the $h''$ mass ($1.5$)
the branching ratio of $h'' \to \mu^+ \mu^-$ as
\begin{equation}
\label{B.19}
B(h'' \to \mu^+ \mu^-) = 3.44 \times 10^{-5}.
\tag{B.19}
\end{equation}
The branching ratio $Z \to \mu^+ \mu^-$ is according to~\cite{ParticleDataGroup:2024cfk}
\begin{equation}
\label{B.20}
B(Z \to \mu^+ \mu^-) = (3.3662 \pm 0.0066)\times 10^{-2}.
\tag{B.20}
\end{equation}
In Fig.~\ref{fig:k3-mupmum}, we present the analogue of Fig.~\ref{fig:k3} for the reactions~\eqref{B.17} and~\eqref{B.18}.
We observe that, in the $\mu^+\mu^-$ channel, the contribution from $Z$ production exceeds that from $h''$ by approximately three orders of magnitude, reflecting the relative difference between the corresponding branching fractions given in ~\eqref{B.19} and~\eqref{B.20}.
Consequently, in this channel the invariant-mass spectrum is entirely dominated by $Z$-boson production (see Fig.~\ref{fig:Mmupmum}).  

\begin{figure}[h]
    \centering
    \includegraphics[width=0.5\textwidth]{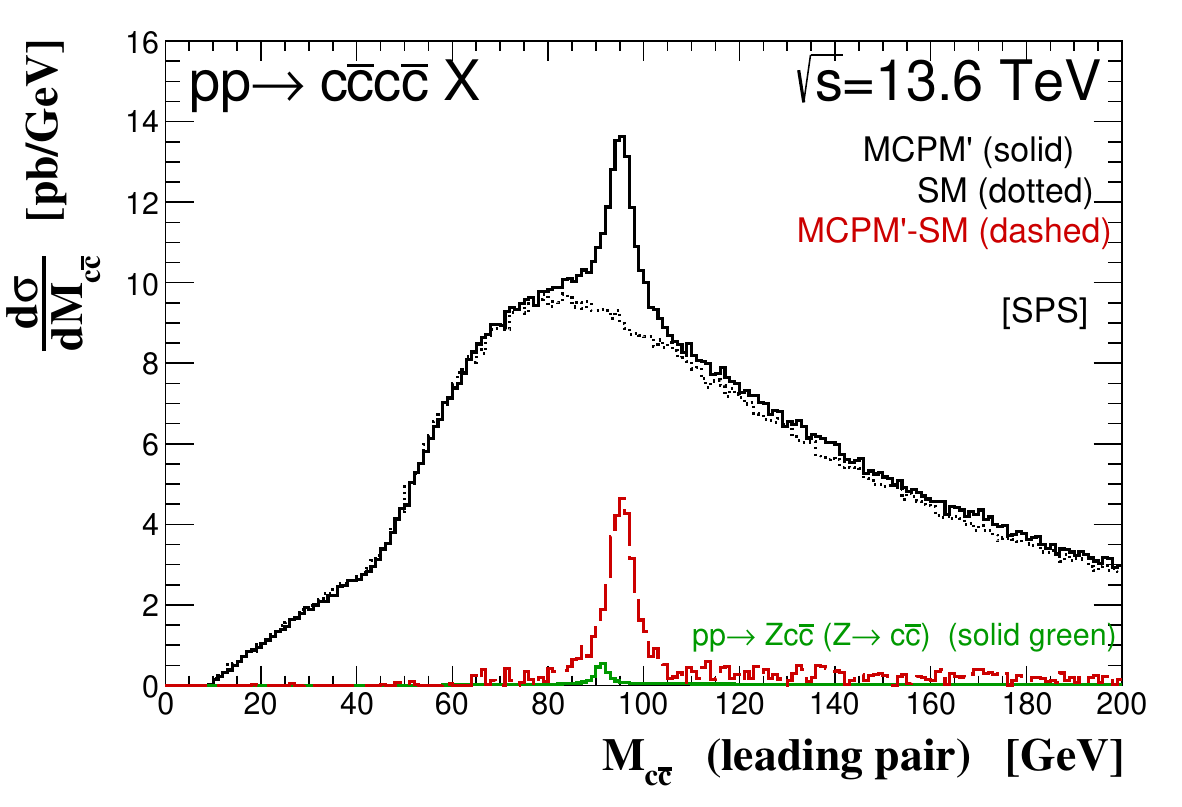}
    \caption{The same as in Fig.~\ref{fig:1-rafal-b} but here extra contribution from the reaction \eqref{B.21} is also shown (green solid histogram; color online).}
    \label{fig:Zccbar}
\end{figure}

In Fig.~\ref{fig:Zccbar} we compare associated $h''$ production~\eqref{eq:1.5}
and the corresponding associated $Z$ production
\begin{equation}
\label{B.21}
p + p \;\to\; (Z \to c + \bar{c}) + c + \bar{c} + X .
\tag{B.21}
\end{equation}

We see that with the cuts \eqref{eq:2.1}
the contribution of~\eqref{B.21} is much smaller than that of
our signal process~\eqref{eq:1.5}.

\section{Double \( h'' \) production}
\label{appB}

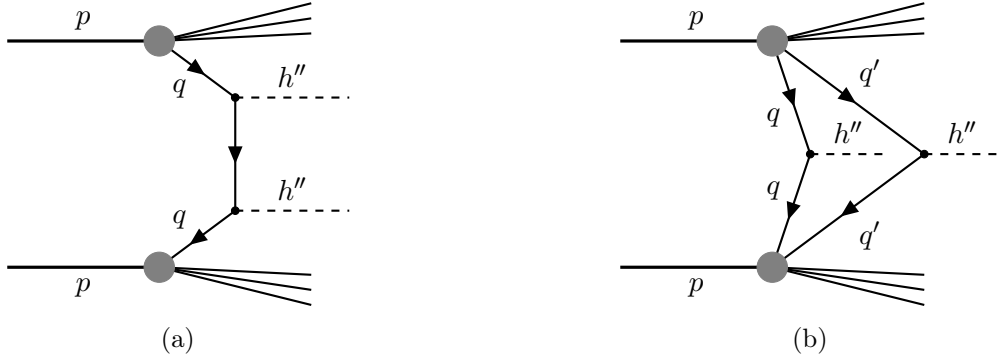
\begin{figure}[h!]
\centering

% Diagram (a) — single parton process
\begin{subfigure}[b]{0.45\textwidth}
\centering
\begin{tikzpicture}
  \begin{feynman}
  %place first vertices
    \vertex (i1) at (0,1.5);
    \vertex (i2) at (0,-1.5);
     \vertex (a) at (2.0, 1.5) ;
     \vertex (b) at (2.0, -1.5) ;
     \vertex (c1) at (3, 0.75);  
     \vertex (c2) at (3, -0.75);

     \vertex (h) at (4,0);

     \vertex (f1) at (4.5, 0.75) ;
     \vertex (f2) at (4.5, -0.75) ;
     \vertex (X1) at (4, 1.8) ;
     \vertex (X1a) at (4, 2.0) ;
     \vertex (X1b) at (4, 1.6) ;
     \vertex (X2) at (4, -1.8) ;
     \vertex (X2a) at (4, -1.6) ;
     \vertex (X2b) at (4, -2.0) ;
    \diagram* {
      (i1) -- [plain, very thick, edge label=\(p\)] (a),
      (i2) -- [plain, very thick,edge label'=\(p\)] (b),
      (a) -- [fermion, thick, edge label'=\(q\)] (c1),
      (c2) -- [fermion, thick, edge label'=\(q\)] (b),
      (c1) -- [fermion, thick, edge label=\(\)] (c2),

      (c2) -- [scalar, thick, edge label=\(h''\)] (f2),

      (c1) -- [scalar, thick, edge label=\(h''\)] (f1),

      (a) -- [plain, thick] (X1),
      (a) -- [plain, thick] (X1a),
      (a) -- [plain, thick] (X1b),
      (b) -- [plain, thick] (X2),      
      (b) -- [plain, thick] (X2a),      
      (b) -- [plain, thick] (X2b),      
    };
   	\draw [fill, gray] (a) circle (.2);
 	\draw [fill, gray] (b) circle (.2);
 	\draw [fill] (c1) circle (.05);
 	 	\draw [fill] (c2) circle (.05);
  \end{feynman}
\end{tikzpicture}
\caption{}
\label{fig:B1-a}
\end{subfigure}
\hfill
% Diagram (b) — double parton scattering
\begin{subfigure}[b]{0.45\textwidth}
\centering
\begin{tikzpicture}
  \begin{feynman}
  %place first vertices
    \vertex (i1) at (0,1.5);
    \vertex (i2) at (0,-1.5);
     \vertex (a) at (2.0, 1.5) ;
     \vertex (b) at (2.0, -1.5) ;
     \vertex (c) at (2.5, 0.0);  
     \vertex (d) at (4., 0.0);

     \vertex (h) at (3.5,0);
          \vertex (i) at (5.,0);

     \vertex (X1) at (4, 1.8) ;
     \vertex (X1a) at (4, 2.0) ;
     \vertex (X1b) at (4, 1.6) ;
     \vertex (X2) at (4, -1.8) ;
     \vertex (X2a) at (4, -1.6) ;
     \vertex (X2b) at (4, -2.0) ;
    \diagram* {
      (i1) -- [plain, very thick, edge label=\(p\)] (a),
      (i2) -- [plain, very thick,edge label'=\(p\)] (b),
      (a) -- [fermion, thick, edge label'=\(q\)] (c),
      (c) -- [fermion, thick, edge label'=\(q\)] (b),

      (a) -- [fermion, thick, edge label=\(q'\)] (d),
      (d) -- [fermion, thick, edge label=\(q'\)] (b),

      (c) -- [scalar, thick, edge label=\(h''\)] (h),

      (d) -- [scalar, thick, edge label=\(h''\)] (i),

      (a) -- [plain, thick] (X1),
      (a) -- [plain, thick] (X1a),
      (a) -- [plain, thick] (X1b),
      (b) -- [plain, thick] (X2),      
      (b) -- [plain, thick] (X2a),      
      (b) -- [plain, thick] (X2b),      
    };
   	\draw [fill, gray] (a) circle (.2);
 	\draw [fill, gray] (b) circle (.2);
 	\draw [fill] (c) circle (.05);
 	 	\draw [fill] (d) circle (.05);
  \end{feynman}
\end{tikzpicture}
\caption{}
\label{fig:B1-b}
\end{subfigure}

\caption{
Diagrams for double \( h'' \) production \eqref{eq:B1}, via the single-parton process (a) and the double-parton process (b). Here \( q, q' \in \{c, s\} \) are the quarks involved.
}
\label{fig:B1}
\end{figure}

Here we study the reaction
\begin{equation}\label{eq:B1}
    p(p_{a}) + p(p_{b}) \rightarrow h''(k_1) + h''(k_2) + X,
    \tag{C.1}
\end{equation}
that is, double \( h'' \) production. This reaction can be due to a single-parton process (Fig.~\ref{fig:B1-a}) and a double-parton process (Fig.~\ref{fig:B1-b}).

The $h''$ decays to nearly 100\% to $c\bar{c}$ for a mass value $m_{h''} = 95.4$~GeV; see Fig.~8 of~\cite{Maniatis:2009vp}. Therefore, the reaction~\eqref{eq:B1} will result in the final state $c\bar{c}c\bar{c}X$ and contribute to the general final states in \eqref{eq:2.1a} and \eqref{eq:2.4}.

In our calculation for \eqref{eq:B1}, we shall restrict ourselves to the diagrams of Fig.~\ref{fig:B1-a} with charm-quark fusion ($q=c$), since the coupling of $h''$ to $c$ quarks is much bigger than to $s$ quarks; see (A.81) of~\cite{Maniatis:2023aww}. We calculate first the reaction
\begin{equation}\label{eq:B2}
    c(q_1) + \bar{c}(q_2) \rightarrow h''(k_1) + h''(k_2). \tag{C.2}
\end{equation}
\noindent We set
\begin{align}
    \hat{s} &= (q_1 + q_2)^2 = (k_1 + k_2)^2, \nonumber \\
    \hat{t} &= (q_1 - k_1)^2 = (q_2 - k_2)^2, \nonumber \\
    \hat{u} &= (q_1 - k_2)^2 = (q_2 - k_1)^2. \tag{C.3}
\end{align}

\noindent A straightforward calculation gives for the transition rate of \eqref{eq:B2} with $V$ the normalization volume
\begin{align}
  d\Gamma\left[c(q_1) + \bar{c}(q_2) \rightarrow h''(k_1) + h''(k_2)\right] 
  &= \frac{1}{V} \cdot \frac{1}{2q_1^0 \, 2q_2^0} \cdot (2\pi)^4 \delta^{(4)}(k_1 + k_2 - q_1 - q_2) \nonumber \\
  &\quad \times \frac{d^3 k_1}{(2\pi)^3 2k_1^0} \cdot \frac{d^3 k_2}{(2\pi)^3 2k_2^0} \cdot F(q_1, q_2, k_1, k_2), \tag{C.4}
\end{align}
\noindent where
\begin{equation}
  F(q_1, q_2, k_1, k_2) = \frac{1}{6} \left( \frac{m_t}{v_0} \right)^4
  \left( \frac{1}{\hat{t} - m_c^2} - \frac{1}{\hat{u} - m_c^2} \right)^2
  \left[ \hat{t} \hat{u} + \hat{s} m_c^2 - (m_{h''}^2 + m_c^2)^2 \right]. \tag{C.5}
\end{equation}
\noindent Now we use this result as input for a standard Drell--Yan type calculation of the diagram of Fig.~\ref{fig:B1-a}; see, for instance, chapter 18.5 of~\cite{Nachtmann:1990ta}. The result reads as follows:
\begin{align}\label{eq:B6}
\, & \mathrm{d} \sigma\Big[ p(p_a) + p(p_b) \rightarrow h''(k_1) + h''(k_2) + X \Big]
=  \frac{1}{2} \sqrt{ \frac{s}{s - 4 m_p^2} }
\frac{ \mathrm{d}^3 k_1 }{ (2\pi)^3 2 k_1^0 }
\frac{ \mathrm{d}^3 k_2 }{ (2\pi)^3 2 k_2^0 } \notag \\
& \times \Bigg\{ \int_0^1 \!\mathrm{d}x_1 \! \int_0^1 \!\mathrm{d}x_2 \,
N_c^p(x_1) N_{\bar{c}}^p(x_2) \, (2\pi)^4 \delta^{(4)}(k_1 + k_2 - x_1 p_a - x_2 p_b) 
\frac{1}{\hat{s}} F(x_1 p_a, x_2 p_b, k_1, k_2) \notag \\
& \quad + \int_0^1 \!\mathrm{d}x_1 \! \int_0^1 \!\mathrm{d}x_2 \,
N_{\bar{c}}^p(x_1) N_c^p(x_2) \, (2\pi)^4 \delta^{(4)}(k_1 + k_2 - x_1 p_a - x_2 p_b)
\frac{1}{\hat{s}} F(x_2 p_b, x_1 p_a, k_1, k_2) \Bigg\}. \tag{C.6}
\end{align}
\noindent Here $N_c^p(x)$, $N_{\bar{c}}^p(x)$ are the $c$ and $\bar{c}$ quark distribution functions of the proton and
\begin{equation}
s = (p_a + p_b)^2 \tag{C.7}
\end{equation}
\noindent is the total c.m. energy squared. The total cross section is obtained by integrating  in \eqref{eq:B6} over $k_1$ and $k_2$
taking the statistical factor $1/2$ for the two identical $h''$ bosons into account. This gives for $\sqrt{s}=13.6$ TeV
\begin{equation}
\sigma_{\text{tot}} \left( p + p \rightarrow h'' + h'' + X \right) = 1.11~\text{pb.} \tag{C.8}
\end{equation}
Compared to the cross section which we found in \eqref{eq:2.1d} this is very small and can be neglected. For the double scattering process of Fig.~\ref{fig:B1-b}
we expect also a small result compared to \eqref{eq:2.1d}.

%\noindent is     publisher = "Dover",    
%    year = "1965"
%    address =   "New York",obtained by integrating in \eqref{eq:B6} over $k_1$ and $k_2$ taking the statistical factor $1/2$ for the two identical $h''$ bosons into account.

\bibliographystyle{JHEP}
\bibliography{references}
\end{document}